\documentclass{aa}  

\usepackage{ulem}
\usepackage{graphicx}
\usepackage{txfonts}
\usepackage[breaklinks=true]{hyperref} 
\usepackage{natbib,twoopt}
\usepackage[dvipsnames]{xcolor}

\bibpunct{(}{)}{;}{a}{}{,} 
\makeatletter
\newcommandtwoopt{\citeads}[3][][]{\href{http://adsabs.harvard.edu/abs/#3}%
{\def\hyper@linkstart##1##2{}%
\let\hyper@linkend\@empty\citealp[#1][#2]{#3}}}
\newcommandtwoopt{\citepads}[3][][]{\href{http://adsabs.harvard.edu/abs/#3}%
{\def\hyper@linkstart##1##2{}%
\let\hyper@linkend\@empty\citep[#1][#2]{#3}}}
\newcommandtwoopt{\citetads}[3][][]{\href{http://adsabs.harvard.edu/abs/#3}%
{\def\hyper@linkstart##1##2{}%
\let\hyper@linkend\@empty\citet[#1][#2]{#3}}}
\newcommandtwoopt{\citeyearads}[3][][]%
{\href{http://adsabs.harvard.edu/abs/#3}
{\def\hyper@linkstart##1##2{}%
\let\hyper@linkend\@empty\citeyear[#1][#2]{#3}}}
\makeatother

\newcommand\gaia{\textit{Gaia}\xspace}
\newcommand\gdrtwo{\gaia~DR2\xspace}
\newcommand\gdrthree{\gaia~EDR3\xspace}
\newcommand\hip{\textit{Hipparcos\xspace}}
\newcommand{\gbp}{{$G_\mathrm{BP}$}\xspace}
\newcommand{\grp}{{$G_\mathrm{RP}$}\xspace}
\newcommand{\NOBJ}{540}   
\newcommand{\NSYS}{339}    
\newcommand{\NSTARS}{375}  
\newcommand{\NBDS}{85}      

\newcommand{\NPLANETS}{77}  
\newcommand{\NWD}{20}       
\newcommand{\NSS}{246}       
\newcommand{\NBS}{69}       
\newcommand{\NTS}{19}       
\newcommand{\orcit}[1]{\protect\href{https://orcid.org/#1}{\protect\includegraphics[width=8pt]{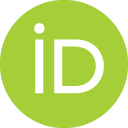}}}

\begin{document}

   \title{The 10 parsec sample in the \gaia era} 
 
    \author{C.~Reyl\'e\orcit{0000-0003-2258-2403}\inst{1}
    \and K.~Jardine\orcit{0000-0001-6068-2734}\inst{2}
    \and P.~Fouqu\'e\orcit{0000-0002-1436-7351}\inst{3}
    \and J.\,A.~Caballero\orcit{0000-0002-7349-1387}\inst{4}
    \and R.\,L.~Smart\orcit{0000-0002-4424-4766}\inst{5}
    \and A.~Sozzetti\orcit{0000-0002-7504-365X}\inst{5}
    }

    \institute{Institut UTINAM, CNRS UMR6213, Univ. Bourgogne Franche-Comt\'e, OSU THETA Franche-Comt\'e-Bourgogne, Observatoire de Besan\c con, BP 1615, 25010 Besan\c con Cedex, France\\
    \email{celine.reyle@obs-besancon.fr}
    \and
    Radagast Solutions, Simon Vestdijkpad 24, 2321 WD Leiden, Netherlands
    \and
    IRAP, Universit\'e de Toulouse, CNRS, 14 av. E. Belin, 31400 Toulouse, France
    \and 
    Centro de Astrobiolog\'ia (CSIC-INTA), ESAC, Camino bajo del castillo 
s/n, 28692 Villanueva de la Ca\~nada, Madrid, Spain
    \and
    INAF - Osservatorio Astrofisico di Torino, via Osservatorio 20, 10025 
Pino Torinese (TO), Italy       
    }

   \date{Received 2 April 2021 / Accepted dd Month 2021}

 
  \abstract
   {The nearest stars provide a fundamental constraint for our understanding of stellar physics and the Galaxy. The nearby sample serves as an anchor where all objects can be seen and understood with precise data. This work is triggered by the most recent data release of the astrometric space mission \gaia and uses its unprecedented high precision parallax measurements to review the census of objects within 10\,pc. }
   {The first aim of this work was to compile all stars and brown dwarfs within 10\,pc observable by \gaia and compare it with the \gaia Catalogue of Nearby Stars as a quality assurance test. We complement the list to get a full 10\,pc census, including bright stars, brown dwarfs, and exoplanets.}
   {We started our compilation from a query on all objects with a parallax larger than 100\,mas  using the Set of Identifications, Measurements, and Bibliography for Astronomical Data database (SIMBAD). We completed the census by adding companions, brown dwarfs with recent parallax measurements not in SIMBAD yet, and vetted exoplanets. The compilation combines astrometry and photometry from the recent \gaia Early Data Release 3 with literature magnitudes, spectral types, and line-of-sight velocities.}
   {We give a description of the astrophysical content of the 10\,pc sample. We find a multiplicity frequency of around 27\%. Among the stars and brown dwarfs, we estimate that around 61\% are M stars and more than half 
of the M stars are within the range from M3.0\,V to M5.0\,V. We give an overview of the brown dwarfs and exoplanets that should be detected in the next 
\gaia data releases along with  future developments.}
   {We provide a catalogue of \NOBJ\ stars, brown dwarfs, and exoplanets in \NSYS\ systems, within 10\,pc from the Sun. This list is as volume-complete as possible from current knowledge and it provides benchmark stars that can be used, for instance, to define calibration samples and to test the quality of the forthcoming \gaia releases. It also has a strong outreach potential.}


    \maketitle
%

\section{Introduction}

Determining the number of stars in the sky must have been in the minds of many people since the dawn of humanity. Ancient astronomers, such as Timocharis  of Alexandria and Hipparchus of Nicaea, started to count 
and catalogue stars visible to the naked eye and built the first magnitude-limited catalogues. Modern astronomers prefer using volume-limited catalogues, with different maximum distance limits \citep[e.g. ][]{1937AJ.....46...95J,1961AJ.....66..528V,2004AJ....128..463R,2015yCat.5035....0G,2018AJ....155..265H}, because any magnitude-limited sample is biased against intrinsically faint (and single) objects \citep{1925MeLuF.106....1M}. 
A good example concerns the low-mass stars (M $\lesssim 0.5$ M$_\odot$). We now know that they constitute an important part of the objects in our Galaxy, while even the brightest of them (\object{AX~Mic}) is invisible to the naked eye. 
Astronomers such as Max Wolf and Frank~E. Ross catalogued stars with a large proper motion to try discovering faint, but nearby stars \citep{1917AN....204..345W,1926AJ.....36..124R}. 
Willem~J. Luyten produced many catalogues \citep[e.g.][]{1979nlcs.book.....L} with different cuts in proper motion and corresponding names (i.e. LFT for five-tenths of an arcsec limit, LTT for two-tenths, and LHS for half a second).

Ever since the first stellar parallaxes were measured 
\citep[][see \citealt{2020AN....341..860R} for a review]{1838MNRAS...4..152B,1839MNRAS...4..168H,1840AN.....17..177V},
astronomers have tried to map out our nearest neighbours.
Individual measurements have been followed by increasingly larger trigonometric parallax catalogues across the 20th century, providing fundamental data for volume-limited catalogues: 
72 stars by \cite{1904MNRAS..64..570N},
1870 stars in the First General Catalogue of trigonometric parallaxes computed by Frank Schlesinger and edited by the Yale University Observatory in 1924, 
6399 stars in the Yale Parallax Catalogue \citep{1963gcts.book.....J}, 
7879 stars in the Fourth General Catalogue of trigonometric parallaxes \citep{1995yCat.1174....0V}, etc. 
The end of the 20th century was marked by the first astrometric space mission, \hip\ \citep[{\it HIgh Precision PARallax COllecting Satellite},][]{1997A&A...323L..49P} providing a catalogue of 117\,955 relatively bright 
stars ($V \lesssim 12.4$\,mag). 
The second astrometric space mission, \gaia \citep{2016A&A...595A...1G}, provides another dramatic increase, both qualitatively and quantitatively, with all sky parallax measurements for about 1.5 billion objects. It offers the means to complete volume-limited samples with larger distance limits. The \gaia Catalogue of Nearby Stars (hereafter GCNS), based on the \gaia Early Data Release 3 \citep[hereafter \gdrthree,][]{2021A&A...649A...1G}, pushes the limit to 100\,pc \citep[][hereafter GSS21]{2021A&A...649A...6G}.

Our first motivation to compile the 10\,pc sample was to use it as a quality assurance test of the GCNS and, therefore, to verify the \gdrthree before its publication. Such information could be derived from the work of the REsearch Consortium On Nearby Stars (RECONS\footnote{\url{http://www.recons.org}}), who have focused on the detection and characterisation of nearby star systems for several decades. They have published their results in a large series of papers. Part of them, as well as statistics, are listed in the RECONS webpage. Yet the compilation of a 10\,pc catalogue from this resource is not straightforward. 

According to RECONS, the 10\,pc sample as of 12~April~2018 included 462 objects in 317 systems \citep{2018AJ....155..265H}. 
The publication of the second \gaia data release  \citep[hereafter \gdrtwo,][]{2018A&A...616A...1G} a few days later provided new, more precise parallaxes that moved some objects inside or outside of the 10\,pc limit. It 
also provided individual parallaxes for components in systems.
It resulted in 418 objects in 305 systems, with eight systems added by \gaia \citep{2019AAS...23325932H}. 
 
However, \gdrtwo also contained a large number of spurious objects: A simple cut at a parallax $\geq100$\,mas  
in \gdrtwo returns 1722 objects.
Using a random forest classifier to disentangle between good and bad astrometric solutions, GSS21 found that 15 sources, although classified as good from the classifier, lie closer than \object{Proxima Centauri} (see their Fig.~12). 
On the contrary, with one more year of observations, better reduction and 
calibration procedures of the \gdrthree, a parallax $\geq100$\,mas selection returns only 315 objects with a very high and improved precision, of which three had an obvious spurious solution and were rejected from the random forest classifier. 
The GCNS essentially offered a reasonably clean sample, with no new discoveries, but with higher precision astrometry and the first individual parallaxes for five objects in systems.

In the framework of the GCNS, the 10\,pc compilation was not exhaustive but restricted to objects that should have been visible to \gaia, given its magnitude limits at the bright ($G \simeq2.5$\,mag) and faint ($G \simeq$ 21\,mag) ends.
In the present work, we give a more complete census of the 10\,pc sample using our knowledge of the nearby objects, including stars and their companions, brown dwarfs, and planets. 
For many of the objects, it also benefits from the exquisite parallaxes obtained from the last data release based on 34 months of operation of  \gaia. 
This list will be used for further \gaia quality assurance. It includes all objects (i.e. planets and unresolved components) as separate entries as many of these will be detected in future \gaia releases. 
We also believe that it could be of general use to the community as it provides a complete list of benchmark and vetted objects and we are making it publicly available. For the foreseeable future, the 10\,pc sphere is the only volume that it will be possible to find and characterise all objects. Finally, the 10\,pc sample has significant outreach potential.

Following in the steps of Louise~F. Jenkins, who published a list of 127 stars with their known companions and gathered the knowledge at that time on the neighbours whose distance is less than 10\,pc from the Sun \citep{1937AJ.....46...95J}, we give here the current 
snapshot of the nearby sample within 10\,pc.
In Sect.~\ref{sec:cat}, we describe the catalogue and how we constructed it. In Sect.~\ref{sec:stats} we explore the content of the catalogue and give a few statistics. Sect.~\ref{sec:future} places the catalogue in the 
context of ongoing and future observational programmes that will impact the 
sample. Sect.~\ref{sec:future} also illustrates the potential of this catalogue for outreach. 
Finally, conclusions are given in Sect.~\ref{sec:ccl}.

\section{The 10\,pc catalogue}
\label{sec:cat}

\begin{table*}
\caption{Example of content of the 10\,pc catalogue (the first object, Proxima Centauri).} 
\label{tab:10pc} 
\small
\begin{tabular}{lllll}
\hline
\hline 
\noalign{\smallskip}
Parameter & Unit & Description & Example  \\
\noalign{\smallskip}
\hline
\noalign{\smallskip}
{\tt NB\_OBJ}                   &...            &Running number for object, ordered by increasing distance  &1\\
{\tt NB\_SYS}                   &...            &Running number for system, ordered by increasing distance  &1\\
{\tt SYSTEM\_NAME}              &...            &Name of the system                                                             &alf Cen\\
{\tt OBJ\_CAT}                  &...            &Star (*), LM (low mass star), BD (brown dwarf), WD (white dwarf), or Planet           &LM\\  
{\tt OBJ\_NAME}         &...            &Name of the object                                                             &Proxima Cen\\
{\tt RA}                           &  deg      & Right ascension (ICRS)   
                                                        &217.392321472009\\        
{\tt DEC}                         &  deg      & Declination (ICRS)        
                                                        &-62.6760751167667\\        
{\tt EPOCH}                    & a            & Epoch for position                                                                    &2016.0\\
{\tt PARALLAX}                  &  mas      & Trigonometric parallax      
                                                        &768.066539187357\\       
      
{\tt PARALLAX\_ERROR}       &  mas     & Parallax uncertainty             
                                                        &0.049872905\\      
{\tt PARALLAX\_BIBCODE} & ...           & Reference for the parallax                                                    & 2020yCat.1350....0G\\          
{\tt PMRA}                        &  mas\,a$^{-1}$  & Proper motion in right ascension                                            & -3781.74100826516\\             
{\tt PMRA\_ERROR}            &  mas\,a$^{-1}$  &  Proper motion uncertainty in right ascension                          & 0.031386077\\            
   
{\tt PMDEC}                    &  mas\,a$^{-1}$  & Proper motion in declination                                                    & 769.465014647862\\             
{\tt PMDEC\_ERROR}        &  mas\,a$^{-1}$  & Proper motion uncertainty in declination                              & 0.050524533\\ 
{\tt PM\_BIBCODE}               &...            &Reference for the proper motion                                          & 2020yCat.1350....0G\\     
{\tt RV}                    &  km\,s$^{-1}$   & Line-of-sight velocity    
                                                & -22.345\\             
{\tt RV\_ERROR}        &  km\,s$^{-1}$  & Line-of-sight velocity uncertainty                                  & 0.006\\ 
{\tt RV\_BIBCODE}               &...            &Reference for the line-of-sight velocity                                                & 
2014MNRAS.439.3094B\\     
{\tt SP\_TYPE}          &...            &Spectral type                                                                          &  M5.5\\
{\tt SP\_BIBCODE}               &...            &Reference for spectral type                                                    &   1995AJ....110.1838R\\
{\tt SP\_METHOD}                &...            &Method used to derive the spectral type (see text) &   Opt Spec\\
{\tt G\_CODE}   &...            &Reference code for the $G$ magnitude (see text) &3\\ 
{\tt G}             &mag        &\gaia $G$ band magnitude measured, given only if G\_CODE is 2 or 3                               & 8.984749\\
{\tt G\_ESTIMATE} &mag  &\gaia $G$ band magnitude estimated, given only if 
G\_CODE is 10 or 20                             & ... \\
{\tt GBP}               &mag    & \gaia $B_P$ band magnitude, given only if G\_CODE is 2 or 3                            & 11.373116\\
{\tt GRP} &mag  & \gaia $R_P$ band magnitude, given only if G\_CODE  is 2 or 3                            & 7.5685353\\
{\tt U}                                 &mag    & $U$ magnitude                                                         & 14.21\\
{\tt B}                                 &mag    & $B$ magnitude                                                         & 12.95\\
{\tt V}                                 &mag    & $V$ magnitude                                                         & 11.13\\
{\tt R}                                 &mag    & $R$ magnitude                                                         & 9.45\\
{\tt I}                                 &mag    & $I$ magnitude                                                         & 7.41\\
{\tt J}                                 &mag    & $J$ magnitude                                                         & 5.357\\
{\tt H}                                 &mag    & $H$ magnitude                                                         & 4.835\\
{\tt K}                                 &mag    & $K_\mathrm{s}$ magnitude                                      & 4.384\\
{\tt SYSTEM\_BIBCODE}   &...&Reference for multiplicity or exoplanets                   & 2018A\&A...615A.172M\\     
{\tt EXOPLANET\_COUNT}  &...            &Number of confirmed exoplanets                                 & 
1\\     
{\tt GAIA\_DR2}                 &  ...         & \gdrtwo identifier                     
                         &   5853498713160606720\\ 
{\tt GAIA\_EDR3}          &  ...         & \gdrthree identifier           
                                     &   5853498713190525696\\ 
{\tt SIMBAD\_NAME}              &...            &Name resolved by SIMBAD                                         &alf Cen C\\
{\tt COMMON\_NAME}              &...            &Common name                                                   &Proxima Cen\\
{\tt GJ}                                        &...            &Gliese \& Jahrei\ss\ catalogue identifier                                         &GJ 551\\
{\tt HD}                                        &...            &Henry Draper catalogue        identifier                                             &...\\
{\tt HIP}                                       &...            &\hip\ catalogue  identifier                                     &HIP 70890\\
{\tt COMMENT}                       &...                &Additional comments on exoplanets, 
multiplicity, etc                       &Proxima Cen c: candidate planet\\
&&& 2019ESS....410203D\\
\noalign{\smallskip}
\hline
\end{tabular}
\end{table*}

\subsection{Catalogue compilation}
\label{sec:compilation}

We started our compilation using the Set of Identifications, Measurements, 
and Bibliography for Astronomical Data (SIMBAD) database\footnote{\url{http://SIMBAD.u-strasbg.fr}} \citepads{2000A&AS..143....9W}.
This database provides information on astronomical objects of interest beyond the Solar System that have been studied and reported in scientific publications. 
We retrieved 378 stars and brown dwarfs with a parallax greater or equal than 100\,mas through the SIMBAD table access protocol (TAP) service\footnote{\url{http://SIMBAD.u-strasbg.fr/SIMBAD/sim-tap}} with the following query, \texttt{
\small
SELECT * FROM basic LEFT JOIN allfluxes on oid=oidref
WHERE plx$\_$value > = 100,} which returns the information on the object type, its astrometry, its photometry in $U$, $B$, $V$, $R$, $I$, $J$, $H$, and $K_s$ bands, and, when available, its spectral type.
To the SIMBAD list we added 21 cool brown dwarfs from recent parallax programmes, but they are not yet included in the database (Sect.~\ref{sec:stats.bd}). 

Since the query is based on the parallax, SIMBAD sometimes returns only one object while the literature papers refer to separate components (e.g. very close astrometric binaries and spectroscopic binaries), so we added these components as explained in Sect.~\ref{sec:stats.bin}. We  removed objects 
whose binarity has been refuted by \gaia parallaxes, from confusion with activity, or, on the contrary, confirmation by high-contrast imaging studies. 
We finally completed the list by adding confirmed exoplanets, starting from existing exoplanet databases and reviewing their status to add only confirmed discoveries (Sect.~\ref{sec:stats.exo}).

We discarded five objects that initially passed our TAP SIMBAD criterion.
Three of them were the brown dwarfs \object{WISE J053516.80--750024.9}, \object{WISE J035934.06--540154.6}, and \object{WISE J154214.00+223005.2}, 
with published parallaxes of $250\pm79$\,mas, $145\pm39$\,mas \citep{2013ApJ...762..119M}, and $96\pm41$\,mas \citep{2014ApJ...796...39T}, respectively, which now have higher-precision parallax measurements from \cite{2021ApJS..253....7K} that move them outside the 10\,pc limit.
The other two objects were also low-mass stars: 
\object{LP~388--55}, an M7.5\,V\,+\,L0 binary that had its parallax re-estimated  
from $110.7\pm5.8$\,mas \citep{2014ApJ...784..156D} to $64.3\pm0.7$\,mas in \gdrthree and 
the binary \object{G~19--15} had a dynamical parallax measured of $74.0\pm0.99$\,mas from \cite{2019MNRAS.482.4096D}.

This compilation resulted in \NOBJ\ objects in \NSYS\ systems that are listed in Table~A.1\footnote{Table~A.1 is only available in electronic form
at the CDS via anonymous ftp to cdsarc.u-strasbg.fr (130.79.128.5) or via 
http://cdsweb.u-strasbg.fr/cgi-bin/qcat?J/A+A/, at \url{https://gruze.org/10pc/}, and at \url{https://gucds.inaf.it/}}. 
It contains \NSTARS\ stars from F to early-L spectral type, including \NWD\ white dwarfs (plus one candidate in a system). 
It also lists \NBDS\ brown dwarfs and \NPLANETS\ confirmed exoplanets. 
We also tabulated, numbered from 1001 and higher in the catalogue, two low-mass star systems, namely \object{G~100-28} (GJ~1083) and \object{Ross~440} (GJ~352),   13 ultra-cool T and Y brown dwarfs whose $1\sigma$ parallax uncertainties will allow them to be located within 10\,pc, and the two components of a brown dwarf binary with a photometric parallax estimate larger than 100 
mas.

The  sample  was  constructed  by  setting  a  strict parallax limit  of  100\,mas. However,  the  parallax  measurements  carry  uncertainties, and objects located within 3-sigma of this limit may not belong to the 10\,pc sample when their measured parallax is larger than this limit or when the value is smaller. We used a SIMBAD query with a 20\,mas parallax cut and replaced the SIMBAD parallax by the more accurate \gdrthree value when available. We find 16 objects with parallaxes within 3-sigma from our 100\,mas parallax limit. This number does not include the 
15 brown dwarfs already identified at a 1-sigma level. However, in general we expect the true distance of an object to be larger than the inversion of the measured parallax so we expect to lose more objects than we gain 
due to errors at this border zone. Indeed, considering the bayesian distances computed in the GCNS, the number of objects with parallaxes > 100\,mas is  312 while the number of objects with median distances < 10\,pc is 310. 

The description of the catalogue is reported in Table~\ref{tab:10pc}, with the first object of the list, \object{Proxima Centauri}, shown as an example. 
The references in the catalogue are given with the bibcode assigned by the SAO/NASA Astrophysics Data System\footnote{\url{https://ui.adsabs.harvard.edu/}}. 
The full references are given in the Appendix.

\subsection{The 10\,pc sample and \gaia}
\label{sec:gaia} 

The ability to fully catalogue and characterise the 10\,pc sample renders 
it a fundamental dataset to test the quality of upcoming \gaia releases. Some example quality checks for \gaia using this dataset are as follows:
(i) Catalogue completeness: To check the completeness of the overall stellar sample and white dwarf population, one can extrapolate from the 
local stellar density as was done by GSS21.
    (ii) Exoplanet detections: While the bulk of the expected large catalogue of exoplanets detected from \gaia astrometry will only appear in the fourth data release, the first sample of exoplanet detections might already be announced in the third data release. It will likely include a subset of those planetary companions within 10\,pc with detectable astrometric signatures (see  Sect.~\ref{sec:stats.exo}). 
    (iii) Magnitude limits: The 10\,pc sample has stars that are too bright for \gaia and brown dwarfs that are too faint. It provides an empirical estimate of the magnitude limits.
    (iv) Binarity detection: there are at least 94 multiple systems in our 10\,pc sample. They cover a wide parameter space in mass ratios, magnitude differences, angular separations, inclinations, and orientations. Binary systems for which we do not find solutions in the \gaia pipeline should be understood.

By comparing the \gdrthree to our 10\,pc sample, we found that there are eight nearby stars too bright to be observed by \gaia and 54 brown dwarfs that are probably too faint. Of the 402 remaining objects, 
90 do not have a full astrometric solution in \gdrthree; they are all in close binary systems.  
Yet 14 of them had a full astrometric solution in \gdrtwo. With twelve more months of observations, the residuals went up and the \gdrthree solution did not meet the restrictive quality cuts \citep[$\texttt{astrometric\_sigma5d\_max} < 1.2$\,mas or 
\texttt{visibility\_periods\_used} $\geq 9$;][] {2021A&A...649A...2L}.  This should no longer be a problem in the next data release with the improved astrometric solution, taking the orbital motion  into account.

\subsection{Astrometry}

We replaced the SIMBAD output astrometry by that of \gdrthree when available (312 stars and brown dwarfs), except for three cases in binary systems from \cite{2016AJ....152..141B} (\object{GJ 831 A}, \object{GJ 791.2 A}, \object{and CD-68 47 A}), who accounted for orbital motion and determined their astrometry with a higher precision. 
Whereas the \hip\ determinations were more precise compared to \gdrtwo values for some bright stars, this was no longer the case compared to \gdrthree values. 

\begin{figure}[]
    \centering
    \includegraphics[width=0.48\textwidth,bb=10 0 320 270,clip=]{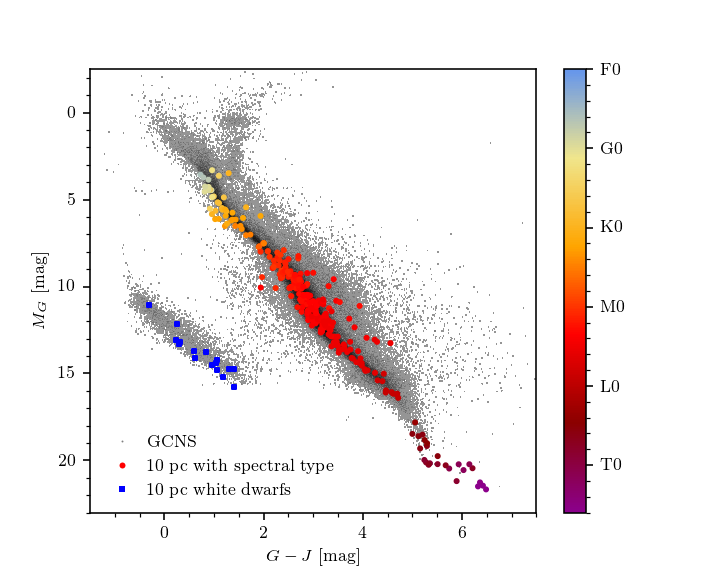}
    \caption{Colour-absolute magnitude diagram of the 10\,pc sample, over-imposed  on the GCNS (grey dots). 
    The colour bar indicates the spectral type.
    White dwarfs are in dark blue. }
    \label{fig:hrd}%
\end{figure}

\subsection{Photometry}

\begin{table}[]
    \caption{{\tt G\_CODE} values for retrieved or estimated $G$ magnitudes.}
    \centering
    \small
    \begin{tabular}{l l c}
    \hline 
    \hline
    \noalign{\smallskip}
{\tt G\_CODE} & $G$ magnitude & $N_{obj}$ \\
    \noalign{\smallskip}
    \hline
    \noalign{\smallskip}
2 &  Retrieved from \gdrtwo & 3 \\
3 &  Retrieved from \gdrthree & 342 \\
10 & Estimated from spectral type$^{a}$, $\sigma \approx$ 0.6\,mag & 39 \\
20 & Estimated from  $M_G =$ 25\,mag$^{b}$, lower limit & 48 \\
    \noalign{\smallskip}
    \hline
    \end{tabular}
    \tablefoot{
\tablefoottext{a}{$G$ magnitude estimated from the absolute magnitude versus spectral type calibration computed as part of the GCNS.}
\tablefoottext{b}{$G$ magnitude of late T and Y brown dwarfs too faint for \gaia computed assuming an arbitrarily absolute $G$ magnitude set to 25\,mag.}    
}
    \label{tab:G_CODE}
\end{table}

In Table~A.1, we also provide the \gaia photometry ($G$, \gbp, and \grp) for 345 objects.
The photometry of all of them is from \gdrthree, except for three brown dwarfs (\object{2MASS J17502484--0016151\,A}, \object{2MASS J08354256--0819237}, and \object{Luhman~16\,B}) that are in \gdrtwo, but not in \gdrthree. 
In unresolved systems, we often tabulate just the primary (or system) magnitudes. 
We included an estimate of the $G$ magnitude ({\tt G\_ESTIMATE}) for the other 87 objects using different procedures, as indicated by the {\tt G\_CODE} value in the catalogue and summarised by Table~\ref{tab:G_CODE}.
The addition of the {\tt G\_ESTIMATE} column  provides a quick way to identify objects that should be detectable by \gaia, but they should not be used for scientific purposes.

\subsection{Spectral type and object category}

We reviewed the output spectral types provided by default from the SIMBAD 
query.
We did not calculate 
an average spectral type from all determinations, but took the most recent 
reliable spectral type based on spectra.
In Table~A.1 we indicate
the method used for the spectral type determination, from photometry or spectroscopy, in the optical or near-infrared. 
Only 40 objects, mainly in close binary systems, have no spectral type. 

We classified all the objects of the 10\,pc sample in five categories ({\tt OBJ\_CAT}\footnote{{\tt OBJ\_TYPE} in SIMBAD.} in Table~A.1):
stars (K and earlier spectral types), low-mass stars (M and early-L types), white dwarfs, brown dwarfs \citep[including the M9-type object \object{BD+16~2708\,Bb} from its dynamical mass determined by][]{2017ApJS..231...15D}, and exoplanets. 
For components in close binaries with no information on the spectral type, we assigned them to the low-mass stars category by default. 
However, this classification should be taken with caution since we know that some of them can actually be brown dwarfs, such as \object{L~768--119\,B}, \object{GJ~867\,D}, and \object{Wolf~227\,B}, which have mass estimates that may place them in the substellar range \citep{2002ApJS..141..503N, 2014AJ....147...26D, 2018AJ....155..125W}.
The probable brown-dwarf nature of the three star candidates is indicated in the  {\tt COMMENT} field of Table~ A.1.

\subsection{Line-of-sight velocities}

The SIMBAD query provides line-of-sight velocities
for 287 objects. 
Among them, 129 come from \gdrtwo or its catalogue of radial velocity standard stars \citep{2018A&A...616A...7S}, and 48 precise measurements come 
from \cite{2020A&A...636A..36L}. Radial velocities of multiple systems may be inaccurate, but we tried to use the same source for the two components of a binary system when available, or we only listed a single measurement 
when this was not the case.

\section{Astrophysical content and statistical exploration}
\label{sec:stats}

\begin{table}[]
    \caption{Summary of the 10\,pc sample$^{a}$.}
    \centering
    \small
    \begin{tabular}{l c}
    \hline 
    \hline
    \noalign{\smallskip}
Type &  Number\\
    \noalign{\smallskip}
    \hline
    \noalign{\smallskip}
O &  0\\
B &  0\\
A &  4\\
F &  8\\
G &  18\\
K &  38\\
M &  249\\
L &  21\\
T &  45\\
Y &  19\\
D &  20\\
N/A & 41 \\
Exoplanets & \NPLANETS\\
    \noalign{\smallskip}
    \hline
    \noalign{\smallskip}
\textbf{Total} &\textbf{\NOBJ}\\
    \noalign{\smallskip}
    \hline
    \noalign{\smallskip}
Single & \NSS \\
Binary &  \NBS\\
Triple$^{b}$ &  \NTS\\
Quadruple$^{b}$ &  3\\
Quintuple$^{b}$ &  2\\
    \noalign{\smallskip}
    \hline
    \end{tabular}
    \tablefoot{
\tablefoottext
{a}{In the column Type, O, B, A... Y stand for stellar and sub-stellar spectral types, D for white dwarfs, and N/A for objects without a spectral type. 
The Sun (G2\,V star) and its eight planets are not included.}
\tablefoottext{b}{The name of the triple, quadruple, and quintuple systems are given in Table~\ref{tab:names}.}
    }
    \label{tab:numbers}
\end{table}
    
In this section we describe the content of the 10\,pc sample in terms of 
astrophysical objects, as illustrated by 
Fig.~\ref{fig:hrd}. 
It shows the $G$-band absolute magnitude as a function of the $G - J$ colour for all objects with \gaia $G$ and 2MASS \citep[Two Micron All Sky Survey;][]{2006AJ....131.1163S} $J$ magnitudes and spectral type determination. 
As a comparison, the GCNS objects with $G$ and $J$ magnitudes are also shown.
In Table~\ref{tab:numbers} we summarise the spectral and multiplicity distribution
of our sample.

\subsection{Multiple systems}
\label{sec:stats.bin}

Multiple systems are reported in large databases such as the Catalog of Components of Double and Multiple Stars \citep{2002yCat.1274....0D} or the 
Washington Double Star catalog\footnote{\url{http://www.astro.gsu.edu/wds/}} \citep[][]{2001AJ....122.3466M}.
Multiplicity is also indicated in the SIMBAD output ({\tt OBJ\_TYPE=**}).
For all multiple system candidates we confirmed that the hypothesis of
being part of that system was consistent with the most
recent parallax determinations.
We discarded five companion candidates:
\object{BD+42 2320} with \object{$\beta$~CVn}, \object{BD+02 521} with \object{$\kappa^{01}$~Cet}, and \object{2MASS J12141817+0037297} with \object{GJ~1154}, based on their \gdrtwo parallax, while the companions of \object{HD~50281~AB} and \object{BD+43 2796} were identified in the \gdrthree with low parallaxes.
In addition, three spectroscopic binary candidates (\object{BD+19~5116~A}, \object{BD+19~5116~B}, and \object{G~13--22}) that are known to be active stars were discarded. 
Details on these discarded components are given in the {\tt COMMENT} column of Table~A.1.

As already stated in Section~\ref{sec:gaia}, the future \gaia data releases will provide solutions for a large number and type of binary (astrometric, spectroscopic, and eclipsing) with periods from 0.2\,d to more than 5\,yr, in amounts of hundreds of thousands in the third \gaia data release (\gaia DR3) and millions in the fourth \gaia data release (\gaia DR4), as predicted by the Gaia Universe Model Snapshot \citep{2012A&A...543A.100R}. 
Within the 10\,pc sphere, one can expect very good forthcoming astrometric solutions, including orbital parameters.
Such new astrometry would complete the characterisation of the systems, even the closest ones; the expected limit is 0.12\,arcsec, but \gaia astrometry will provide information on binarity even for objects it cannot resolve.

\subsection{White dwarfs}
\label{sec:stats.wd}

Twenty objects are white dwarfs, six of which are part of multiple systems. Their spectral type distribution is nine DA, five DQ, four DZ, and two 
DC. 
They all have a precise parallax from \gdrthree except for Procyon~B, most likely due to the short current separation and brightness difference of 
about 10\,mag with respect to Procyon~A. 
With a more eccentric orbit than Procyon~B and a similar brightness ratio with the primary, Sirius~B offered, however, a more favourable situation to be detected by \gaia. 

Our 10\,pc sample may be supplemented with new faint, dark, white dwarfs in the future, in particular in unresolved multiple systems. 
For example, we found two candidates in our list. \object{G~203--47} is a 
spectroscopic binary \cite{1997AJ....113.2246R} with one possible white dwarf component: \cite{1999A&A...344..897D} argued that the companion’s mass is too large ($M > 0.5 M_\odot$) to be something other than a degenerate star.
Likewise, \object{CD--32~5613} was quoted as an unresolved double white dwarf by \cite{2017A&A...602A..16T}.

\subsection{Brown dwarfs}
\label{sec:stats.bd}

According to \cite{2017MNRAS.469..401S}, \gaia can detect L5 dwarfs to 29\,pc, T0 dwarfs to\,14 pc, T6 dwarfs to 10\,pc, and T9 dwarfs to 2\,pc, assuming a magnitude limit $G$ of 20.7\,mag. 
These predictions are consistent with the 10\,pc sample: The latest object with a \gaia parallax determination is just T6.
There are, however, a few examples for which \gaia does not determine an astrometric solution or even a detection. This is the case for the nearest pair of brown dwarfs, \object{Luhman~16\,AB} \citep[L7.5+T0.5;][]{2013ApJ...767L...1L}. 
Whereas both components are in \gdrtwo, \gdrthree tabulates only the A component and neither of the two releases provides a solution with a parallax. 
The other cases where \gaia failed to acquire astrometry are listed in Table~\ref{tab:bd}. 
Except for three objects close to the faint limit of \gaia, all are in multiple systems for which the current \gaia astrometric solution is applying a single star solution.
For nearby objects in multiple systems, the orbital motion induces large residuals in the single star solution that the pipeline marks as errors and a full solution is not provided. Future \gaia data release will employ 
multiple star solutions so we expect full astrometric solutions for these brown dwarfs. 

\begin{table}
    \caption{Brown dwarfs expected to have a full astrometric solution in 
future \gaia data releases.}
    \label{tab:bd} 
    \small
    \centering
    \begin{tabular}{l c c}
    \hline 
    \hline
    \noalign{\smallskip}
    Name & Parallax  & Spectral \\
    ~ & (mas)  & type \\
    \noalign{\smallskip}
    \hline
    \noalign{\smallskip}
\object{Luhman 16 A} & 501.6\,$\pm$\,0.1 & L7.5 \\
\object{Luhman 16 B} & 501.6\,$\pm$\,0.1 & T0.5 \\
\object{$\epsilon$ Ind C} & 270.7\,$\pm$\,0.7 & T6.0 \\
\object{SCR J1845--6357 B}  & 249.7\,$\pm$\,0.1 & T6.0 \\
\object{Scholz's Star B} & 147.1\,$\pm$\,1.2 & T5.0 \\
\object{SCR J1546--5534 B} & 119.1\,$\pm$\,0.7 & T6.0 \\
\object{2MASS J16471580+5632057} & 116.0\,$\pm$\,29.0 & L9 pec \\
\object{WISE J223617.59+510551.9} & 102.8\,$\pm$\,1.9 & T5.5 \\
\object{CFBDS J213926+022023 A} & 101.5\,$\pm$\,2.0 & L8.5 \\
\object{CFBDS J213926+022023 B} & 101.5\,$\pm$\,2.0 & T3.5 \\
\object{2MASS J07584037+3247245} & 101.3\,$\pm$\,3.3 & T2.5 \\
\object{BD+16 2708 B} & 100.5\,$\pm$\,0.1 & M9 \\
    \noalign{\smallskip}
    \hline
    \end{tabular}
\end{table}

We completed our compilation of brown dwarfs with those presented by \citet{2019ApJS..240...19K}, \citet{2020AJ....159..257B}, and, especially, \citet{2021ApJS..253....7K}, from where we added 38 ultra-cool objects with 
new or more precise parallaxes.
We 
expect the 10\,pc census to be further supplemented with cool T- and Y-type dwarfs in the near future. 
Of the 19 candidates with {\tt NB\_OBJ} $\geq 1001$ (Sect.~\ref{sec:compilation}),
15 are T and Y dwarfs, while \cite{2021ApJS..253....7K} report additional candidates with only parallax estimates that could fill in the 10\,pc sample (e.g. the binary CWISE J061741.79+194512.8 AB with a parallax estimate of 133 mas).

\subsection{Exoplanets}
\label{sec:stats.exo}

The existing exoplanet catalogues, such as the Extrasolar Planets Encyclop{\ae}dia\footnote{\url{http://exoplanet.eu/}}, the NASA Exoplanet Archive\footnote{\url{https://exoplanetarchive.ipac.caltech.edu/}}, the Exoplanet Orbit Database\footnote{\url{http://exoplanets.org/}}, or the Open Exoplanet Catalogue\footnote{\url{http://openexoplanetcatalogue.com/}}, are not fully consistent. 
Discrepancies are partly due to different selection criteria, notations, and diligence in updating their data bases, and they are also due to the heterogeneity 
of information provided in discovery papers that different catalogues capture in different ways. 
As a consequence, it is almost impossible to achieve full homogeneity, and any direct comparison between catalogues can be difficult depending on the specific application. This is the case even for the sample of exoplanetary systems nearest to the Sun. 

We cross-matched the Extrasolar Planets Encyclop{\ae}dia and the NASA Exoplanet Archive in order to select the most reliable set of stars with exoplanets within 10\,pc, and we added those that we considered to be confirmed to our catalogue. 
The most recent discovery added to our catalogue 
is the transiting rocky planet \object{GJ~486}\,b \citep{2021Sci...371.1038T}.
The astrometry given in Table~A.1 is the one of the host star, and the discovery reference is given in the {\tt SYSTEM\_BIBCODE} field. 

Non-listed candidate, unconfirmed, or controversial exoplanets are enumerated, though, in the corresponding {\tt COMMENT} field of the host star.
For them, we employ the term `candidate' when the publication reported the companion with that 
terminology or when the statistical evidence was not strong for the presence of the signal. 
A notable example is the second long-period planet candidate orbiting Proxima Centauri c \citep{2020SciA....6.7467D}.
We use the term `unconfirmed' or `controversial' when the radial-velocity 
or imaging signal 
has not been seen by different groups analysing the same datasets, when different groups use different datasets and do not find the same signals, or when the radial-velocity signal can be explained in terms of stellar activity variations. 
This nomenclature is also used to point at discovery announcements in papers that have only appeared on the arXiv open-access repository of electronic preprints, but have not been accepted for publication after a reasonable amount of time. 
Noteworthy examples include radial-velocity signals of an unclear nature and 
period in the time series of the nearby K dwarf \object{HD~219134} \citep{2015A&A...584A..72M,2015ApJ...814...12V,2017NatAs...1E..56G}, the putative directly imaged planet around \object{Fomalhaut} \citep[e.g.][and references therein]{2008Sci...322.1345K,2020A&A...640A..93J,2021MNRAS.tmp..759P}, and a number of terrestrial-mass companions tentatively detected inside and outside the temperate zones of nearby M dwarfs, such as \object{GJ~581} \citep[which harbours the most highly debated habitable-zone system, see,][and references therein]{2018A&A...609A.117T},
\object{$\tau$~Ceti} \citep{2013A&A...551A..79T,2017A&A...605A.103F}, \object{GJ~667\,C} \citep{2013A&A...553A...8D,2013A&A...556A.126A,2014MNRAS.437.3540F}, and \object{HD~40307} \citep{2009A&A...493..639M,2013A&A...549A..48T,2016A&A...591A.146D}. 

After applying the filters above, we came
up with a total of \NPLANETS\ known and confirmed planets within 10\,pc. 
In the case of a circular orbit, their true (for the few that are seen in 
transit) or minimum (for the radial-velocity-detected companions) astrometric signal in arcsec is $\alpha = (M_p/M_\star)\times(a_p/d)$, with $M_p$ and $M_\star$ in the same units,
$d$ in pc, and $a_p$ in au. 
The astrometric signature of the vast majority of short-period ($P<100$\,d) super-Earths and sub-Neptunes within 10\,pc are not expected to be detected by \gaia, as their amplitude will usually fall well below the end-of-nominal-mission systematic noise floor for the along-scan astrometric measurements in the bright star regime \citep[$\sim50~\mu$as for a single CCD crossing, see e.g.][] {2021A&A...649A...2L}. Indeed, the much expected catalogue of tens of thousands of exoplanets will be mostly populated by gas giants in the 1--4\,au separation regime \citep[e.g.][and references therein]{2018haex.bookE..81S}.
However, there are over a dozen exoplanet candidates for which we expect to see the astrometric signature. Some of these exoplanets may have already been detected with the three-year time baseline of \gaia DR3, and most of them should be detected in \gaia DR4.
However, their actual detectability in future \gaia data releases relies on the effectiveness of a successful calibration of the astrometric data in 
the very bright star regime ($G\lesssim9$\,mag). 
We list below the radial-velocity exoplanet candidates that should be detected by \gaia: 

\begin{itemize}
    \item \object{GJ 15\,Ac} is expected to induce $\alpha > 570\,\mu$as, 
but it has a period in the neighbourhood of 20\,yr \citep{2018A&A...617A.104P}. Even a 10-yr \gaia mission will not see more than half of the orbit. The planet's motion should be detected as a curvature effect in the stellar proper motion and described in terms of an acceleration solution.
    \item \object{$\epsilon$\, Eri b}, with an orbital period of $\sim7.4$\,yr \citep{2000ApJ...544L.145H,2019AJ....157...33M} and an expected $\alpha \sim1000\,\mu$as, should be easily detectable. The host star is, however, very bright ($G=3.46$\,mag) and the major source of uncertainty is 
the effective calibration of the astrometric time-series.
    \item \object{$\epsilon$~Ind A\,b} has a semi-major axis of $\sim10$\,au, but the minimum mass of a massive super-Jupiter \citep{ 2019MNRAS.490.5002F}. It should be detected as an acceleration solution; however, simlar to $\epsilon$~Eri, the host star is very bright ($G=4.32$\,mag).
    \item \object{GJ 649\,b}, with $a_p=1.13$\,au \citep{2010PASP..122..149J}, should induce $\alpha > 65\,\mu$as. It might be detectable if its true inclination is small.
    \item \object{GJ 3512\,b} has $a_p=0.33$\,au \citep{2019Sci...365.1441M}, but it orbits a mid-M dwarf; therefore with $\alpha > 130\,\mu$as, it should be detectable.
    \item \object{GJ 849\,b} with $a_p=2.35$\,au \citep{2006PASP..118.1685B} and $\alpha > 500\,\mu$as should be easily detectable.
    \item \object{GJ 849\,c} has a long period $P\sim$ 15--20\,yr \citep{2015ApJ...800...22F}, so it might be detectable as an acceleration solution on top of the signal induced by \object{GJ 849\,b}.
    \item \object{GJ 433 c} with $\alpha > 100\,\mu$as is, in principle, detectable; however, its period is $>10$\,yr \citep{2020ApJS..246...11F}, so it might be described in terms of an acceleration solution.
    \item \object{HD 219134\,g} is possibly a sub-Saturn-mass object with 
an unclear but long ($P>5$\,yr) period \citep{2015A&A...584A..72M,2015ApJ...814...12V,2017NatAs...1E..56G}, likely inducing $\alpha > 100\,\mu$as. It is in principle detectable, but the K-dwarf host star is very bright ($G=5.23$\,mag), so the same calibration issues as in the case of $\epsilon$~Eri and $\epsilon$~Ind\,A will need to be successfully addressed.
    \item \object{GJ 876\,b}, with $\alpha\sim$ 250--350\,$\mu$as \citep[depending on the actual inclination angle, see][] {2010A&A...511A..21C}, was detected by the {\it Hubble Space Telescope} \citep{2002ApJ...581L.115B} and  it is expected to be clearly identified by \gaia.
    \item \object{GJ 876\,c} has an expected $\alpha\sim70$\,$\mu$as, but 
with a period of only 30\,d \citep{2001ApJ...556..296M} it will likely be 
very difficult for \gaia due to the possible degeneracy with periodic aliases of the scanning law.
    \item \object{GJ 832\,b} as $a_p\sim$ 3.5--4.0\,au \citep{2009ApJ...690..743B}, and with $\alpha > 1000\,\mu$as should be clearly detectable by \gaia, either as acceleration or a full orbital solution.
    \item \object{GJ 9066\,c} has $a_p=0.87$\,au \citep{2020ApJS..250...29F}, and with $\alpha > 200\,\mu$as it is expected to be detectable by \gaia.
    \item The candidate \object{Proxima Cen c} with $a_p=1.5$\,au is expected to induce  $\alpha>170\,\mu$as \citep{2020SciA....6.7467D}. A confirmation of its existence by \gaia should be possible. 
\end{itemize}

\subsection{Statistics}

In terms of statistical studies, the 10\,pc sphere is two-fold. 
Only in this nearby volume one can expect to detect and characterise all objects, but it also probes a small volume and, thus, offers small statistics. 
As a result, the 10\,pc sphere is complementary to statistically more significant samples with larger volumes, but that suffer from incompleteness. 
Keeping that in mind, below, we provide a few numbers on the multiplicity rate, spectral type distribution, and luminosity class, which give an overall 
picture of the immediate vicinity to our Sun.

There is no giant star within 10\,pc and only four evolved stars, which are all sub-giants. These are \object{ $\beta$~Hyi}, \object{$\mu$~Her}~Aa, \object{$\delta$~Pav}, and \object{$\delta$~Eri}. There are only about five pre-main-sequence stars within 10\,pc: the triple system \object{AT Mic A}, \object{AT Mic B}, \object{AU Mic} being a bona-fide member, and \object{YZ~CMi} being a candidate member of the $\sim24$\,Myr $\beta$ Pictoris association \citep{2001ApJ...562L..87Z, 2015A&A...583A..85A,2014MNRAS.445.2169M},
and \object{AP Col}, which may belong to the $\sim 50$\,Myr Argus / IC~2391 association \citep[][but see \citealt{2015MNRAS.454..593B} about the existence of the Argus association]{2011AJ....142..104R}.

Almost half of the stars and brown dwarfs are in multiple systems. As summarised in the bottom part of Table~\ref{tab:numbers}, our 10\,pc sample contains \NSS\ single, \NBS\ double, \NTS\ triple, three quadruple, and two quintuple systems ({\tt NB\_SYS} in Table~A.1).
Following the definitions of \cite{1997AJ....113.2246R}, for example, and adding the Sun as a single star, these numbers translate into a multiplicity frequency (which quantifies the number of multiple systems within the sample) 
and a companion frequency (which quantifies the total number of companions) of $27.4\pm2.3$\% and $36.5\pm3.2$\%, respectively. In Table~\ref{tab:names}, we give the names of the triple, quadruple, and quintuple systems 
for convenience.

\begin{table}[]
    \caption{Names of triple and higher order systems.}
    \centering
    \small
    \begin{tabular}{l l }
    \hline 
    \hline
    \noalign{\smallskip}
{Multiplicity} & Name  \\
    \noalign{\smallskip}
    \hline
    \noalign{\smallskip}
Triple &  $\alpha$ Cen; EZ Aqr; $\epsilon$ Ind; GJ 1245; $o^{02}$ Eri; 36 
Oph;  \\
    & G 41--14; BD-17 588; HD 16160; HD 156384;  \\
    & HD 50281; 41 Ara; $\alpha$ PsA; LP 881-64; HD 115953; \\
    & AT Mic; BD+66 34; G 184-19; BD+16 2708 \\
Quadruple &  GJ 570; $\mu$ Her; GJ 867 \\
Quintuple & $\xi$ UMa; HD 152751 \\
    \noalign{\smallskip}
    \hline
    \end{tabular}
    \label{tab:names}
\end{table}

\begin{figure}
   \centering
   \includegraphics[width=0.48\textwidth,bb=10 0 340 270,clip=]{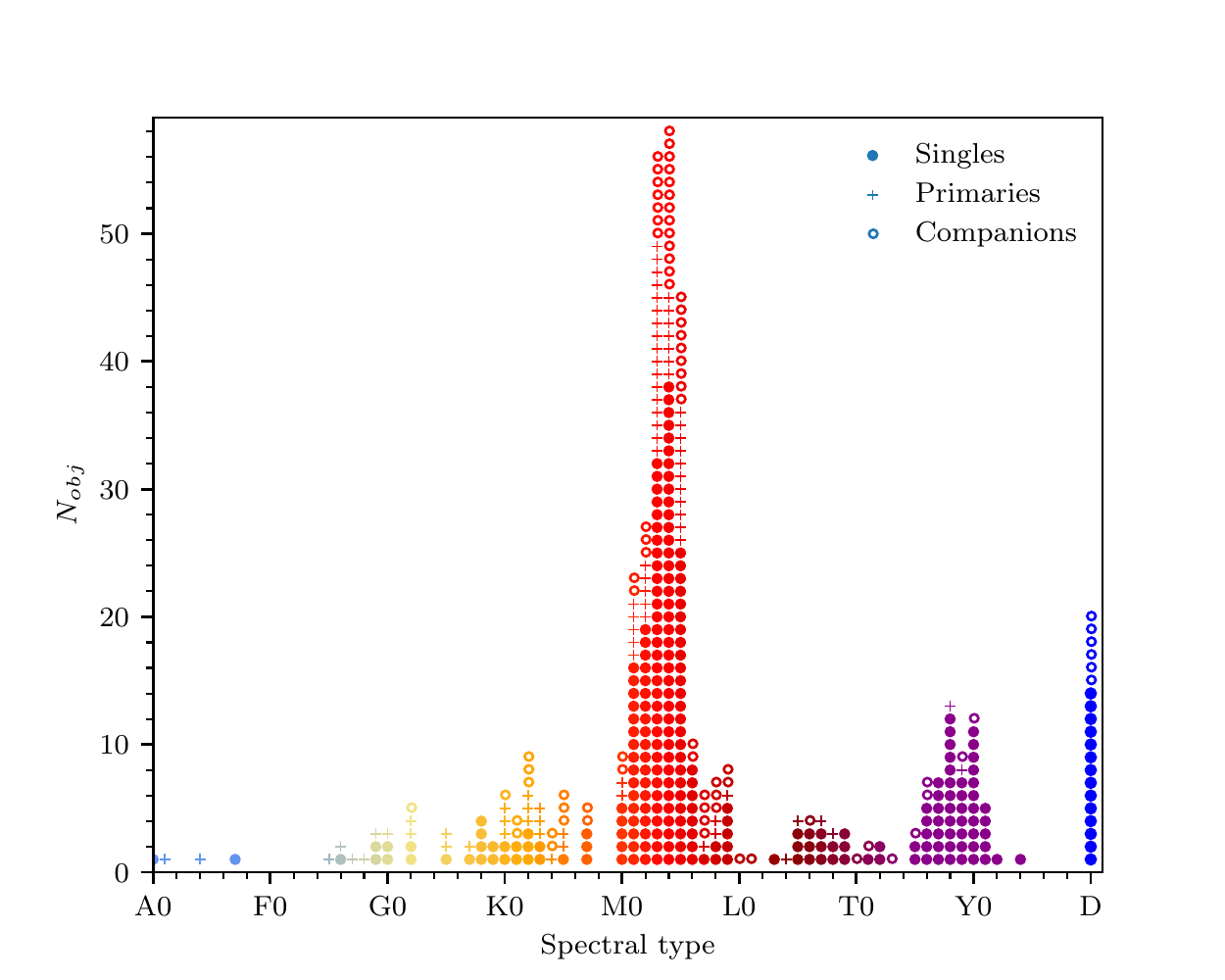}
   \caption{Spectral type distribution of the 10\,pc sample. D are white dwarfs. The different symbols indicate 
   single stars, primaries, and companions. }
              \label{fig:spt}%
    \end{figure}

The spectral type distribution is shown in Figure~\ref{fig:spt}. 
We found 249 M stars among the 423 objects with a measured spectral type, 
which translates into a ratio of $58.9\pm5.8$\%. 
This relatively small value is in contrast with other higher previous determinations of the order of 70\% \citep[e.g.][]{2006AJ....132.2360H, 2010AJ....139.2679B}.
This small value probably comes from a more complete sample of brown dwarfs compared to older studies. 
In the substellar regime, L-type objects amount to only half the number of T-type ones.

There are 41 objects without a spectral type measurement, all being secondary components of close binaries. 
They could slightly bias these proportions, so we used published individual masses, either computed from orbit fitting or estimated from adaptive optics contrast measurements, to estimate their spectral types. 
We found 36 possible M stars, four possible L dwarfs, and one possible white dwarf (all of them marked in the column {\tt COMMENT} in Table~A.1). 
The proportion of M stars now becomes $61.3\pm5.9$\%, which is not significantly different from the ratio derived from measured spectral types only.

More than half of the M dwarfs ($57.0\pm7.3$\%) have spectral types M3.0\,V to M5.0\,V. 
This proportion remains stable when including the estimated spectral types of the unresolved secondary components ($57.4\pm6.9$\%).
Translating these numbers into an observed mass function requires some care, but it seems to indicate that the number of stars increases up to about 0.3\,$M_\odot$ \citep[$\sim$ M4.0\,V;][]{2020A&A...642A.115C} and decreases for later M spectral subtypes. 
This maximum of the mass function, similar to other slope changes observed in very young open clusters \citep[e.g.][]{2012ApJ...754...30P}, corresponds to the fully-convective transition in the main sequence.

\section{The 10\,pc sample in the future}
\label{sec:future}

\subsection{Science cases and the next upgrades}

Apart from multiplicity studies, mass function analyses, and long-term exoplanet surveys, 
there is a number of science topics that can be covered with the 10\,pc catalogue.
One of them is kinematics and membership in thin and thick disc populations and stellar moving groups and associations.
There are on-going efforts to relate precise Galactocentric space velocities to youth features for a sample of over 2000 nearby M dwarfs (M.~Cort\'es-Contreras, priv.~comm.) and to measure, for the first time, radial velocities of a number of late-type ultracool dwarfs (W.~Cooper, priv.~comm.).
These works will be presented in forthcoming publications and will complement future releases of the 10\,pc catalogue.

Further improvements of the 10\,pc catalogue include adding the new {\it Gaia} DR3 astro-photometric data (expected in 2022), 
updating spectral types for poorly investigated companions and radial velocities of the faintest brown dwarfs, 
and adding more parameters useful for a variety of topics, such as atmospheric astrophysical parameters ($T_{\rm eff}$, $\log{g}$, [Fe/H]), chromospheric (equivalent widths of H$\alpha$ and Ca~{\sc ii}) and coronal (X-rays) activity indicators, and rotational velocity.
Some novel parameters, for instance the exozodi level, will also be useful for 
future space missions such as the {\it Large Interferometer for exoplanets} \citep{2021arXiv210107500L}.

\subsection{Obsolescence}

This catalogue will inevitably need to be updated when \gaia and other surveys 
issue their next data releases. Apart from extremely cool objects similar 
to WISEA~J085510.74--071442.5, new objects probably hide in the Milky Way 
plane \citep[see the discoveries by][]{2013A&A...557L...8B,2014A&A...561A.113S,2018RNAAS...2...33S,2018ApJ...868...44F}.
Such new objects, in spite of their expected large proper motions, will likely be detected by state-of-the-art photometric surveys from the ground 
such as the Panoramic Survey Telescope and Rapid Response System \citep{2002SPIE.4836..154K}, J-PLUS/J-PAS \citep{2019A&A...627A..29S}, and, specially, the Legacy Survey of Space and Time \citep{2009arXiv0912.0201L}, as well as from the space. 
In particular, the ESA medium-class {\it Euclid} space mission will cover 
more than 35\,\% of the celestial sphere in the red optical and near-infrared $Y$, $J$, and $H$ bands with an unprecedented depth and spatial resolution \citep{2011arXiv1110.3193L}. 
The {\it Euclid} Legacy Science on Ultracool Dwarfs will be particularly sensitive to low Galactic-latitude, high proper-motion, very red late-type dwarfs that have not been identified yet \citep{2020sea..confE.157M}.
The NASA {\it SPHEREx} space mission, with its all-sky, low-spectral-resolution capabilities in the 0.75--5.0\,$\mu$m range \citep{2020SPIE11443E..0IC}, will also help to discover new ultracool objects.

In addition to the yield from these photometric surveys, we expect that most of the new additions to the 10\,pc sample will be very close companions to our targets.
First, current and future spectroscopic surveys and adaptive optics observations will probably resolve some of the single stars into multiple components \citep[e.g.][]{2018A&A...619A..32B,2018MNRAS.475.1960F,2019AJ....157..216W}. 
Second, the component of the 10\,pc catalogue that will see the largest increase in number corresponds to new exoplanets that will be discovered or confirmed in the coming years as  
most stars are 
orbited by at least one exoplanet.
For instance, \cite{2015ApJ...807...45D} predicted $2.5\pm0.2$ small and close-by planets per M star, so we could expect more than 600 new exoplanets to be discovered, outnumbering the number of stellar and sub-stellar objects within 10\,pc.
Such an optimistic estimation is in line with the recent discovery of small planets around the closest stars, such as \object{Proxima Centauri} \citep[two planets:][]{2016Natur.536..437A,2020SciA....6.7467D,2020A&A...635L..14K}, 
\object{Barnard's star} \citep[one planet:][]{2018Natur.563..365R}, or
\object{Lalande~21185} \citep[one planet:][]{2019A&A...625A..17D,2020A&A...643A.112S}\footnote{\object{CN~Leo}, the third closest stellar system 
to the Sun, is an active M dwarf with a large radial-velocity scatter in the visible range that prevented the discovery of small planets until now 
 \citep{2018A&A...614A.122T}, but near-infrared observations are less sensitive to activity and may reveal a planetary system in the future.}.
However, even if these new planets are predicted from {\it Kepler}'s results, we will probably not detect more than a fraction of them in the coming years for several reasons:
($i$) Planets with periods close to that of stellar rotation will mostly go undetected;
($ii$) stellar activity will prevent others from being detected; and 
($iii$) close in planets on highly inclined orbits with small $\sin{i}$ values imply small radial-velocity semi-amplitudes.

While the closest planets have, in general, been discovered with precision radial-velocity spectrographs working in the red optical and/or near-infrared (especially designed for M-dwarf surveys; e.g. CARMENES, ESPRESSO, GIANO-B+HARPS-N, HPF, IRD, MAROON-X, SPIRou, 
and, in the future, NIRPS and CRIRES+)\footnote{\url{ https://carmenes.caha.es/ext/instrument/index.html}; \url{https://www.eso.org/sci/facilities/paranal/instruments/espresso.html};  \url{http://www.tng.iac.es/instruments/giano-b/}; \url{https://plone.unige.ch/HARPS-N/}; \url{https://hpf.psu.edu/}; \url{http://ird.mtk.nao.ac.jp/IRDpub/index_tmp.html};  \url{https://www.gemini.edu/instrumentation/maroon-x/}; \url{http://spirou.irap.omp.eu/Instrument/Cryogenic-spectrograph}; \url{https://www.unige.ch/sciences/astro/exoplanets/en/projects/nirps/}; \url{https://www.eso.org/sci/facilities/develop/instruments/crires_up.html}},
the NASA {\it TESS} space mission \citep{2015JATIS...1a4003R} is also discovering small transiting planets at less than 10\,pc, such as GJ~357\,b and GJ~486\,b 
(\citealt{2019A&A...628A..39L,2021Sci...371.1038T}; see also GJ~436\,b, a 
Neptune-sized planet at 9.8\,pc discovered by \citealt{2004ApJ...617..580B} and \citealt{2007A&A...472L..13G}).
A few more transiting planets in the immediate vicinity might also be detected in the near future with SPECULOOS from the ground \citep{2021A&A...645A.100S} and {\it PLATO} from space \citep{2014ExA....38..249R}. Finally, global astrometry with \gaia, particularly in the case of a fully extended 10 year mission, might unveil the presence of $\sim10-20$ new cold giant planets up to Jupiter-like orbital separations\citep[][see Sect.~\ref{sec:stats.exo}]{2018haex.bookE..81S}.

\subsection{Didactics}
\label{sec:outreach}

The 10\,pc sample has tremendous outreach potential. The objects are our nearest neighbours, they cover a large range of stellar and brown dwarfs parameter space, and many of them have a significant historical story that can be shared. 
If we just consider the first few objects, \object{$\alpha$~Cen\,A} is almost a solar twin, \object{$\alpha$~Cen\,C} (Proxima) harbours the nearest terrestrial habitable-zone planet and has another candidate planet, \object{Barnard's star} is an old thick-disc dwarf with the largest proper motion on the sky, \object{Luhman~16}\,AB is a brown dwarf binary, and \object{WISEA J085510.74--071442.5} is the coolest brown dwarf known to date. 
Among the first ten objects, only one system, $\alpha$~Cen, is visible to 
the naked eye; the brightest star, \object{Sirius}, is 12th in distance; and, the object with a first measured parallax, \object{61~Cyg}, is 28th in distance. 

To aid in this outreach, we produced some divulgative material that we
release with this contribution. 
The three-dimensional nature of the dataset makes creating maps more of a 
challenge than for more traditional terrestrial cartography. 
We generated maps in several different formats from the data, including a rotating animation of all the objects in the catalogue, a 3D fly-through JavaScript web application, a top down poster (see Fig.~\ref{fig:map}), and two 5\,pc and 10\,pc maps with `star columns' showing distance above and 
below the galactic plane. All the resources are available online\footnote{\url{https://gruze.org/10pc/resources}}.

\section{Conclusions}
\label{sec:ccl}

We provide a catalogue of all objects closer than 10\,pc from the Sun. 
It contains \NOBJ\ objects divided between \NSTARS\ stars, including 20 confirmed white dwarfs and one candidate white dwarf, \NBDS\ confirmed and 
three candidate brown dwarfs, and \NPLANETS\ confirmed exoplanets in \NSYS\ systems made up of \NBS\ binaries, \NTS\ triplets, three quadruplets, and two quintuplets.

During the catalogue compilation, we extensively checked all individual entries 
from what is available in the published literature. 
In particular, it contains the most recent astrometry from the last \gaia 
data release when available. 

The catalogue will be used to assess the quality of the forthcoming \gaia 
releases
to place limits on the frequency of planets and other components within multiple systems, as well as providing targets for focused planetary searches. The 10\,pc sample is incredibly varied: Our first ten neighbouring systems include two confirmed and two  candidate planets, a thick-disc object, a white dwarf, and four brown dwarfs. We recognise the didactic value of this sample and have provided various materials for that exploitation. 

The latest addition to the 10\,pc sample is the planet \object{GJ~486}\,b 
\citep{2021Sci...371.1038T}, but the last free floating objects have been discovered using the {\it WISE} \citep[{\it Wide-field Infrared Survey Explorer};][]{2010AJ....140.1868W} survey. The coolest and lowest-mass object  \object{WISEA J085510.74--071442.5,} a $>$Y4-type ultra-cool dwarf, was 
discovered by \citet{2014ApJ...786L..18L} as the result of significant data-mining, and we concur with the result of \cite{2021ApJS..253....7K} that the 10\,pc volume is probably still not complete for objects later than spectral type Y2. The distribution of these lowest mass objects will indicate the minimum mass cutoff for stellar formation; therefore, finding all objects in this local volume will provide  an important constraint for formation mechanisms. In addition, as the latest addition attests, the discovery of planets and other components within known systems is on the increase as our detection ability improves.  Hence, while we expect the number of very low mass objects, planets, and low mass components with 10\,pc within systems to increase, we do not expect to add any more higher mass, isolated, earlier type objects to the 10\,pc census.

\begin{acknowledgements}
The authors thank the referee for useful and prompt comments.
C.R. thanks Catherine Turon for fruitful discussions on the parallax measurements across the history. C.R. thanks Will J. Cooper for his contribution to plotting. This work was supported by the "Programme National de Physique Stellaire" (PNPS) of CNRS/INSU co-funded by CEA and CNES. 
R.L.S. and A.S. acknowledge support from the Italian Space Agency (ASI) under contract 2018-24-HH.0. A.S. acknowledges the financial contribution from the agreement ASI-INAF n.2018-16-HH.0.
J.A.C. acknowledges financial support from the Agencia Estatal de Investigaci\'on of the Ministerio de Ciencia, Innovaci\'on y Universidades and the ERDF through project PID2019-109522GB-C51/AEI/10.13039/501100011033.
This work has made use of data from the European Space Agency (ESA) mission {\it Gaia} (\url{https://www.cosmos.esa.int/gaia}), processed by the {\it Gaia} Data Processing and Analysis Consortium (DPAC, \url{https://www.cosmos.esa.int/web/gaia/dpac/consortium}). Funding for the DPAC has been provided by national institutions, in particular the institutions participating in the {\it Gaia} Multilateral Agreement.
This research has made use of the SIMBAD database,
operated at CDS, Strasbourg, France. This research has made use of the NASA Exoplanet Archive, which is operated by the California Institute of Technology, under contract with the National Aeronautics and Space Administration under the Exoplanet Exploration Program. This research has made use of data obtained from or tools provided by the portal \url{exoplanet.eu} of The Extrasolar Planets Encyclop{\ae}dia. This research has made use of the Washington Double Star Catalog maintained at the U.S. Naval Observatory. This research has made use of the TOPCAT\footnote{http://www.starlink.ac.uk/topcat/} tool \citep{Taylor2005}.
\end{acknowledgements}

%
%

 \bibliographystyle{aa}
 \bibliography{ref}

\begin{thebibliography}{332}
\expandafter\ifx\csname natexlab\endcsname\relax\def\natexlab#1{#1}\fi

\bibitem[{{Aannestad} {et~al.}(1993){Aannestad}, {Kenyon}, {Hammond}, \&
  {Sion}}]{1993AJ....105.1033A}
{Aannestad}, P.~A., {Kenyon}, S.~J., {Hammond}, G.~L., \& {Sion}, E.~M. 1993,
  \aj, 105, 1033

\bibitem[{{Abt} \& {Levy}(1976)}]{1976ApJS...30..273A}
{Abt}, H.~A. \& {Levy}, S.~G. 1976, \apjs, 30, 273

\bibitem[{{Adelman-McCarthy} \& {et al.}(2009)}]{2009yCat.2294....0A}
{Adelman-McCarthy}, J.~K. \& {et al.} 2009, VizieR Online Data Catalog, II/294

\bibitem[{{Alonso-Floriano} {et~al.}(2015){Alonso-Floriano}, {Caballero},
  {Cort{\'e}s-Contreras}, {Solano}, \& {Montes}}]{2015A&A...583A..85A}
{Alonso-Floriano}, F.~J., {Caballero}, J.~A., {Cort{\'e}s-Contreras}, M.,
  {Solano}, E., \& {Montes}, D. 2015, \aap, 583, A85

\bibitem[{{Anglada-Escud{\'e}} {et~al.}(2016){Anglada-Escud{\'e}}, {Amado},
  {Barnes}, {Berdi{\~n}as}, {Butler}, {Coleman}, {de La Cueva}, {Dreizler},
  {Endl}, {Giesers}, {Jeffers}, {Jenkins}, {Jones}, {Kiraga}, {K{\\"u}rster},
  {L{\'o}pez-Gonz{\'a}lez}, {Marvin}, {Morales}, {Morin}, {Nelson}, {Ortiz},
  {Ofir}, {Paardekooper}, {Reiners}, {Rodr{\'\i}guez},
  {Rodr{\'\i}guez-L{\'o}pez}, {Sarmiento}, {Strachan}, {Tsapras}, {Tuomi}, \&
  {Zechmeister}}]{2016Natur.536..437A}
{Anglada-Escud{\'e}}, G., {Amado}, P.~J., {Barnes}, J., {et~al.} 2016, \ at,
  536, 437

\bibitem[{{Anglada-Escude} {et~al.}(2014){Anglada-Escude}, {Arriagada},
  {Tuomi}, {Zechmeister}, {Jenkins}, {Ofir}, {Dreizler}, {Gerlach}, {Marvin},
  {Reiners}, {Jeffers}, {Butler}, {Vogt}, {Amado}, {Rodriguez-Lopez},
  {Berdinas}, {Morin}, {Crane}, {Shectman}, {Thompson}, {Diaz}, {Rivera},
  {Sarmiento}, \& {Jones}}]{2014MNRAS.443L..89A}
{Anglada-Escude}, G., {Arriagada}, P., {Tuomi}, M., {et~al.} 2014, \mnras, 443,
  L89

\bibitem[{{Anglada-Escud{\'e}} {et~al.}(2012){Anglada-Escud{\'e}}, {Arriagada},
  {Vogt}, {Rivera}, {Butler}, {Crane}, {Shectman}, {Thompson}, {Minniti},
  {Haghighipour}, {Carter}, {Tinney}, {Wittenmyer}, {Bailey}, {O'Toole},
  {Jones}, \& {Jenkins}}]{2012ApJ...751L..16A}
{Anglada-Escud{\'e}}, G., {Arriagada}, P., {Vogt}, S.~S., {et~al.} 2012, \apjl,
  751, L16

\bibitem[{{Anglada-Escud{\'e}} {et~al.}(2013){Anglada-Escud{\'e}}, {Tuomi},
  {Gerlach}, {Barnes}, {Heller}, {Jenkins}, {Wende}, {Vogt}, {Butler},
  {Reiners}, \& {Jones}}]{2013A&A...556A.126A}
{Anglada-Escud{\'e}}, G., {Tuomi}, M., {Gerlach}, E., {et~al.} 2013, \aap, 556,
  A126

\bibitem[{{Astudillo-Defru} {et~al.}(2017{\natexlab{a}}){Astudillo-Defru},
  {D{\'\i}az}, {Bonfils}, {Almenara}, {Delisle}, {Bouchy}, {Delfosse},
  {Forveille}, {Lovis}, {Mayor}, {Murgas}, {Pepe}, {Santos}, {S{\'e}gransan},
  {Udry}, \& {W{\\"u}nsche}}]{2017A&A...605L..11A}
{Astudillo-Defru}, N., {D{\'\i}az}, R.~F., {Bonfils}, X., {et~al.}
  2017{\natexlab{a}}, \aap, 605, L11

\bibitem[{{Astudillo-Defru} {et~al.}(2017{\natexlab{b}}){Astudillo-Defru},
  {Forveille}, {Bonfils}, {S{\'e}gransan}, {Bouchy}, {Delfosse}, {Lovis},
  {Mayor}, {Murgas}, {Pepe}, {Santos}, {Udry}, \&
  {W{\\"u}nsche}}]{2017A&A...602A..88A}
{Astudillo-Defru}, N., {Forveille}, T., {Bonfils}, X., {et~al.}
  2017{\natexlab{b}}, \aap, 602, A88

\bibitem[{{Bailey} {et~al.}(2009){Bailey}, {Butler}, {Tinney}, {Jones},
  {O'Toole}, {Carter}, \& {Marcy}}]{2009ApJ...690..743B}
{Bailey}, J., {Butler}, R.~P., {Tinney}, C.~G., {et~al.} 2009, \apj, 690, 743

\bibitem[{{Ball} {et~al.}(2005){Ball}, {Drake}, {Lin}, {Kashyap}, {Laming}, \&
  {Garc{\'i}a-Alvarez}}]{2005ApJ...634.1336B}
{Ball}, B., {Drake}, J.~J., {Lin}, L., {et~al.} 2005, \apj, 634, 1336

\bibitem[{{Barnes} {et~al.}(2014){Barnes}, {Jenkins}, {Jones}, {Jeffers},
  {Rojo}, {Arriagada}, {Jord{\'a}n}, {Minniti}, {Tuomi}, {Pinfield}, \&
  {Anglada-Escud{\'e}}}]{2014MNRAS.439.3094B}
{Barnes}, J.~R., {Jenkins}, J.~S., {Jones}, H.~R.~A., {et~al.} 2014, \mnras,
  439, 3094

\bibitem[{{Baroch} {et~al.}(2018){Baroch}, {Morales}, {Ribas}, {Tal-Or},
  {Zechmeister}, {Reiners}, {Caballero}, {Quirrenbach}, {Amado}, {Dreizler},
  {Lalitha}, {Jeffers}, {Lafarga}, {B{\'e}jar}, {Colom{\'e}},
  {Cort{\'e}s-Contreras}, {D{\'\i}ez-Alonso}, {Galad{\'\i}-Enr{\'\i}quez},
  {Guenther}, {Hagen}, {Henning}, {Herrero}, {K{\\"u}rster}, {Montes}, {Nagel},
  {Passegger}, {Perger}, {Rosich}, {Schweitzer}, \&
  {Seifert}}]{2018A&A...619A..32B}
{Baroch}, D., {Morales}, J.~C., {Ribas}, I., {et~al.} 2018, \aap, 619, A32

\bibitem[{{Barry} {et~al.}(2012){Barry}, {Demory}, {S{\'e}gransan},
  {Forveille}, {Danchi}, {Di Folco}, {Queloz}, {Spooner}, {Torres}, {Traub},
  {Delfosse}, {Mayor}, {Perrier}, \& {Udry}}]{2012ApJ...760...55B}
{Barry}, R.~K., {Demory}, B.~O., {S{\'e}gransan}, D., {et~al.} 2012, \apj, 760,
  55

\bibitem[{{Bartlett}(2007)}]{2007PhDT.........2B}
{Bartlett}, J.~L. 2007, PhD thesis, University of Virginia

\bibitem[{{Bauer} {et~al.}(2020){Bauer}, {Zechmeister}, {Kaminski},
  {Rodr{\'\i}guez L{\'o}pez}, {Caballero}, {Azzaro}, {Stahl}, {Kossakowski},
  {Quirrenbach}, {Becerril Jarque}, {Rodr{\'\i}guez}, {Amado}, {Seifert},
  {Reiners}, {Sch{\\"a}fer}, {Ribas}, {B{\'e}jar}, {Cort{\'e}s-Contreras},
  {Dreizler}, {Hatzes}, {Henning}, {Jeffers}, {K{\\"u}rster}, {Lafarga},
  {Montes}, {Morales}, {Schmitt}, {Schweitzer}, \&
  {Solano}}]{2020A&A...640A..50B}
{Bauer}, F.~F., {Zechmeister}, M., {Kaminski}, A., {et~al.} 2020, \aap, 640,
  A50

\bibitem[{{Beam{\'\i}n} {et~al.}(2013){Beam{\'\i}n}, {Minniti}, {Gromadzki},
  {Kurtev}, {Ivanov}, {Beletsky}, {Lucas}, {Saito}, \&
  {Borissova}}]{2013A&A...557L...8B}
{Beam{\'\i}n}, J.~C., {Minniti}, D., {Gromadzki}, M., {et~al.} 2013, \aap, 557,
  L8

\bibitem[{{Bell} {et~al.}(2015){Bell}, {Mamajek}, \&
  {Naylor}}]{2015MNRAS.454..593B}
{Bell}, C. P.~M., {Mamajek}, E.~E., \& {Naylor}, T. 2015, \mnras, 454, 593

\bibitem[{{Benedict} {et~al.}(2016){Benedict}, {Henry}, {Franz}, {McArthur},
  {Wasserman}, {Jao}, {Cargile}, {Dieterich}, {Bradley}, {Nelan}, \&
  {Whipple}}]{2016AJ....152..141B}
{Benedict}, G.~F., {Henry}, T.~J., {Franz}, O.~G., {et~al.} 2016, \aj, 152, 141

\bibitem[{{Benedict} {et~al.}(2002){Benedict}, {McArthur}, {Forveille},
  {Delfosse}, {Nelan}, {Butler}, {Spiesman}, {Marcy}, {Goldman}, {Perrier},
  {Jefferys}, \& {Mayor}}]{2002ApJ...581L.115B}
{Benedict}, G.~F., {McArthur}, B.~E., {Forveille}, T., {et~al.} 2002, \apjl,
  581, L115

\bibitem[{{Berdi{\~n}as} {et~al.}(2016){Berdi{\~n}as}, {Amado},
  {Anglada-Escud{\'e}}, {Rodr{\'\i}guez-L{\'o}pez}, \&
  {Barnes}}]{2016MNRAS.459.3551B}
{Berdi{\~n}as}, Z.~M., {Amado}, P.~J., {Anglada-Escud{\'e}}, G.,
  {Rodr{\'\i}guez-L{\'o}pez}, C., \& {Barnes}, J. 2016, \mnras, 459, 3551

\bibitem[{{Bessel}(1838)}]{1838MNRAS...4..152B}
{Bessel}, F.~W. 1838, \mnras, 4, 152

\bibitem[{{Best} {et~al.}(2020){Best}, {Liu}, {Magnier}, \&
  {Dupuy}}]{2020AJ....159..257B}
{Best}, W. M.~J., {Liu}, M.~C., {Magnier}, E.~A., \& {Dupuy}, T.~J. 2020, \aj,
  159, 257

\bibitem[{{Beuzit} {et~al.}(2004){Beuzit}, {S{\'e}gransan}, {Forveille},
  {Udry}, {Delfosse}, {Mayor}, {Perrier}, {Hainaut}, {Roddier}, {Roddier}, \&
  {Mart{\'\i}n}}]{2004A&A...425..997B}
{Beuzit}, J.~L., {S{\'e}gransan}, D., {Forveille}, T., {et~al.} 2004, \aap,
  425, 997

\bibitem[{{Bidelman}(1985)}]{1985ApJS...59..197B}
{Bidelman}, W.~P. 1985, \apjs, 59, 197

\bibitem[{{Blake} {et~al.}(2010){Blake}, {Charbonneau}, \&
  {White}}]{2010ApJ...723..684B}
{Blake}, C.~H., {Charbonneau}, D., \& {White}, R.~J. 2010, \apj, 723, 684

\bibitem[{{Blazit} {et~al.}(1987){Blazit}, {Bonneau}, \&
  {Foy}}]{1987A&AS...71...57B}
{Blazit}, A., {Bonneau}, D., \& {Foy}, R. 1987, \aaps, 71, 57

\bibitem[{{Bochanski} {et~al.}(2010){Bochanski}, {Hawley}, {Covey}, {West},
  {Reid}, {Golimowski}, \& {Ivezi{\'c}}}]{2010AJ....139.2679B}
{Bochanski}, J.~J., {Hawley}, S.~L., {Covey}, K.~R., {et~al.} 2010, \aj, 139,
  2679

\bibitem[{{Bonavita} \& {Desidera}(2007)}]{2007A&A...468..721B}
{Bonavita}, M. \& {Desidera}, S. 2007, \aap, 468, 721

\bibitem[{{Bond} {et~al.}(2015){Bond}, {Gilliland}, {Schaefer}, {Demarque},
  {Girard}, {Holberg}, {Gudehus}, {Mason}, {Kozhurina-Platais}, {Burleigh},
  {Barstow}, \& {Nelan}}]{2015ApJ...813..106B}
{Bond}, H.~E., {Gilliland}, R.~L., {Schaefer}, G.~H., {et~al.} 2015, \apj, 813,
  106

\bibitem[{{Bonfils} {et~al.}(2018){Bonfils}, {Astudillo-Defru}, {D{\'\i}az},
  {Almenara}, {Forveille}, {Bouchy}, {Delfosse}, {Lovis}, {Mayor}, {Murgas},
  {Pepe}, {Santos}, {S{\'e}gransan}, {Udry}, \&
  {W{\\"u}nsche}}]{2018A&A...613A..25B}
{Bonfils}, X., {Astudillo-Defru}, N., {D{\'\i}az}, R., {et~al.} 2018, \aap,
  613, A25

\bibitem[{{Bonfils} {et~al.}(2013){Bonfils}, {Delfosse}, {Udry}, {Forveille},
  {Mayor}, {Perrier}, {Bouchy}, {Gillon}, {Lovis}, {Pepe}, {Queloz}, {Santos},
  {S{\'e}gransan}, \& {Bertaux}}]{2013A&A...549A.109B}
{Bonfils}, X., {Delfosse}, X., {Udry}, S., {et~al.} 2013, \aap, 549, A109

\bibitem[{{Bonfils} {et~al.}(2007){Bonfils}, {Mayor}, {Delfosse}, {Forveille},
  {Gillon}, {Perrier}, {Udry}, {Bouchy}, {Lovis}, {Pepe}, {Queloz}, {Santos},
  \& {Bertaux}}]{2007A&A...474..293B}
{Bonfils}, X., {Mayor}, M., {Delfosse}, X., {et~al.} 2007, \aap, 474, 293

\bibitem[{{Bowler} {et~al.}(2010){Bowler}, {Liu}, \&
  {Dupuy}}]{2010APJ...710...45B}
{Bowler}, B.~P., {Liu}, M.~C., \& {Dupuy}, T.~J. 2010, \apj, 710, 45

\bibitem[{{Bowler} {et~al.}(2015){Bowler}, {Liu}, {Shkolnik}, \&
  {Tamura}}]{2015ApJS..216....7B}
{Bowler}, B.~P., {Liu}, M.~C., {Shkolnik}, E.~L., \& {Tamura}, M. 2015, \apjs,
  216, 7

\bibitem[{{Burgasser} {et~al.}(2010{\natexlab{a}}){Burgasser}, {Cruz},
  {Cushing}, {Gelino}, {Looper}, {Faherty}, {Kirkpatrick}, \&
  {Reid}}]{2010ApJ...710.1142B}
{Burgasser}, A.~J., {Cruz}, K.~L., {Cushing}, M., {et~al.} 2010{\natexlab{a}},
  \apj, 710, 1142

\bibitem[{{Burgasser} {et~al.}(2006){Burgasser}, {Geballe}, {Leggett},
  {Kirkpatrick}, \& {Golimowski}}]{2006ApJ...637.1067B}
{Burgasser}, A.~J., {Geballe}, T.~R., {Leggett}, S.~K., {Kirkpatrick}, J.~D.,
  \& {Golimowski}, D.~A. 2006, \apj, 637, 1067

\bibitem[{{Burgasser} {et~al.}(2015{\natexlab{a}}){Burgasser}, {Gillon},
  {Melis}, {Bowler}, {Michelsen}, {Bardalez Gagliuffi}, {Gelino}, {Jehin},
  {Delrez}, {Manfroid}, \& {Blake}}]{2015AJ....149..104B}
{Burgasser}, A.~J., {Gillon}, M., {Melis}, C., {et~al.} 2015{\natexlab{a}},
  \aj, 149, 104

\bibitem[{{Burgasser} {et~al.}(2000){Burgasser}, {Kirkpatrick}, {Cutri},
  {McCallon}, {Kopan}, {Gizis}, {Liebert}, {Reid}, {Brown}, {Monet}, {Dahn},
  {Beichman}, \& {Skrutskie}}]{2000ApJ...531L..57B}
{Burgasser}, A.~J., {Kirkpatrick}, J.~D., {Cutri}, R.~M., {et~al.} 2000, \apjl,
  531, L57

\bibitem[{{Burgasser} {et~al.}(2003){Burgasser}, {Kirkpatrick}, {McElwain},
  {Cutri}, {Burgasser}, \& {Skrutskie}}]{2003AJ....125..850B}
{Burgasser}, A.~J., {Kirkpatrick}, J.~D., {McElwain}, M.~W., {et~al.} 2003,
  \aj, 125, 850

\bibitem[{{Burgasser} {et~al.}(2015{\natexlab{b}}){Burgasser}, {Logsdon},
  {Gagn{\'e}}, {Bochanski}, {Faherty}, {West}, {Mamajek}, {Schmidt}, \&
  {Cruz}}]{2015ApJS..220...18B}
{Burgasser}, A.~J., {Logsdon}, S.~E., {Gagn{\'e}}, J., {et~al.}
  2015{\natexlab{b}}, \apjs, 220, 18

\bibitem[{{Burgasser} {et~al.}(2010{\natexlab{b}}){Burgasser}, {Looper}, \&
  {Rayner}}]{2010AJ....139.2448B}
{Burgasser}, A.~J., {Looper}, D., \& {Rayner}, J.~T. 2010{\natexlab{b}}, \aj,
  139, 2448

\bibitem[{{Burgasser} {et~al.}(2008){Burgasser}, {Tinney}, {Cushing}, {Saumon},
  {Marley}, {Bennett}, \& {Kirkpatrick}}]{2008ApJ...689L..53B}
{Burgasser}, A.~J., {Tinney}, C.~G., {Cushing}, M.~C., {et~al.} 2008, \apjl,
  689, L53

\bibitem[{{Burningham} {et~al.}(2010){Burningham}, {Leggett}, {Lucas},
  {Pinfield}, {Smart}, {Day-Jones}, {Jones}, {Murray}, {Nickson}, {Tamura},
  {Zhang}, {Lodieu}, {Tinney}, \& {Zapatero Osorio}}]{2010MNRAS.404.1952B}
{Burningham}, B., {Leggett}, S.~K., {Lucas}, P.~W., {et~al.} 2010, \mnras, 404,
  1952

\bibitem[{{Burt} {et~al.}(2021){Burt}, {Feng}, {Holden}, {Mamajek}, {Huang},
  {Rosenthal}, {Wang}, {Butler}, {Vogt}, {Laughlin}, {Henry}, {Teske}, {Wang},
  {Crane}, \& {Shectman}}]{2021AJ....161...10B}
{Burt}, J., {Feng}, F., {Holden}, B., {et~al.} 2021, \aj, 161, 10

\bibitem[{{Butler} {et~al.}(2006{\natexlab{a}}){Butler}, {Johnson}, {Marcy},
  {Wright}, {Vogt}, \& {Fischer}}]{2006PASP..118.1685B}
{Butler}, R.~P., {Johnson}, J.~A., {Marcy}, G.~W., {et~al.} 2006{\natexlab{a}},
  \pasp, 118, 1685

\bibitem[{{Butler} {et~al.}(2004){Butler}, {Vogt}, {Marcy}, {Fischer},
  {Wright}, {Henry}, {Laughlin}, \& {Lissauer}}]{2004ApJ...617..580B}
{Butler}, R.~P., {Vogt}, S.~S., {Marcy}, G.~W., {et~al.} 2004, \apj, 617, 580

\bibitem[{{Butler} {et~al.}(2006{\natexlab{b}}){Butler}, {Wright}, {Marcy},
  {Fischer}, {Vogt}, {Tinney}, {Jones}, {Carter}, {Johnson}, {McCarthy}, \&
  {Penny}}]{2006ApJ...646..505B}
{Butler}, R.~P., {Wright}, J.~T., {Marcy}, G.~W., {et~al.} 2006{\natexlab{b}},
  \apj, 646, 505

\bibitem[{{Caballero}(2009)}]{2009A&A...507..251C}
{Caballero}, J.~A. 2009, \aap, 507, 251

\bibitem[{{Cayrel de Strobel} {et~al.}(1989){Cayrel de Strobel}, {Perrin},
  {Cayrel}, \& {Lebreton}}]{1989A&A...225..369C}
{Cayrel de Strobel}, G., {Perrin}, M.~N., {Cayrel}, R., \& {Lebreton}, Y. 1989,
  \aap, 225, 369

\bibitem[{{Chiu} {et~al.}(2006){Chiu}, {Fan}, {Leggett}, {Golimowski}, {Zheng},
  {Geballe}, {Schneider}, \& {Brinkmann}}]{2006AJ....131.2722C}
{Chiu}, K., {Fan}, X., {Leggett}, S.~K., {et~al.} 2006, \aj, 131, 2722

\bibitem[{{Cifuentes} {et~al.}(2020){Cifuentes}, {Caballero},
  {Cort{\'e}s-Contreras}, {Montes}, {Abell{\'a}n}, {Dorda}, {Holgado},
  {Zapatero Osorio}, {Morales}, {Amado}, {Passegger}, {Quirrenbach}, {Reiners},
  {Ribas}, {Sanz-Forcada}, {Schweitzer}, {Seifert}, \&
  {Solano}}]{2020A&A...642A.115C}
{Cifuentes}, C., {Caballero}, J.~A., {Cort{\'e}s-Contreras}, M., {et~al.} 2020,
  \aap, 642, A115

\bibitem[{{Corbally}(1984)}]{1984ApJS...55..657C}
{Corbally}, C.~J. 1984, \apjs, 55, 657

\bibitem[{{Correia} {et~al.}(2010){Correia}, {Couetdic}, {Laskar}, {Bonfils},
  {Mayor}, {Bertaux}, {Bouchy}, {Delfosse}, {Forveille}, {Lovis}, {Pepe},
  {Perrier}, {Queloz}, \& {Udry}}]{2010A&A...511A..21C}
{Correia}, A.~C.~M., {Couetdic}, J., {Laskar}, J., {et~al.} 2010, \aap, 511,
  A21

\bibitem[{{Cort{\'e}s-Contreras} {et~al.}(2017){Cort{\'e}s-Contreras},
  {B{\'e}jar}, {Caballero}, {Gauza}, {Montes}, {Alonso-Floriano}, {Jeffers},
  {Morales}, {Reiners}, {Ribas}, {Sch{\\"o}fer}, {Quirrenbach}, {Amado},
  {Mundt}, \& {Seifert}}]{2017A&A...597A..47C}
{Cort{\'e}s-Contreras}, M., {B{\'e}jar}, V.~J.~S., {Caballero}, J.~A., {et~al.}
  2017, \aap, 597, A47

\bibitem[{{Cowley} {et~al.}(1967){Cowley}, {Hiltner}, \&
  {Witt}}]{1967AJ.....72.1334C}
{Cowley}, A.~P., {Hiltner}, W.~A., \& {Witt}, A.~N. 1967, \aj, 72, 1334

\bibitem[{{Crill} {et~al.}(2020){Crill}, {Werner}, {Akeson}, {Ashby}, {Bleem},
  {Bock}, {Bryan}, {Burnham}, {Byunh}, {Chang}, {Chiang}, {Cook}, {Cooray},
  {Davis}, {Dor{\'e}}, {Dowell}, {Dubois-Felsmann}, {Eifler}, {Faisst},
  {Habib}, {Heinrich}, {Heitmann}, {Heaton}, {Hirata}, {Hristov}, {Hui},
  {Jeong}, {Kang}, {Kecman}, {Kirkpatrick}, {Korngut}, {Krause}, {Lee},
  {Lisse}, {Masters}, {Mauskopf}, {Melnick}, {Miyasaka}, {Nayyeri}, {Nguyen},
  {{\"O}berg}, {Padin}, {Paladini}, {Pourrahmani}, {Pyo}, {Smith}, {Song},
  {Symons}, {Teplitz}, {Tolls}, {Unwin}, {Windhorst}, {Yang}, \&
  {Zemcov}}]{2020SPIE11443E..0IC}
{Crill}, B.~P., {Werner}, M., {Akeson}, R., {et~al.} 2020, in Society of
  Photo-Optical Instrumentation Engineers (SPIE) Conference Series, Vol. 11443,
  Society of Photo-Optical Instrumentation Engineers (SPIE) Conference Series,
  114430I

\bibitem[{{Crosley} \& {Osten}(2018)}]{2018ApJ...856...39C}
{Crosley}, M.~K. \& {Osten}, R.~A. 2018, \apj, 856, 39

\bibitem[{{Cruz} {et~al.}(2007){Cruz}, {Reid}, {Kirkpatrick}, {Burgasser},
  {Liebert}, {Solomon}, {Schmidt}, {Allen}, {Hawley}, \&
  {Covey}}]{2007AJ....133..439C}
{Cruz}, K.~L., {Reid}, I.~N., {Kirkpatrick}, J.~D., {et~al.} 2007, \aj, 133,
  439

\bibitem[{{Cuartas-Restrepo} {et~al.}(2016){Cuartas-Restrepo}, {Melita},
  {Zuluaga}, {Portilla-Revelo}, {Sucerquia}, \& {Miloni}}]{2016MNRAS.463.1592C}
{Cuartas-Restrepo}, P.~A., {Melita}, M., {Zuluaga}, J.~I., {et~al.} 2016,
  \mnras, 463, 1592

\bibitem[{{Damasso}(2019)}]{2019ESS.....410203D}
{Damasso}, M. 2019, in AAS/Division for Extreme Solar Systems Abstracts,
  Vol.~51, AAS/Division for Extreme Solar Systems Abstracts, 102.03

\bibitem[{{Damasso} {et~al.}(2020){Damasso}, {Del Sordo}, {Anglada-Escud{\'e}},
  {Giacobbe}, {Sozzetti}, {Morbidelli}, {Pojmanski}, {Barbato}, {Butler},
  {Jones}, {Hambsch}, {Jenkins}, {L{\'o}pez-Gonz{\'a}lez}, {Morales}, {Pe{\~n}a
  Rojas}, {Rodr{\'\i}guez-L{\'o}pez}, {Rodr{\'\i}guez}, {Amado}, {Anglada},
  {Feng}, \& {G{\'o}mez}}]{2020SciA....6.7467D}
{Damasso}, M., {Del Sordo}, F., {Anglada-Escud{\'e}}, G., {et~al.} 2020,
  Science Advances, 6, eaax7467

\bibitem[{{Davison} {et~al.}(2015){Davison}, {White}, {Henry}, {Riedel}, {Jao},
  {Bailey}, {Quinn}, {Cantrell}, {Subasavage}, \&
  {Winters}}]{2015AJ....149..106D}
{Davison}, C.~L., {White}, R.~J., {Henry}, T.~J., {et~al.} 2015, \aj, 149, 106

\bibitem[{{Davison} {et~al.}(2014){Davison}, {White}, {Jao}, {Henry}, {Bailey},
  {Quinn}, {Cantrell}, {Riedel}, {Subasavage}, {Winters}, \&
  {Crockett}}]{2014AJ....147...26D}
{Davison}, C.~L., {White}, R.~J., {Jao}, W.~C., {et~al.} 2014, \aj, 147, 26

\bibitem[{{Deka-Szymankiewicz} {et~al.}(2018){Deka-Szymankiewicz},
  {Niedzielski}, {Adamczyk}, {Adam{\'o}w}, {Nowak}, \&
  {Wolszczan}}]{2018A&A...615A..31D}
{Deka-Szymankiewicz}, B., {Niedzielski}, A., {Adamczyk}, M., {et~al.} 2018,
  \aap, 615, A31

\bibitem[{{Delfosse} {et~al.}(2013){Delfosse}, {Bonfils}, {Forveille}, {Udry},
  {Mayor}, {Bouchy}, {Gillon}, {Lovis}, {Neves}, {Pepe}, {Perrier}, {Queloz},
  {Santos}, \& {S{\'e}gransan}}]{2013A&A...553A...8D}
{Delfosse}, X., {Bonfils}, X., {Forveille}, T., {et~al.} 2013, \aap, 553, A8

\bibitem[{{Delfosse} {et~al.}(1999){Delfosse}, {Forveille}, {Beuzit}, {Udry},
  {Mayor}, \& {Perrier}}]{1999A&A...344..897D}
{Delfosse}, X., {Forveille}, T., {Beuzit}, J.~L., {et~al.} 1999, \aap, 344, 897

\bibitem[{{Delorme} {et~al.}(2010){Delorme}, {Albert}, {Forveille}, {Artigau},
  {Delfosse}, {Reyl{\'e}}, {Willott}, {Bertin}, {Wilkins}, {Allard}, \&
  {Arzoumanian}}]{2010A&A...518A..39D}
{Delorme}, P., {Albert}, L., {Forveille}, T., {et~al.} 2010, \aap, 518, A39

\bibitem[{{D{\'\i}az} {et~al.}(2019){D{\'\i}az}, {Delfosse}, {Hobson},
  {Boisse}, {Astudillo-Defru}, {Bonfils}, {Henry}, {Arnold}, {Bouchy},
  {Bourrier}, {Brugger}, {Dalal}, {Deleuil}, {Demangeon}, {Dolon}, {Dumusque},
  {Forveille}, {Hara}, {H{\'e}brard}, {Kiefer}, {Lopez}, {Mignon}, {Moreau},
  {Mousis}, {Moutou}, {Pepe}, {Perruchot}, {Richaud}, {Santerne}, {Santos},
  {Sottile}, {Stalport}, {S{\'e}gransan}, {Udry}, {Unger}, \&
  {Wilson}}]{2019A&A...625A..17D}
{D{\'\i}az}, R.~F., {Delfosse}, X., {Hobson}, M.~J., {et~al.} 2019, \aap, 625,
  A17

\bibitem[{{D{\'\i}az} {et~al.}(2016){D{\'\i}az}, {Rey}, {Demangeon},
  {H{\'e}brard}, {Boisse}, {Arnold}, {Astudillo-Defru}, {Beuzit}, {Bonfils},
  {Borgniet}, {Bouchy}, {Bourrier}, {Courcol}, {Deleuil}, {Delfosse},
  {Ehrenreich}, {Forveille}, {Lagrange}, {Mayor}, {Moutou}, {Pepe}, {Queloz},
  {Santerne}, {Santos}, {Sahlmann}, {S{\'e}gransan}, {Udry}, \&
  {Wilson}}]{2016A&A...591A.146D}
{D{\'\i}az}, R.~F., {Rey}, J., {Demangeon}, O., {et~al.} 2016, \aap, 591, A146

\bibitem[{{Dieterich} {et~al.}(2012){Dieterich}, {Henry}, {Golimowski},
  {Krist}, \& {Tanner}}]{2012AJ....144...64D}
{Dieterich}, S.~B., {Henry}, T.~J., {Golimowski}, D.~A., {Krist}, J.~E., \&
  {Tanner}, A.~M. 2012, \aj, 144, 64

\bibitem[{{Dittmann} {et~al.}(2014){Dittmann}, {Irwin}, {Charbonneau}, \&
  {Berta-Thompson}}]{2014ApJ...784..156D}
{Dittmann}, J.~A., {Irwin}, J.~M., {Charbonneau}, D., \& {Berta-Thompson},
  Z.~K. 2014, \apj, 784, 156

\bibitem[{{Docobo} {et~al.}(2019){Docobo}, {Gomez}, {Campo}, {Andrade},
  {Horch}, {Costa}, \& {Mendez}}]{2019MNRAS.482.4096D}
{Docobo}, J.~A., {Gomez}, J., {Campo}, P.~P., {et~al.} 2019, \mnras, 482, 4096

\bibitem[{{Docobo} {et~al.}(2006){Docobo}, {Tamazian}, {Balega}, \&
  {Melikian}}]{2006AJ....132..994D}
{Docobo}, J.~A., {Tamazian}, V.~S., {Balega}, Y.~Y., \& {Melikian}, N.~D. 2006,
  \aj, 132, 994

\bibitem[{{Dommanget} \& {Nys}(2002)}]{2002yCat.1274....0D}
{Dommanget}, J. \& {Nys}, O. 2002, VizieR Online Data Catalog, I/274

\bibitem[{{Dreizler} {et~al.}(2020){Dreizler}, {Jeffers}, {Rodr{\'\i}guez},
  {Zechmeister}, {Barnes}, {Haswell}, {Coleman}, {Lalitha}, {Hidalgo Soto},
  {Strachan}, {Hambsch}, {L{\'o}pez-Gonz{\'a}lez}, {Morales}, {Rodr{\'\i}guez
  L{\'o}pez}, {Berdi{\~n}as}, {Ribas}, {Pall{\'e}}, {Reiners}, \&
  {Anglada-Escud{\'e}}}]{2020MNRAS.493..536D}
{Dreizler}, S., {Jeffers}, S.~V., {Rodr{\'\i}guez}, E., {et~al.} 2020, \mnras,
  493, 536

\bibitem[{{Dressing} \& {Charbonneau}(2015)}]{2015ApJ...807...45D}
{Dressing}, C.~D. \& {Charbonneau}, D. 2015, \apj, 807, 45

\bibitem[{{Dupuy} \& {Liu}(2012)}]{2012ApJS..201...19D}
{Dupuy}, T.~J. \& {Liu}, M.~C. 2012, \apjs, 201, 19

\bibitem[{{Dupuy} \& {Liu}(2017)}]{2017ApJS..231...15D}
{Dupuy}, T.~J. \& {Liu}, M.~C. 2017, \apjs, 231, 15

\bibitem[{{Eggen}(1956)}]{1956AJ.....61..405E}
{Eggen}, O.~J. 1956, \aj, 61, 405

\bibitem[{{Eggenberger} {et~al.}(2008){Eggenberger}, {Miglio}, {Carrier},
  {Fernandes}, \& {Santos}}]{2008A&A...482..631E}
{Eggenberger}, P., {Miglio}, A., {Carrier}, F., {Fernandes}, J., \& {Santos},
  N.~C. 2008, \aap, 482, 631

\bibitem[{{Eggleton} \& {Tokovinin}(2008)}]{2008MNRAS.389..869E}
{Eggleton}, P.~P. \& {Tokovinin}, A.~A. 2008, \mnras, 389, 869

\bibitem[{{Evans} {et~al.}(1957){Evans}, {Menzies}, \&
  {Stoy}}]{1957MNRAS.117..534E}
{Evans}, D.~S., {Menzies}, A., \& {Stoy}, R.~H. 1957, \mnras, 117, 534

\bibitem[{{Faherty} {et~al.}(2012){Faherty}, {Burgasser}, {Walter}, {Van der
  Bliek}, {Shara}, {Cruz}, {West}, {Vrba}, \&
  {Anglada-Escud{\'e}}}]{2012ApJ...752...56F}
{Faherty}, J.~K., {Burgasser}, A.~J., {Walter}, F.~M., {et~al.} 2012, \apj,
  752, 56

\bibitem[{{Faherty} {et~al.}(2018){Faherty}, {Gagn{\'e}}, {Burgasser},
  {Mamajek}, {Gonzales}, {Bardalez Gagliuffi}, \&
  {Marocco}}]{2018ApJ...868...44F}
{Faherty}, J.~K., {Gagn{\'e}}, J., {Burgasser}, A.~J., {et~al.} 2018, \apj,
  868, 44

\bibitem[{{Faherty} {et~al.}(2016){Faherty}, {Riedel}, {Cruz}, {Gagne},
  {Filippazzo}, {Lambrides}, {Fica}, {Weinberger}, {Thorstensen}, {Tinney},
  {Baldassare}, {Lemonier}, \& {Rice}}]{2016ApJS..225...10F}
{Faherty}, J.~K., {Riedel}, A.~R., {Cruz}, K.~L., {et~al.} 2016, \apjs, 225, 10

\bibitem[{{Farrington} {et~al.}(2010){Farrington}, {ten Brummelaar}, {Mason},
  {Hartkopf}, {McAlister}, {Raghavan}, {Turner}, {Sturmann}, {Sturmann}, \&
  {Ridgway}}]{2010AJ....139.2308F}
{Farrington}, C.~D., {ten Brummelaar}, T.~A., {Mason}, B.~D., {et~al.} 2010,
  \aj, 139, 2308

\bibitem[{{Feng} {et~al.}(2019){Feng}, {Anglada-Escud{\'e}}, {Tuomi}, {Jones},
  {Chanam{\'e}}, {Butler}, \& {Janson}}]{2019MNRAS.490.5002F}
{Feng}, F., {Anglada-Escud{\'e}}, G., {Tuomi}, M., {et~al.} 2019, \mnras, 490,
  5002

\bibitem[{{Feng} {et~al.}(2020{\natexlab{a}}){Feng}, {Butler}, {Shectman},
  {Crane}, {Vogt}, {Chambers}, {Jones}, {Xuesong Wang}, {Teske}, {Burt},
  {D{\'\i}az}, \& {Thompson}}]{2020ApJS..246...11F}
{Feng}, F., {Butler}, R.~P., {Shectman}, S.~A., {et~al.} 2020{\natexlab{a}},
  \apjs, 246, 11

\bibitem[{{Feng} {et~al.}(2020{\natexlab{b}}){Feng}, {Shectman}, {Clement},
  {Vogt}, {Tuomi}, {Teske}, {Burt}, {Crane}, {Holden}, {Wang}, {Thompson},
  {D{\'\i}az}, \& {Butler}}]{2020ApJS..250...29F}
{Feng}, F., {Shectman}, S.~A., {Clement}, M.~S., {et~al.} 2020{\natexlab{b}},
  \apjs, 250, 29

\bibitem[{{Feng} {et~al.}(2017{\natexlab{a}}){Feng}, {Tuomi}, \&
  {Jones}}]{2017A&A...605A.103F}
{Feng}, F., {Tuomi}, M., \& {Jones}, H.~R.~A. 2017{\natexlab{a}}, \aap, 605,
  A103

\bibitem[{{Feng} {et~al.}(2017{\natexlab{b}}){Feng}, {Tuomi}, {Jones},
  {Barnes}, {Anglada-Escud{\'e}}, {Vogt}, \& {Butler}}]{2017AJ....154..135F}
{Feng}, F., {Tuomi}, M., {Jones}, H.~R.~A., {et~al.} 2017{\natexlab{b}}, \aj,
  154, 135

\bibitem[{{Feng} {et~al.}(2015){Feng}, {Wright}, {Nelson}, {Wang}, {Ford},
  {Marcy}, {Isaacson}, \& {Howard}}]{2015ApJ...800...22F}
{Feng}, Y.~K., {Wright}, J.~T., {Nelson}, B., {et~al.} 2015, \apj, 800, 22

\bibitem[{{Fernandes} {et~al.}(1998){Fernandes}, {Lebreton}, {Baglin}, \&
  {Morel}}]{1998A&A...338..455F}
{Fernandes}, J., {Lebreton}, Y., {Baglin}, A., \& {Morel}, P. 1998, \aap, 338,
  455

\bibitem[{{Feroz} \& {Hobson}(2014)}]{2014MNRAS.437.3540F}
{Feroz}, F. \& {Hobson}, M.~P. 2014, \mnras, 437, 3540

\bibitem[{{Fleischer}(1957)}]{1957AJ.....62..379F}
{Fleischer}, R. 1957, \aj, 62, 379

\bibitem[{{Fouqu{\'e}} {et~al.}(2018){Fouqu{\'e}}, {Moutou}, {Malo},
  {Martioli}, {Lim}, {Rajpurohit}, {Artigau}, {Delfosse}, {Donati},
  {Forveille}, {Morin}, {Allard}, {Delage}, {Doyon}, {H{\'e}brard}, \&
  {Neves}}]{2018MNRAS.475.1960F}
{Fouqu{\'e}}, P., {Moutou}, C., {Malo}, L., {et~al.} 2018, \mnras, 475, 1960

\bibitem[{{Gaia Collaboration}(2018)}]{2018yCat.1345....0G}
{Gaia Collaboration}. 2018, VizieR Online Data Catalog, I/345

\bibitem[{{Gaia Collaboration}(2020)}]{2020yCat.1350....0G}
{Gaia Collaboration}. 2020, VizieR Online Data Catalog, I/350

\bibitem[{{Gaia Collaboration} {et~al.}(2018){Gaia Collaboration}, {Brown},
  {Vallenari}, {Prusti}, {de Bruijne}, {Babusiaux}, {Bailer-Jones}, {Biermann},
  {Evans}, {Eyer}, {Jansen}, {Jordi}, {Klioner}, {Lammers}, {Lindegren},
  {Luri}, {Mignard}, {Panem}, {Pourbaix}, {Randich}, {Sartoretti}, {Siddiqui},
  {Soubiran}, {van Leeuwen}, {Walton}, {Arenou}, {Bastian}, {Cropper},
  {Drimmel}, {Katz}, {Lattanzi}, {Bakker}, {Cacciari}, {Casta{\~n}eda},
  {Chaoul}, {Cheek}, {De Angeli}, {Fabricius}, {Guerra}, {Holl}, {Masana},
  {Messineo}, {Mowlavi}, {Nienartowicz}, {Panuzzo}, {Portell}, {Riello},
  {Seabroke}, {Tanga}, {Th{\'e}venin}, {Gracia-Abril}, {Comoretto},
  {Garcia-Reinaldos}, {Teyssier}, {Altmann}, {Andrae}, {Audard},
  {Bellas-Velidis}, {Benson}, {Berthier}, {Blomme}, {Burgess}, {Busso},
  {Carry}, {Cellino}, {Clementini}, {Clotet}, {Creevey}, {Davidson}, {De
  Ridder}, {Delchambre}, {Dell'Oro}, {Ducourant},
  {Fern{\'a}ndez-Hern{\'a}ndez}, {Fouesneau}, {Fr{\'e}mat}, {Galluccio},
  {Garc{\'\i}a-Torres}, {Gonz{\'a}lez-N{\'u}{\~n}ez}, {Gonz{\'a}lez-Vidal},
  {Gosset}, {Guy}, {Halbwachs}, {Hambly}, {Harrison}, {Hern{\'a}ndez},
  {Hestroffer}, {Hodgkin}, {Hutton}, {Jasniewicz}, {Jean-Antoine-Piccolo},
  {Jordan}, {Korn}, {Krone-Martins}, {Lanzafame}, {Lebzelter}, {L{\"o}ffler},
  {Manteiga}, {Marrese}, {Mart{\'\i}n-Fleitas}, {Moitinho}, {Mora}, {Muinonen},
  {Osinde}, {Pancino}, {Pauwels}, {Petit}, {Recio-Blanco}, {Richards},
  {Rimoldini}, {Robin}, {Sarro}, {Siopis}, {Smith}, {Sozzetti}, {S{\"u}veges},
  {Torra}, {van Reeven}, {Abbas}, {Abreu Aramburu}, {Accart}, {Aerts},
  {Altavilla}, {{\'A}lvarez}, {Alvarez}, {Alves}, {Anderson}, {Andrei},
  {Anglada Varela}, {Antiche}, {Antoja}, {Arcay}, {Astraatmadja}, {Bach},
  {Baker}, {Balaguer-N{\'u}{\~n}ez}, {Balm}, {Barache}, {Barata}, {Barbato},
  {Barblan}, {Barklem}, {Barrado}, {Barros}, {Barstow}, {Bartholom{\'e}
  Mu{\~n}oz}, {Bassilana}, {Becciani}, {Bellazzini}, {Berihuete}, {Bertone},
  {Bianchi}, {Bienaym{\'e}}, {Blanco-Cuaresma}, {Boch}, {Boeche}, {Bombrun},
  {Borrachero}, {Bossini}, {Bouquillon}, {Bourda}, {Bragaglia}, {Bramante},
  {Breddels}, {Bressan}, {Brouillet}, {Br{\"u}semeister}, {Brugaletta},
  {Bucciarelli}, {Burlacu}, {Busonero}, {Butkevich}, {Buzzi}, {Caffau},
  {Cancelliere}, {Cannizzaro}, {Cantat-Gaudin}, {Carballo}, {Carlucci},
  {Carrasco}, {Casamiquela}, {Castellani}, {Castro-Ginard}, {Charlot},
  {Chemin}, {Chiavassa}, {Cocozza}, {Costigan}, {Cowell}, {Crifo}, {Crosta},
  {Crowley}, {Cuypers}, {Dafonte}, {Damerdji}, {Dapergolas}, {David}, {David},
  {de Laverny}, {De Luise}, {De March}, {de Martino}, {de Souza}, {de Torres},
  {Debosscher}, {del Pozo}, {Delbo}, {Delgado}, {Delgado}, {Di Matteo},
  {Diakite}, {Diener}, {Distefano}, {Dolding}, {Drazinos}, {Dur{\'a}n},
  {Edvardsson}, {Enke}, {Eriksson}, {Esquej}, {Eynard Bontemps}, {Fabre},
  {Fabrizio}, {Faigler}, {Falc{\~a}o}, {Farr{\`a}s Casas}, {Federici},
  {Fedorets}, {Fernique}, {Figueras}, {Filippi}, {Findeisen}, {Fonti},
  {Fraile}, {Fraser}, {Fr{\'e}zouls}, {Gai}, {Galleti}, {Garabato},
  {Garc{\'\i}a-Sedano}, {Garofalo}, {Garralda}, {Gavel}, {Gavras}, {Gerssen},
  {Geyer}, {Giacobbe}, {Gilmore}, {Girona}, {Giuffrida}, {Glass}, {Gomes},
  {Granvik}, {Gueguen}, {Guerrier}, {Guiraud}, {Guti{\'e}rrez-S{\'a}nchez},
  {Haigron}, {Hatzidimitriou}, {Hauser}, {Haywood}, {Heiter}, {Helmi}, {Heu},
  {Hilger}, {Hobbs}, {Hofmann}, {Holland}, {Huckle}, {Hypki}, {Icardi},
  {Jan{\ss}en}, {Jevardat de Fombelle}, {Jonker}, {Juh{\'a}sz}, {Julbe},
  {Karampelas}, {Kewley}, {Klar}, {Kochoska}, {Kohley}, {Kolenberg},
  {Kontizas}, {Kontizas}, {Koposov}, {Kordopatis}, {Kostrzewa-Rutkowska},
  {Koubsky}, {Lambert}, {Lanza}, {Lasne}, {Lavigne}, {Le Fustec}, {Le
  Poncin-Lafitte}, {Lebreton}, {Leccia}, {Leclerc}, {Lecoeur-Taibi},
  {Lenhardt}, {Leroux}, {Liao}, {Licata}, {Lindstr{\o}m}, {Lister}, {Livanou},
  {Lobel}, {L{\'o}pez}, {Managau}, {Mann}, {Mantelet}, {Marchal}, {Marchant},
  {Marconi}, {Marinoni}, {Marschalk{\'o}}, {Marshall}, {Martino}, {Marton},
  {Mary}, {Massari}, {Matijevi{\v{c}}}, {Mazeh}, {McMillan}, {Messina},
  {Michalik}, {Millar}, {Molina}, {Molinaro}, {Moln{\'a}r}, {Montegriffo},
  {Mor}, {Morbidelli}, {Morel}, {Morris}, {Mulone}, {Muraveva}, {Musella},
  {Nelemans}, {Nicastro}, {Noval}, {O'Mullane}, {Ord{\'e}novic},
  {Ord{\'o}{\~n}ez-Blanco}, {Osborne}, {Pagani}, {Pagano}, {Pailler},
  {Palacin}, {Palaversa}, {Panahi}, {Pawlak}, {Piersimoni}, {Pineau}, {Plachy},
  {Plum}, {Poggio}, {Poujoulet}, {Pr{\v{s}}a}, {Pulone}, {Racero}, {Ragaini},
  {Rambaux}, {Ramos-Lerate}, {Regibo}, {Reyl{\'e}}, {Riclet}, {Ripepi}, {Riva},
  {Rivard}, {Rixon}, {Roegiers}, {Roelens}, {Romero-G{\'o}mez}, {Rowell},
  {Royer}, {Ruiz-Dern}, {Sadowski}, {Sagrist{\`a} Sell{\'e}s}, {Sahlmann},
  {Salgado}, {Salguero}, {Sanna}, {Santana-Ros}, {Sarasso}, {Savietto},
  {Schultheis}, {Sciacca}, {Segol}, {Segovia}, {S{\'e}gransan}, {Shih},
  {Siltala}, {Silva}, {Smart}, {Smith}, {Solano}, {Solitro}, {Sordo}, {Soria
  Nieto}, {Souchay}, {Spagna}, {Spoto}, {Stampa}, {Steele},
  {Steidelm{\"u}ller}, {Stephenson}, {Stoev}, {Suess}, {Surdej}, {Szabados},
  {Szegedi-Elek}, {Tapiador}, {Taris}, {Tauran}, {Taylor}, {Teixeira},
  {Terrett}, {Teyssandier}, {Thuillot}, {Titarenko}, {Torra Clotet}, {Turon},
  {Ulla}, {Utrilla}, {Uzzi}, {Vaillant}, {Valentini}, {Valette}, {van Elteren},
  {Van Hemelryck}, {van Leeuwen}, {Vaschetto}, {Vecchiato}, {Veljanoski},
  {Viala}, {Vicente}, {Vogt}, {von Essen}, {Voss}, {Votruba}, {Voutsinas},
  {Walmsley}, {Weiler}, {Wertz}, {Wevers}, {Wyrzykowski}, {Yoldas},
  {{\v{Z}}erjal}, {Ziaeepour}, {Zorec}, {Zschocke}, {Zucker}, {Zurbach}, \&
  {Zwitter}}]{2018A&A...616A...1G}
{Gaia Collaboration}, {Brown}, A.~G.~A., {Vallenari}, A., {et~al.} 2018, \aap,
  616, A1

\bibitem[{{Gaia Collaboration} {et~al.}(2021{\natexlab{a}}){Gaia
  Collaboration}, {Brown}, {Vallenari}, {Prusti}, {de Bruijne}, {Babusiaux},
  {Biermann}, {Creevey}, {Evans}, {Eyer}, {Hutton}, {Jansen}, {Jordi},
  {Klioner}, {Lammers}, {Lindegren}, {Luri}, {Mignard}, {Panem}, {Pourbaix},
  {Randich}, {Sartoretti}, {Soubiran}, {Walton}, {Arenou}, {Bailer-Jones},
  {Bastian}, {Cropper}, {Drimmel}, {Katz}, {Lattanzi}, {van Leeuwen}, {Bakker},
  {Cacciari}, {Casta{\~n}eda}, {De Angeli}, {Ducourant}, {Fabricius},
  {Fouesneau}, {Fr{\'e}mat}, {Guerra}, {Guerrier}, {Guiraud}, {Jean-Antoine
  Piccolo}, {Masana}, {Messineo}, {Mowlavi}, {Nicolas}, {Nienartowicz},
  {Pailler}, {Panuzzo}, {Riclet}, {Roux}, {Seabroke}, {Sordo}, {Tanga},
  {Th{\'e}venin}, {Gracia-Abril}, {Portell}, {Teyssier}, {Altmann}, {Andrae},
  {Bellas-Velidis}, {Benson}, {Berthier}, {Blomme}, {Brugaletta}, {Burgess},
  {Busso}, {Carry}, {Cellino}, {Cheek}, {Clementini}, {Damerdji}, {Davidson},
  {Delchambre}, {Dell'Oro}, {Fern{\'a}ndez-Hern{\'a}ndez}, {Galluccio},
  {Garc{\'\i}a-Lario}, {Garcia-Reinaldos}, {Gonz{\'a}lez-N{\'u}{\~n}ez},
  {Gosset}, {Haigron}, {Halbwachs}, {Hambly}, {Harrison}, {Hatzidimitriou},
  {Heiter}, {Hern{\'a}ndez}, {Hestroffer}, {Hodgkin}, {Holl}, {Jan{\ss}en},
  {Jevardat de Fombelle}, {Jordan}, {Krone-Martins}, {Lanzafame},
  {L{\"o}ffler}, {Lorca}, {Manteiga}, {Marchal}, {Marrese}, {Moitinho}, {Mora},
  {Muinonen}, {Osborne}, {Pancino}, {Pauwels}, {Petit}, {Recio-Blanco},
  {Richards}, {Riello}, {Rimoldini}, {Robin}, {Roegiers}, {Rybizki}, {Sarro},
  {Siopis}, {Smith}, {Sozzetti}, {Ulla}, {Utrilla}, {van Leeuwen}, {van
  Reeven}, {Abbas}, {Abreu Aramburu}, {Accart}, {Aerts}, {Aguado}, {Ajaj},
  {Altavilla}, {{\'A}lvarez}, {{\'A}lvarez Cid-Fuentes}, {Alves}, {Anderson},
  {Anglada Varela}, {Antoja}, {Audard}, {Baines}, {Baker},
  {Balaguer-N{\'u}{\~n}ez}, {Balbinot}, {Balog}, {Barache}, {Barbato},
  {Barros}, {Barstow}, {Bartolom{\'e}}, {Bassilana}, {Bauchet},
  {Baudesson-Stella}, {Becciani}, {Bellazzini}, {Bernet}, {Bertone}, {Bianchi},
  {Blanco-Cuaresma}, {Boch}, {Bombrun}, {Bossini}, {Bouquillon}, {Bragaglia},
  {Bramante}, {Breedt}, {Bressan}, {Brouillet}, {Bucciarelli}, {Burlacu},
  {Busonero}, {Butkevich}, {Buzzi}, {Caffau}, {Cancelliere}, {C{\'a}novas},
  {Cantat-Gaudin}, {Carballo}, {Carlucci}, {Carnerero}, {Carrasco},
  {Casamiquela}, {Castellani}, {Castro-Ginard}, {Castro Sampol}, {Chaoul},
  {Charlot}, {Chemin}, {Chiavassa}, {Cioni}, {Comoretto}, {Cooper}, {Cornez},
  {Cowell}, {Crifo}, {Crosta}, {Crowley}, {Dafonte}, {Dapergolas}, {David},
  {David}, {de Laverny}, {De Luise}, {De March}, {De Ridder}, {de Souza}, {de
  Teodoro}, {de Torres}, {del Peloso}, {del Pozo}, {Delbo}, {Delgado},
  {Delgado}, {Delisle}, {Di Matteo}, {Diakite}, {Diener}, {Distefano},
  {Dolding}, {Eappachen}, {Edvardsson}, {Enke}, {Esquej}, {Fabre}, {Fabrizio},
  {Faigler}, {Fedorets}, {Fernique}, {Fienga}, {Figueras}, {Fouron},
  {Fragkoudi}, {Fraile}, {Franke}, {Gai}, {Garabato}, {Garcia-Gutierrez},
  {Garc{\'\i}a-Torres}, {Garofalo}, {Gavras}, {Gerlach}, {Geyer}, {Giacobbe},
  {Gilmore}, {Girona}, {Giuffrida}, {Gomel}, {Gomez}, {Gonzalez-Santamaria},
  {Gonz{\'a}lez-Vidal}, {Granvik}, {Guti{\'e}rrez-S{\'a}nchez}, {Guy},
  {Hauser}, {Haywood}, {Helmi}, {Hidalgo}, {Hilger}, {H{\l}adczuk}, {Hobbs},
  {Holland}, {Huckle}, {Jasniewicz}, {Jonker}, {Juaristi Campillo}, {Julbe},
  {Karbevska}, {Kervella}, {Khanna}, {Kochoska}, {Kontizas}, {Kordopatis},
  {Korn}, {Kostrzewa-Rutkowska}, {Kruszy{\'n}ska}, {Lambert}, {Lanza}, {Lasne},
  {Le Campion}, {Le Fustec}, {Lebreton}, {Lebzelter}, {Leccia}, {Leclerc},
  {Lecoeur-Taibi}, {Liao}, {Licata}, {Lindstr{\o}m}, {Lister}, {Livanou},
  {Lobel}, {Madrero Pardo}, {Managau}, {Mann}, {Marchant}, {Marconi}, {Marcos
  Santos}, {Marinoni}, {Marocco}, {Marshall}, {Martin Polo},
  {Mart{\'\i}n-Fleitas}, {Masip}, {Massari}, {Mastrobuono-Battisti}, {Mazeh},
  {McMillan}, {Messina}, {Michalik}, {Millar}, {Mints}, {Molina}, {Molinaro},
  {Moln{\'a}r}, {Montegriffo}, {Mor}, {Morbidelli}, {Morel}, {Morris},
  {Mulone}, {Munoz}, {Muraveva}, {Murphy}, {Musella}, {Noval}, {Ord{\'e}novic},
  {Orr{\`u}}, {Osinde}, {Pagani}, {Pagano}, {Palaversa}, {Palicio}, {Panahi},
  {Pawlak}, {Pe{\~n}alosa Esteller}, {Penttil{\"a}}, {Piersimoni}, {Pineau},
  {Plachy}, {Plum}, {Poggio}, {Poretti}, {Poujoulet}, {Pr{\v{s}}a}, {Pulone},
  {Racero}, {Ragaini}, {Rainer}, {Raiteri}, {Rambaux}, {Ramos}, {Ramos-Lerate},
  {Re Fiorentin}, {Regibo}, {Reyl{\'e}}, {Ripepi}, {Riva}, {Rixon}, {Robichon},
  {Robin}, {Roelens}, {Rohrbasser}, {Romero-G{\'o}mez}, {Rowell}, {Royer},
  {Rybicki}, {Sadowski}, {Sagrist{\`a} Sell{\'e}s}, {Sahlmann}, {Salgado},
  {Salguero}, {Samaras}, {Sanchez Gimenez}, {Sanna}, {Santove{\~n}a},
  {Sarasso}, {Schultheis}, {Sciacca}, {Segol}, {Segovia}, {S{\'e}gransan},
  {Semeux}, {Shahaf}, {Siddiqui}, {Siebert}, {Siltala}, {Slezak}, {Smart},
  {Solano}, {Solitro}, {Souami}, {Souchay}, {Spagna}, {Spoto}, {Steele},
  {Steidelm{\"u}ller}, {Stephenson}, {S{\"u}veges}, {Szabados}, {Szegedi-Elek},
  {Taris}, {Tauran}, {Taylor}, {Teixeira}, {Thuillot}, {Tonello}, {Torra},
  {Torra}, {Turon}, {Unger}, {Vaillant}, {van Dillen}, {Vanel}, {Vecchiato},
  {Viala}, {Vicente}, {Voutsinas}, {Weiler}, {Wevers}, {Wyrzykowski}, {Yoldas},
  {Yvard}, {Zhao}, {Zorec}, {Zucker}, {Zurbach}, \&
  {Zwitter}}]{2021A&A...649A...1G}
{Gaia Collaboration}, {Brown}, A.~G.~A., {Vallenari}, A., {et~al.}
  2021{\natexlab{a}}, \aap, 649, A1

\bibitem[{{Gaia Collaboration} {et~al.}(2016){Gaia Collaboration}, {Prusti},
  {de Bruijne}, {Brown}, {Vallenari}, {Babusiaux}, {Bailer-Jones}, {Bastian},
  {Biermann}, {Evans}, {Eyer}, {Jansen}, {Jordi}, {Klioner}, {Lammers},
  {Lindegren}, {Luri}, {Mignard}, {Milligan}, {Panem}, {Poinsignon},
  {Pourbaix}, {Randich}, {Sarri}, {Sartoretti}, {Siddiqui}, {Soubiran},
  {Valette}, {van Leeuwen}, {Walton}, {Aerts}, {Arenou}, {Cropper}, {Drimmel},
  {H{\o}g}, {Katz}, {Lattanzi}, {O'Mullane}, {Grebel}, {Holland}, {Huc},
  {Passot}, {Bramante}, {Cacciari}, {Casta{\~n}eda}, {Chaoul}, {Cheek}, {De
  Angeli}, {Fabricius}, {Guerra}, {Hern{\'a}ndez}, {Jean-Antoine-Piccolo},
  {Masana}, {Messineo}, {Mowlavi}, {Nienartowicz}, {Ord{\'o}{\~n}ez-Blanco},
  {Panuzzo}, {Portell}, {Richards}, {Riello}, {Seabroke}, {Tanga},
  {Th{\'e}venin}, {Torra}, {Els}, {Gracia-Abril}, {Comoretto},
  {Garcia-Reinaldos}, {Lock}, {Mercier}, {Altmann}, {Andrae}, {Astraatmadja},
  {Bellas-Velidis}, {Benson}, {Berthier}, {Blomme}, {Busso}, {Carry},
  {Cellino}, {Clementini}, {Cowell}, {Creevey}, {Cuypers}, {Davidson}, {De
  Ridder}, {de Torres}, {Delchambre}, {Dell'Oro}, {Ducourant}, {Fr{\'e}mat},
  {Garc{\'\i}a-Torres}, {Gosset}, {Halbwachs}, {Hambly}, {Harrison}, {Hauser},
  {Hestroffer}, {Hodgkin}, {Huckle}, {Hutton}, {Jasniewicz}, {Jordan},
  {Kontizas}, {Korn}, {Lanzafame}, {Manteiga}, {Moitinho}, {Muinonen},
  {Osinde}, {Pancino}, {Pauwels}, {Petit}, {Recio-Blanco}, {Robin}, {Sarro},
  {Siopis}, {Smith}, {Smith}, {Sozzetti}, {Thuillot}, {van Reeven}, {Viala},
  {Abbas}, {Abreu Aramburu}, {Accart}, {Aguado}, {Allan}, {Allasia},
  {Altavilla}, {{\'A}lvarez}, {Alves}, {Anderson}, {Andrei}, {Anglada Varela},
  {Antiche}, {Antoja}, {Ant{\'o}n}, {Arcay}, {Atzei}, {Ayache}, {Bach},
  {Baker}, {Balaguer-N{\'u}{\~n}ez}, {Barache}, {Barata}, {Barbier}, {Barblan},
  {Baroni}, {Barrado y Navascu{\'e}s}, {Barros}, {Barstow}, {Becciani},
  {Bellazzini}, {Bellei}, {Bello Garc{\'\i}a}, {Belokurov}, {Bendjoya},
  {Berihuete}, {Bianchi}, {Bienaym{\'e}}, {Billebaud}, {Blagorodnova},
  {Blanco-Cuaresma}, {Boch}, {Bombrun}, {Borrachero}, {Bouquillon}, {Bourda},
  {Bouy}, {Bragaglia}, {Breddels}, {Brouillet}, {Br{\"u}semeister},
  {Bucciarelli}, {Budnik}, {Burgess}, {Burgon}, {Burlacu}, {Busonero}, {Buzzi},
  {Caffau}, {Cambras}, {Campbell}, {Cancelliere}, {Cantat-Gaudin}, {Carlucci},
  {Carrasco}, {Castellani}, {Charlot}, {Charnas}, {Charvet}, {Chassat},
  {Chiavassa}, {Clotet}, {Cocozza}, {Collins}, {Collins}, {Costigan}, {Crifo},
  {Cross}, {Crosta}, {Crowley}, {Dafonte}, {Damerdji}, {Dapergolas}, {David},
  {David}, {De Cat}, {de Felice}, {de Laverny}, {De Luise}, {De March}, {de
  Martino}, {de Souza}, {Debosscher}, {del Pozo}, {Delbo}, {Delgado},
  {Delgado}, {di Marco}, {Di Matteo}, {Diakite}, {Distefano}, {Dolding}, {Dos
  Anjos}, {Drazinos}, {Dur{\'a}n}, {Dzigan}, {Ecale}, {Edvardsson}, {Enke},
  {Erdmann}, {Escolar}, {Espina}, {Evans}, {Eynard Bontemps}, {Fabre},
  {Fabrizio}, {Faigler}, {Falc{\~a}o}, {Farr{\`a}s Casas}, {Faye}, {Federici},
  {Fedorets}, {Fern{\'a}ndez-Hern{\'a}ndez}, {Fernique}, {Fienga}, {Figueras},
  {Filippi}, {Findeisen}, {Fonti}, {Fouesneau}, {Fraile}, {Fraser}, {Fuchs},
  {Furnell}, {Gai}, {Galleti}, {Galluccio}, {Garabato}, {Garc{\'\i}a-Sedano},
  {Gar{\'e}}, {Garofalo}, {Garralda}, {Gavras}, {Gerssen}, {Geyer}, {Gilmore},
  {Girona}, {Giuffrida}, {Gomes}, {Gonz{\'a}lez-Marcos},
  {Gonz{\'a}lez-N{\'u}{\~n}ez}, {Gonz{\'a}lez-Vidal}, {Granvik}, {Guerrier},
  {Guillout}, {Guiraud}, {G{\'u}rpide}, {Guti{\'e}rrez-S{\'a}nchez}, {Guy},
  {Haigron}, {Hatzidimitriou}, {Haywood}, {Heiter}, {Helmi}, {Hobbs},
  {Hofmann}, {Holl}, {Holland}, {Hunt}, {Hypki}, {Icardi}, {Irwin}, {Jevardat
  de Fombelle}, {Jofr{\'e}}, {Jonker}, {Jorissen}, {Julbe}, {Karampelas},
  {Kochoska}, {Kohley}, {Kolenberg}, {Kontizas}, {Koposov}, {Kordopatis},
  {Koubsky}, {Kowalczyk}, {Krone-Martins}, {Kudryashova}, {Kull}, {Bachchan},
  {Lacoste-Seris}, {Lanza}, {Lavigne}, {Le Poncin-Lafitte}, {Lebreton},
  {Lebzelter}, {Leccia}, {Leclerc}, {Lecoeur-Taibi}, {Lemaitre}, {Lenhardt},
  {Leroux}, {Liao}, {Licata}, {Lindstr{\o}m}, {Lister}, {Livanou}, {Lobel},
  {L{\"o}ffler}, {L{\'o}pez}, {Lopez-Lozano}, {Lorenz}, {Loureiro},
  {MacDonald}, {Magalh{\~a}es Fernandes}, {Managau}, {Mann}, {Mantelet},
  {Marchal}, {Marchant}, {Marconi}, {Marie}, {Marinoni}, {Marrese},
  {Marschalk{\'o}}, {Marshall}, {Mart{\'\i}n-Fleitas}, {Martino}, {Mary},
  {Matijevi{\v{c}}}, {Mazeh}, {McMillan}, {Messina}, {Mestre}, {Michalik},
  {Millar}, {Miranda}, {Molina}, {Molinaro}, {Molinaro}, {Moln{\'a}r},
  {Moniez}, {Montegriffo}, {Monteiro}, {Mor}, {Mora}, {Morbidelli}, {Morel},
  {Morgenthaler}, {Morley}, {Morris}, {Mulone}, {Muraveva}, {Musella},
  {Narbonne}, {Nelemans}, {Nicastro}, {Noval}, {Ord{\'e}novic},
  {Ordieres-Mer{\'e}}, {Osborne}, {Pagani}, {Pagano}, {Pailler}, {Palacin},
  {Palaversa}, {Parsons}, {Paulsen}, {Pecoraro}, {Pedrosa}, {Pentik{\"a}inen},
  {Pereira}, {Pichon}, {Piersimoni}, {Pineau}, {Plachy}, {Plum}, {Poujoulet},
  {Pr{\v{s}}a}, {Pulone}, {Ragaini}, {Rago}, {Rambaux}, {Ramos-Lerate},
  {Ranalli}, {Rauw}, {Read}, {Regibo}, {Renk}, {Reyl{\'e}}, {Ribeiro},
  {Rimoldini}, {Ripepi}, {Riva}, {Rixon}, {Roelens}, {Romero-G{\'o}mez},
  {Rowell}, {Royer}, {Rudolph}, {Ruiz-Dern}, {Sadowski}, {Sagrist{\`a}
  Sell{\'e}s}, {Sahlmann}, {Salgado}, {Salguero}, {Sarasso}, {Savietto},
  {Schnorhk}, {Schultheis}, {Sciacca}, {Segol}, {Segovia}, {Segransan},
  {Serpell}, {Shih}, {Smareglia}, {Smart}, {Smith}, {Solano}, {Solitro},
  {Sordo}, {Soria Nieto}, {Souchay}, {Spagna}, {Spoto}, {Stampa}, {Steele},
  {Steidelm{\"u}ller}, {Stephenson}, {Stoev}, {Suess}, {S{\"u}veges}, {Surdej},
  {Szabados}, {Szegedi-Elek}, {Tapiador}, {Taris}, {Tauran}, {Taylor},
  {Teixeira}, {Terrett}, {Tingley}, {Trager}, {Turon}, {Ulla}, {Utrilla},
  {Valentini}, {van Elteren}, {Van Hemelryck}, {van Leeuwen}, {Varadi},
  {Vecchiato}, {Veljanoski}, {Via}, {Vicente}, {Vogt}, {Voss}, {Votruba},
  {Voutsinas}, {Walmsley}, {Weiler}, {Weingrill}, {Werner}, {Wevers},
  {Whitehead}, {Wyrzykowski}, {Yoldas}, {{\v{Z}}erjal}, {Zucker}, {Zurbach},
  {Zwitter}, {Alecu}, {Allen}, {Allende Prieto}, {Amorim},
  {Anglada-Escud{\'e}}, {Arsenijevic}, {Azaz}, {Balm}, {Beck}, {Bernstein},
  {Bigot}, {Bijaoui}, {Blasco}, {Bonfigli}, {Bono}, {Boudreault}, {Bressan},
  {Brown}, {Brunet}, {Bunclark}, {Buonanno}, {Butkevich}, {Carret}, {Carrion},
  {Chemin}, {Ch{\'e}reau}, {Corcione}, {Darmigny}, {de Boer}, {de Teodoro}, {de
  Zeeuw}, {Delle Luche}, {Domingues}, {Dubath}, {Fodor}, {Fr{\'e}zouls},
  {Fries}, {Fustes}, {Fyfe}, {Gallardo}, {Gallegos}, {Gardiol}, {Gebran},
  {Gomboc}, {G{\'o}mez}, {Grux}, {Gueguen}, {Heyrovsky}, {Hoar}, {Iannicola},
  {Isasi Parache}, {Janotto}, {Joliet}, {Jonckheere}, {Keil}, {Kim},
  {Klagyivik}, {Klar}, {Knude}, {Kochukhov}, {Kolka}, {Kos}, {Kutka}, {Lainey},
  {LeBouquin}, {Liu}, {Loreggia}, {Makarov}, {Marseille}, {Martayan},
  {Martinez-Rubi}, {Massart}, {Meynadier}, {Mignot}, {Munari}, {Nguyen},
  {Nordlander}, {Ocvirk}, {O'Flaherty}, {Olias Sanz}, {Ortiz}, {Osorio},
  {Oszkiewicz}, {Ouzounis}, {Palmer}, {Park}, {Pasquato}, {Peltzer}, {Peralta},
  {P{\'e}turaud}, {Pieniluoma}, {Pigozzi}, {Poels}, {Prat}, {Prod'homme},
  {Raison}, {Rebordao}, {Risquez}, {Rocca-Volmerange}, {Rosen}, {Ruiz-Fuertes},
  {Russo}, {Sembay}, {Serraller Vizcaino}, {Short}, {Siebert}, {Silva},
  {Sinachopoulos}, {Slezak}, {Soffel}, {Sosnowska}, {Strai{\v{z}}ys}, {ter
  Linden}, {Terrell}, {Theil}, {Tiede}, {Troisi}, {Tsalmantza}, {Tur},
  {Vaccari}, {Vachier}, {Valles}, {Van Hamme}, {Veltz}, {Virtanen}, {Wallut},
  {Wichmann}, {Wilkinson}, {Ziaeepour}, \& {Zschocke}}]{2016A&A...595A...1G}
{Gaia Collaboration}, {Prusti}, T., {de Bruijne}, J.~H.~J., {et~al.} 2016,
  \aap, 595, A1

\bibitem[{{Gaia Collaboration} {et~al.}(2021{\natexlab{b}}){Gaia
  Collaboration}, {Smart}, {Sarro}, {Rybizki}, {Reyl{\'e}}, {Robin}, {Hambly},
  {Abbas}, {Barstow}, {de Bruijne}, {Bucciarelli}, {Carrasco}, {Cooper},
  {Hodgkin}, {Masana}, {Michalik}, {Sahlmann}, {Sozzetti}, {Brown},
  {Vallenari}, {Prusti}, {Babusiaux}, {Biermann}, {Creevey}, {Evans}, {Eyer},
  {Hutton}, {Jansen}, {Jordi}, {Klioner}, {Lammers}, {Lindegren}, {Luri},
  {Mignard}, {Panem}, {Pourbaix}, {Randich}, {Sartoretti}, {Soubiran},
  {Walton}, {Arenou}, {Bailer-Jones}, {Bastian}, {Cropper}, {Drimmel}, {Katz},
  {Lattanzi}, {van Leeuwen}, {Bakker}, {Casta{\~n}eda}, {De Angeli},
  {Ducourant}, {Fabricius}, {Fouesneau}, {Fr{\'e}mat}, {Guerra}, {Guerrier},
  {Guiraud}, {Jean-Antoine Piccolo}, {Messineo}, {Mowlavi}, {Nicolas},
  {Nienartowicz}, {Pailler}, {Panuzzo}, {Riclet}, {Roux}, {Seabroke}, {Sordo},
  {Tanga}, {Th{\'e}venin}, {Gracia-Abril}, {Portell}, {Teyssier}, {Altmann},
  {Andrae}, {Bellas-Velidis}, {Benson}, {Berthier}, {Blomme}, {Brugaletta},
  {Burgess}, {Busso}, {Carry}, {Cellino}, {Cheek}, {Clementini}, {Damerdji},
  {Davidson}, {Delchambre}, {Dell'Oro}, {Fern{\'a}ndez-Hern{\'a}ndez},
  {Galluccio}, {Garc{\'\i}a-Lario}, {Garcia-Reinaldos},
  {Gonz{\'a}lez-N{\'u}{\~n}ez}, {Gosset}, {Haigron}, {Halbwachs}, {Harrison},
  {Hatzidimitriou}, {Heiter}, {Hern{\'a}ndez}, {Hestroffer}, {Holl},
  {Jan{\ss}en}, {Jevardat de Fombelle}, {Jordan}, {Krone-Martins}, {Lanzafame},
  {L{\"o}ffler}, {Lorca}, {Manteiga}, {Marchal}, {Marrese}, {Moitinho}, {Mora},
  {Muinonen}, {Osborne}, {Pancino}, {Pauwels}, {Recio-Blanco}, {Richards},
  {Riello}, {Rimoldini}, {Roegiers}, {Siopis}, {Smith}, {Ulla}, {Utrilla}, {van
  Leeuwen}, {van Reeven}, {Abreu Aramburu}, {Accart}, {Aerts}, {Aguado},
  {Ajaj}, {Altavilla}, {{\'A}lvarez}, {{\'A}lvarez Cid-Fuentes}, {Alves},
  {Anderson}, {Anglada Varela}, {Antoja}, {Audard}, {Baines}, {Baker},
  {Balaguer-N{\'u}{\~n}ez}, {Balbinot}, {Balog}, {Barache}, {Barbato},
  {Barros}, {Bartolom{\'e}}, {Bassilana}, {Bauchet}, {Baudesson-Stella},
  {Becciani}, {Bellazzini}, {Bernet}, {Bertone}, {Bianchi}, {Blanco-Cuaresma},
  {Boch}, {Bombrun}, {Bossini}, {Bouquillon}, {Bragaglia}, {Bramante},
  {Breedt}, {Bressan}, {Brouillet}, {Burlacu}, {Busonero}, {Butkevich},
  {Buzzi}, {Caffau}, {Cancelliere}, {C{\'a}novas}, {Cantat-Gaudin}, {Carballo},
  {Carlucci}, {Carnerero}, {Casamiquela}, {Castellani}, {Castro-Ginard},
  {Castro Sampol}, {Chaoul}, {Charlot}, {Chemin}, {Chiavassa}, {Cioni},
  {Comoretto}, {Cornez}, {Cowell}, {Crifo}, {Crosta}, {Crowley}, {Dafonte},
  {Dapergolas}, {David}, {David}, {de Laverny}, {De Luise}, {De March}, {De
  Ridder}, {de Souza}, {de Teodoro}, {de Torres}, {del Peloso}, {del Pozo},
  {Delgado}, {Delgado}, {Delisle}, {Di Matteo}, {Diakite}, {Diener},
  {Distefano}, {Dolding}, {Eappachen}, {Edvardsson}, {Enke}, {Esquej}, {Fabre},
  {Fabrizio}, {Faigler}, {Fedorets}, {Fernique}, {Fienga}, {Figueras},
  {Fouron}, {Fragkoudi}, {Fraile}, {Franke}, {Gai}, {Garabato},
  {Garcia-Gutierrez}, {Garc{\'\i}a-Torres}, {Garofalo}, {Gavras}, {Gerlach},
  {Geyer}, {Giacobbe}, {Gilmore}, {Girona}, {Giuffrida}, {Gomel}, {Gomez},
  {Gonzalez-Santamaria}, {Gonz{\'a}lez-Vidal}, {Granvik},
  {Guti{\'e}rrez-S{\'a}nchez}, {Guy}, {Hauser}, {Haywood}, {Helmi}, {Hidalgo},
  {Hilger}, {H{\l}adczuk}, {Hobbs}, {Holland}, {Huckle}, {Jasniewicz},
  {Jonker}, {Juaristi Campillo}, {Julbe}, {Karbevska}, {Kervella}, {Khanna},
  {Kochoska}, {Kontizas}, {Kordopatis}, {Korn}, {Kostrzewa-Rutkowska},
  {Kruszy{\'n}ska}, {Lambert}, {Lanza}, {Lasne}, {Le Campion}, {Le Fustec},
  {Lebreton}, {Lebzelter}, {Leccia}, {Leclerc}, {Lecoeur-Taibi}, {Liao},
  {Licata}, {Lindstr{\o}m}, {Lister}, {Livanou}, {Lobel}, {Madrero Pardo},
  {Managau}, {Mann}, {Marchant}, {Marconi}, {Marcos Santos}, {Marinoni},
  {Marocco}, {Marshall}, {Martin Polo}, {Mart{\'\i}n-Fleitas}, {Masip},
  {Massari}, {Mastrobuono-Battisti}, {Mazeh}, {McMillan}, {Messina}, {Millar},
  {Mints}, {Molina}, {Molinaro}, {Moln{\'a}r}, {Montegriffo}, {Mor},
  {Morbidelli}, {Morel}, {Morris}, {Mulone}, {Munoz}, {Muraveva}, {Murphy},
  {Musella}, {Noval}, {Ord{\'e}novic}, {Orr{\`u}}, {Osinde}, {Pagani},
  {Pagano}, {Palaversa}, {Palicio}, {Panahi}, {Pawlak}, {Pe{\~n}alosa
  Esteller}, {Penttil{\"a}}, {Piersimoni}, {Pineau}, {Plachy}, {Plum},
  {Poggio}, {Poretti}, {Poujoulet}, {Pr{\v{s}}a}, {Pulone}, {Racero},
  {Ragaini}, {Rainer}, {Raiteri}, {Rambaux}, {Ramos}, {Ramos-Lerate}, {Re
  Fiorentin}, {Regibo}, {Ripepi}, {Riva}, {Rixon}, {Robichon}, {Robin},
  {Roelens}, {Rohrbasser}, {Romero-G{\'o}mez}, {Rowell}, {Royer}, {Rybicki},
  {Sadowski}, {Sagrist{\`a} Sell{\'e}s}, {Salgado}, {Salguero}, {Samaras},
  {Sanchez Gimenez}, {Sanna}, {Santove{\~n}a}, {Sarasso}, {Schultheis},
  {Sciacca}, {Segol}, {Segovia}, {S{\'e}gransan}, {Semeux}, {Shahaf},
  {Siddiqui}, {Siebert}, {Siltala}, {Slezak}, {Solano}, {Solitro}, {Souami},
  {Souchay}, {Spagna}, {Spoto}, {Steele}, {Steidelm{\"u}ller}, {Stephenson},
  {S{\"u}veges}, {Szabados}, {Szegedi-Elek}, {Taris}, {Tauran}, {Taylor},
  {Teixeira}, {Thuillot}, {Tonello}, {Torra}, {Torra}, {Turon}, {Unger},
  {Vaillant}, {van Dillen}, {Vanel}, {Vecchiato}, {Viala}, {Vicente},
  {Voutsinas}, {Weiler}, {Wevers}, {Wyrzykowski}, {Yoldas}, {Yvard}, {Zhao},
  {Zorec}, {Zucker}, {Zurbach}, \& {Zwitter}}]{2021A&A...649A...6G}
{Gaia Collaboration}, {Smart}, R.~L., {Sarro}, L.~M., {et~al.}
  2021{\natexlab{b}}, \aap, 649, A6

\bibitem[{{Garc{\'i}a-Alvarez} {et~al.}(2002){Garc{\'i}a-Alvarez},
  {Jevremovi{\'c}}, {Doyle}, \& {Butler}}]{2002A&A...383..548G}
{Garc{\'i}a-Alvarez}, D., {Jevremovi{\'c}}, D., {Doyle}, J.~G., \& {Butler},
  C.~J. 2002, \aap, 383, 548

\bibitem[{{Gatewood} {et~al.}(2003){Gatewood}, {Coban}, \&
  {Han}}]{2003AJ....125.1530G}
{Gatewood}, G., {Coban}, L., \& {Han}, I. 2003, \aj, 125, 1530

\bibitem[{{Geballe} {et~al.}(2002){Geballe}, {Knapp}, {Leggett}, {Fan},
  {Golimowski}, {Anderson}, {Brinkmann}, {Csabai}, {Gunn}, {Hawley},
  {Hennessy}, {Henry}, {Hill}, {Hindsley}, {Ivezi{\'c}}, {Lupton}, {McDaniel},
  {Munn}, {Narayanan}, {Peng}, {Pier}, {Rockosi}, {Schneider}, {Smith},
  {Strauss}, {Tsvetanov}, {Uomoto}, {York}, \& {Zheng}}]{2002ApJ...564..466G}
{Geballe}, T.~R., {Knapp}, G.~R., {Leggett}, S.~K., {et~al.} 2002, \apj, 564,
  466

\bibitem[{{Gelino} {et~al.}(2011){Gelino}, {Kirkpatrick}, {Cushing},
  {Eisenhardt}, {Griffith}, {Mainzer}, {Marsh}, {Skrutskie}, \&
  {Wright}}]{2011AJ....142...57G}
{Gelino}, C.~R., {Kirkpatrick}, J.~D., {Cushing}, M.~C., {et~al.} 2011, \aj,
  142, 57

\bibitem[{{Geyer} {et~al.}(1988){Geyer}, {Harrington}, \&
  {Worley}}]{1988AJ.....95.1841G}
{Geyer}, D.~W., {Harrington}, R.~S., \& {Worley}, C.~E. 1988, \aj, 95, 1841

\bibitem[{{Gillon} {et~al.}(2017){Gillon}, {Demory}, {Van Grootel}, {Motalebi},
  {Lovis}, {Collier Cameron}, {Charbonneau}, {Latham}, {Molinari}, {Pepe},
  {S{\'e}gransan}, {Sasselov}, {Udry}, {Mayor}, {Micela}, {Piotto}, \&
  {Sozzetti}}]{2017NatAs...1E..56G}
{Gillon}, M., {Demory}, B.-O., {Van Grootel}, V., {et~al.} 2017, Nature
  Astronomy, 1, 0056

\bibitem[{{Gillon} {et~al.}(2007){Gillon}, {Pont}, {Demory}, {Mallmann},
  {Mayor}, {Mazeh}, {Queloz}, {Shporer}, {Udry}, \&
  {Vuissoz}}]{2007A&A...472L..13G}
{Gillon}, M., {Pont}, F., {Demory}, B.~O., {et~al.} 2007, \aap, 472, L13

\bibitem[{{Gizis} {et~al.}(2015){Gizis}, {Burgasser}, \&
  {Vrba}}]{2015AJ....150..179G}
{Gizis}, J.~E., {Burgasser}, A.~J., \& {Vrba}, F.~J. 2015, \aj, 150, 179

\bibitem[{{Gizis} {et~al.}(2002){Gizis}, {Reid}, \&
  {Hawley}}]{2002AJ....123.3356G}
{Gizis}, J.~E., {Reid}, I.~N., \& {Hawley}, S.~L. 2002, \aj, 123, 3356

\bibitem[{{Gizis} \& {Reid}(1996)}]{1996AJ....111..365G}
{Gizis}, J.~E. \& {Reid}, N.~I. 1996, \aj, 111, 365

\bibitem[{{Gizis} {et~al.}(2016){Gizis}, {Williams}, {Burgasser}, {Libralato},
  {Nardiello}, {Piotto}, {Bedin}, {Berger}, \& {Paudel}}]{2016AJ....152..123G}
{Gizis}, J.~E., {Williams}, P. K.~G., {Burgasser}, A.~J., {et~al.} 2016, \aj,
  152, 123

\bibitem[{{Gliese} \& {Jahreiss}(2015)}]{2015yCat.5035....0G}
{Gliese}, W. \& {Jahreiss}, H. 2015, VizieR Online Data Catalog, V/35

\bibitem[{{Golimowski} {et~al.}(1998){Golimowski}, {Burrows}, {Kulkarni},
  {Oppenheimer}, \& {Brukardt}}]{1998AJ....115.2579G}
{Golimowski}, D.~A., {Burrows}, C.~J., {Kulkarni}, S.~R., {Oppenheimer}, B.~R.,
  \& {Brukardt}, R.~A. 1998, \aj, 115, 2579

\bibitem[{{Golimowski} {et~al.}(2004){Golimowski}, {Henry}, {Krist},
  {Dieterich}, {Ford}, {Illingworth}, {Ardila}, {Clampin}, {Franz},
  {Wasserman}, {Benedict}, {McArthur}, \& {Nelan}}]{2004AJ....128.1733G}
{Golimowski}, D.~A., {Henry}, T.~J., {Krist}, J.~E., {et~al.} 2004, \aj, 128,
  1733

\bibitem[{{Golimowski} {et~al.}(2000){Golimowski}, {Henry}, {Krist},
  {Schroeder}, {Marcy}, {Fischer}, \& {Butler}}]{2000AJ....120.2082G}
{Golimowski}, D.~A., {Henry}, T.~J., {Krist}, J.~E., {et~al.} 2000, \aj, 120,
  2082

\bibitem[{{Golimowski} {et~al.}(1995){Golimowski}, {Nakajima}, {Kulkarni}, \&
  {Oppenheimer}}]{1995ApJ...444L.101G}
{Golimowski}, D.~A., {Nakajima}, T., {Kulkarni}, S.~R., \& {Oppenheimer}, B.~R.
  1995, \apjl, 444, L101

\bibitem[{{Gontcharov}(2006)}]{2006AstL...32..759G}
{Gontcharov}, G.~A. 2006, Astronomy Letters, 32, 759

\bibitem[{{Gonz{\'a}lez-{\'A}lvarez} {et~al.}(2020){Gonz{\'a}lez-{\'A}lvarez},
  {Zapatero Osorio}, {Caballero}, {Sanz-Forcada}, {B{\'e}jar},
  {Gonz{\'a}lez-Cuesta}, {Dreizler}, {Bauer}, {Rodr{\'\i}guez}, {Tal-Or},
  {Zechmeister}, {Montes}, {L{\'o}pez-Gonz{\'a}lez}, {Ribas}, {Reiners},
  {Quirrenbach}, {Amado}, {Anglada-Escud{\'e}}, {Azzaro},
  {Cort{\'e}s-Contreras}, {Hatzes}, {Henning}, {Jeffers}, {Kaminski},
  {K{\\"u}rster}, {Lafarga}, {Morales}, {Pall{\'e}}, {Perger}, \&
  {Schmitt}}]{2020A&A...637A..93G}
{Gonz{\'a}lez-{\'A}lvarez}, E., {Zapatero Osorio}, M.~R., {Caballero}, J.~A.,
  {et~al.} 2020, \aap, 637, A93

\bibitem[{{Gray} {et~al.}(2006){Gray}, {Corbally}, {Garrison}, {McFadden},
  {Bubar}, {McGahee}, {O'Donoghue}, \& {Knox}}]{2006AJ....132..161G}
{Gray}, R.~O., {Corbally}, C.~J., {Garrison}, R.~F., {et~al.} 2006, \aj, 132,
  161

\bibitem[{{Gray} {et~al.}(2003){Gray}, {Corbally}, {Garrison}, {McFadden}, \&
  {Robinson}}]{2003AJ....126.2048G}
{Gray}, R.~O., {Corbally}, C.~J., {Garrison}, R.~F., {McFadden}, M.~T., \&
  {Robinson}, P.~E. 2003, \aj, 126, 2048

\bibitem[{{Gray} {et~al.}(2001){Gray}, {Napier}, \&
  {Winkler}}]{2001AJ....121.2148G}
{Gray}, R.~O., {Napier}, M.~G., \& {Winkler}, L.~I. 2001, \aj, 121, 2148

\bibitem[{{Halbwachs} {et~al.}(2018){Halbwachs}, {Mayor}, \&
  {Udry}}]{2018A&A...619A..81H}
{Halbwachs}, J.~L., {Mayor}, M., \& {Udry}, S. 2018, \aap, 619, A81

\bibitem[{{Hambaryan} {et~al.}(2004){Hambaryan}, {Staude}, {Schwope}, {Scholz},
  {Kimeswenger}, \& {Neuh{\\"a}user}}]{2004A&A...415..265H}
{Hambaryan}, V., {Staude}, A., {Schwope}, A.~D., {et~al.} 2004, \aap, 415, 265

\bibitem[{{Han} \& {Gatewood}(2002)}]{2002PASP..114..224H}
{Han}, I. \& {Gatewood}, G. 2002, \pasp, 114, 224

\bibitem[{{Hartkopf} {et~al.}(2012){Hartkopf}, {Tokovinin}, \&
  {Mason}}]{2012AJ....143...42H}
{Hartkopf}, W.~I., {Tokovinin}, A., \& {Mason}, B.~D. 2012, \aj, 143, 42

\bibitem[{{Hatzes} {et~al.}(2000){Hatzes}, {Cochran}, {McArthur}, {Baliunas},
  {Walker}, {Campbell}, {Irwin}, {Yang}, {K{\\"u}rster}, {Endl}, {Els},
  {Butler}, \& {Marcy}}]{2000ApJ...544L.145H}
{Hatzes}, A.~P., {Cochran}, W.~D., {McArthur}, B., {et~al.} 2000, \apjl, 544,
  L145

\bibitem[{{Hawley} {et~al.}(1996){Hawley}, {Gizis}, \&
  {Reid}}]{1996AJ....112.2799H}
{Hawley}, S.~L., {Gizis}, J.~E., \& {Reid}, I.~N. 1996, \aj, 112, 2799

\bibitem[{{Henderson}(1839)}]{1839MNRAS...4..168H}
{Henderson}, T. 1839, \mnras, 4, 168

\bibitem[{{Henry} {et~al.}(2019){Henry}, {Jao}, {Riedel}, {Slatten}, \&
  {Winters}}]{2019AAS...23325932H}
{Henry}, T., {Jao}, W.-C., {Riedel}, A.~R., {Slatten}, K.~J., \& {Winters}, J.
  2019, in American Astronomical Society Meeting Abstracts, Vol. 233, American
  Astronomical Society Meeting Abstracts \#233, 259.32

\bibitem[{{Henry} {et~al.}(2006){Henry}, {Jao}, {Subasavage}, {Beaulieu},
  {Ianna}, {Costa}, \& {M{\'e}ndez}}]{2006AJ....132.2360H}
{Henry}, T.~J., {Jao}, W.-C., {Subasavage}, J.~P., {et~al.} 2006, \aj, 132,
  2360

\bibitem[{{Henry} {et~al.}(2018){Henry}, {Jao}, {Winters}, {Dieterich},
  {Finch}, {Ianna}, {Riedel}, {Silverstein}, {Subasavage}, \&
  {Vrijmoet}}]{2018AJ....155..265H}
{Henry}, T.~J., {Jao}, W.-C., {Winters}, J.~G., {et~al.} 2018, \aj, 155, 265

\bibitem[{{Henry} {et~al.}(2002){Henry}, {Walkowicz}, {Barto}, \&
  {Golimowski}}]{2002AJ....123.2002H}
{Henry}, T.~J., {Walkowicz}, L.~M., {Barto}, T.~C., \& {Golimowski}, D.~A.
  2002, \aj, 123, 2002

\bibitem[{{Herschel}(1803)}]{1803RSPT...93..339H}
{Herschel}, W. 1803, Philosophical Transactions of the Royal Society of London
  Series I, 93, 339

\bibitem[{{Hershey} \& {Taff}(1998)}]{1998AJ....116.1440H}
{Hershey}, J.~L. \& {Taff}, L.~G. 1998, \aj, 116, 1440

\bibitem[{{Holberg} {et~al.}(1998){Holberg}, {Barstow}, {Bruhweiler}, {Cruise},
  \& {Penny}}]{1998ApJ...497..935H}
{Holberg}, J.~B., {Barstow}, M.~A., {Bruhweiler}, F.~C., {Cruise}, A.~M., \&
  {Penny}, A.~J. 1998, \apj, 497, 935

\bibitem[{{Howard} {et~al.}(2014){Howard}, {Marcy}, {Fischer}, {Isaacson},
  {Muirhead}, {Henry}, {Boyajian}, {von Braun}, {Becker}, {Wright}, \&
  {Johnson}}]{2014ApJ...794...51H}
{Howard}, A.~W., {Marcy}, G.~W., {Fischer}, D.~A., {et~al.} 2014, \apj, 794, 51

\bibitem[{{Janson} {et~al.}(2012){Janson}, {Hormuth}, {Bergfors}, {Brandner},
  {Hippler}, {Daemgen}, {Kudryavtseva}, {Schmalzl}, {Schnupp}, \&
  {Henning}}]{2012ApJ...754...44J}
{Janson}, M., {Hormuth}, F., {Bergfors}, C., {et~al.} 2012, \apj, 754, 44

\bibitem[{{Janson} {et~al.}(2020){Janson}, {Wu}, {Cataldi}, \&
  {Brandeker}}]{2020A&A...640A..93J}
{Janson}, M., {Wu}, Y., {Cataldi}, G., \& {Brandeker}, A. 2020, \aap, 640, A93

\bibitem[{{Jeffers} {et~al.}(2020){Jeffers}, {Dreizler}, {Barnes}, {Haswell},
  {Nelson}, {Rodr{\'\i}guez}, {L{\'o}pez-Gonz\u2027lez}, {Morales}, {Luque},
  {Zechmeister}, {Vogt}, {Jenkins}, {Palle}, {Berdi {\~n}as}, {Coleman},
  {D{\'\i}az}, {Ribas}, {Jones}, {Butler}, {Tinney}, {Bailey}, {Carter},
  {O'Toole}, {Wittenmyer}, {Crane}, {Feng}, {Shectman}, {Teske}, {Reiners},
  {Amado}, \& {Anglada-Escud{\'e}}}]{2020Sci...368.1477J}
{Jeffers}, S.~V., {Dreizler}, S., {Barnes}, J.~R., {et~al.} 2020, Science, 368,
  1477

\bibitem[{{Jeffers} {et~al.}(2018){Jeffers}, {Sch{\\"o}fer}, {Lamert},
  {Reiners}, {Montes}, {Caballero}, {Cort{\'e}s-Contreras}, {Marvin},
  {Passegger}, {Zechmeister}, {Quirrenbach}, {Alonso-Floriano}, {Amado},
  {Bauer}, {Casal}, {Diez Alonso}, {Herrero}, {Morales}, {Mundt}, {Ribas}, \&
  {Sarmiento}}]{2018A&A...614A..76J}
{Jeffers}, S.~V., {Sch{\\"o}fer}, P., {Lamert}, A., {et~al.} 2018, \aap, 614,
  A76

\bibitem[{{Jenkins} {et~al.}(2015){Jenkins}, {D{\'\i}az}, {Jones}, {Butler},
  {Tinney}, {O'Toole}, {Carter}, {Wittenmyer}, \&
  {Pinfield}}]{2015MNRAS.453.1439J}
{Jenkins}, J.~S., {D{\'\i}az}, M., {Jones}, H.~R.~A., {et~al.} 2015, \mnras,
  453, 1439

\bibitem[{{Jenkins} {et~al.}(2019){Jenkins}, {Harrington}, {Challener},
  {Kurtovic}, {Ramirez}, {Pe{\~n}a}, {McIntyre}, {Himes}, {Rodr{\'\i}guez},
  {Anglada-Escud{\'e}}, {Dreizler}, {Ofir}, {Pe{\~n}a Rojas}, {Ribas}, {Rojo},
  {Kipping}, {Butler}, {Amado}, {Rodr{\'\i}guez-L{\'o}pez}, {Kempton}, {Palle},
  \& {Murgas}}]{2019MNRAS.487..268J}
{Jenkins}, J.~S., {Harrington}, J., {Challener}, R.~C., {et~al.} 2019, \mnras,
  487, 268

\bibitem[{{Jenkins} {et~al.}(2009){Jenkins}, {Ramsey}, {Jones}, {Pavlenko},
  {Gallardo}, {Barnes}, \& {Pinfield}}]{2009ApJ...704..975J}
{Jenkins}, J.~S., {Ramsey}, L.~W., {Jones}, H.~R.~A., {et~al.} 2009, \apj, 704,
  975

\bibitem[{{Jenkins}(1937)}]{1937AJ.....46...95J}
{Jenkins}, L.~F. 1937, \aj, 46, 95

\bibitem[{{Jenkins}(1963)}]{1963gcts.book.....J}
{Jenkins}, L.~F. 1963, {General catalogue of trigonometric stellar parallaxes}

\bibitem[{{J{\'o}dar} {et~al.}(2013){J{\'o}dar}, {P{\'e}rez-Garrido},
  {D{\'\i}az-S{\'a}nchez}, {Vill{\'o}}, {Rebolo}, \&
  {P{\'e}rez-Prieto}}]{2013MNRAS.429..859J}
{J{\'o}dar}, E., {P{\'e}rez-Garrido}, A., {D{\'\i}az-S{\'a}nchez}, A., {et~al.}
  2013, \mnras, 429, 859

\bibitem[{{Johnson} {et~al.}(2010){Johnson}, {Howard}, {Marcy}, {Bowler},
  {Henry}, {Fischer}, {Apps}, {Isaacson}, \& {Wright}}]{2010PASP..122..149J}
{Johnson}, J.~A., {Howard}, A.~W., {Marcy}, G.~W., {et~al.} 2010, \pasp, 122,
  149

\bibitem[{{Johnson} {et~al.}(2016){Johnson}, {Endl}, {Cochran}, {Meschiari},
  {Robertson}, {MacQueen}, {Brugamyer}, {Caldwell}, {Hatzes}, {Ram{\'\i}rez},
  \& {Wittenmyer}}]{2016ApJ...821...74J}
{Johnson}, M.~C., {Endl}, M., {Cochran}, W.~D., {et~al.} 2016, \apj, 821, 74

\bibitem[{{Kaiser} {et~al.}(2002){Kaiser}, {Aussel}, {Burke}, {Boesgaard},
  {Chambers}, {Chun}, {Heasley}, {Hodapp}, {Hunt}, {Jedicke}, {Jewitt},
  {Kudritzki}, {Luppino}, {Maberry}, {Magnier}, {Monet}, {Onaka}, {Pickles},
  {Rhoads}, {Simon}, {Szalay}, {Szapudi}, {Tholen}, {Tonry}, {Waterson}, \&
  {Wick}}]{2002SPIE.4836..154K}
{Kaiser}, N., {Aussel}, H., {Burke}, B.~E., {et~al.} 2002, in Society of
  Photo-Optical Instrumentation Engineers (SPIE) Conference Series, Vol. 4836,
  Survey and Other Telescope Technologies and Discoveries, ed. J.~A. {Tyson} \&
  S.~{Wolff}, 154--164

\bibitem[{{Kalas} {et~al.}(2008){Kalas}, {Graham}, {Chiang}, {Fitzgerald},
  {Clampin}, {Kite}, {Stapelfeldt}, {Marois}, \& {Krist}}]{2008Sci...322.1345K}
{Kalas}, P., {Graham}, J.~R., {Chiang}, E., {et~al.} 2008, Science, 322, 1345

\bibitem[{{Kaminski} {et~al.}(2018){Kaminski}, {Trifonov}, {Caballero},
  {Quirrenbach}, {Ribas}, {Reiners}, {Amado}, {Zechmeister}, {Dreizler},
  {Perger}, {Tal-Or}, {Bonfils}, {Mayor}, {Astudillo-Defru}, {Bauer},
  {B{\'e}jar}, {Cifuentes}, {Colom{\'e}}, {Cort{\'e}s-Contreras}, {Delfosse},
  {D{\'\i}ez-Alonso}, {Forveille}, {Guenther}, {Hatzes}, {Henning}, {Jeffers},
  {K{\\"u}rster}, {Lafarga}, {Luque}, {Mandel}, {Montes}, {Morales},
  {Passegger}, {Pedraz}, {Reffert}, {Sadegi}, {Schweitzer}, {Seifert}, {Stahl},
  \& {Udry}}]{2018A&A...618A.115K}
{Kaminski}, A., {Trifonov}, T., {Caballero}, J.~A., {et~al.} 2018, \aap, 618,
  A115

\bibitem[{{Kammer} {et~al.}(2014){Kammer}, {Knutson}, {Howard}, {Laughlin},
  {Deming}, {Todorov}, {Desert}, {Agol}, {Burrows}, {Fortney}, {Showman}, \&
  {Lewis}}]{2014ApJ...781..103K}
{Kammer}, J.~A., {Knutson}, H.~A., {Howard}, A.~W., {et~al.} 2014, \apj, 781,
  103

\bibitem[{{Karata{\c{s}}} {et~al.}(2004){Karata{\c{s}}}, {Bilir}, {Eker}, \&
  {Demircan}}]{2004MNRAS.349.1069K}
{Karata{\c{s}}}, Y., {Bilir}, S., {Eker}, Z., \& {Demircan}, O. 2004, \mnras,
  349, 1069

\bibitem[{{Keenan} \& {McNeil}(1989)}]{1989ApJS...71..245K}
{Keenan}, P.~C. \& {McNeil}, R.~C. 1989, \apjs, 71, 245

\bibitem[{{Kellett} \& {Tsikoudi}(1999)}]{1999MNRAS.308..111K}
{Kellett}, B.~J. \& {Tsikoudi}, V. 1999, \mnras, 308, 111

\bibitem[{{Kervella} {et~al.}(2020){Kervella}, {Arenou}, \&
  {Schneider}}]{2020A&A...635L..14K}
{Kervella}, P., {Arenou}, F., \& {Schneider}, J. 2020, \aap, 635, L14

\bibitem[{{Kervella} {et~al.}(2016){Kervella}, {M{\'e}rand}, {Ledoux},
  {Demory}, \& {Le Bouquin}}]{2016A&A...593A.127K}
{Kervella}, P., {M{\'e}rand}, A., {Ledoux}, C., {Demory}, B.~O., \& {Le
  Bouquin}, J.~B. 2016, \aap, 593, A127

\bibitem[{{Kervella} {et~al.}(2008){Kervella}, {M{\'e}rand}, {Pichon},
  {Th{\'e}venin}, {Heiter}, {Bigot}, {ten Brummelaar}, {McAlister}, {Ridgway},
  {Turner}, {Sturmann}, {Sturmann}, {Goldfinger}, \&
  {Farrington}}]{2008A&A...488..667K}
{Kervella}, P., {M{\'e}rand}, A., {Pichon}, B., {et~al.} 2008, \aap, 488, 667

\bibitem[{{Khandrika} {et~al.}(2013){Khandrika}, {Burgasser}, {Melis}, {Luk},
  {Bowsher}, \& {Swift}}]{2013AJ....145...71K}
{Khandrika}, H., {Burgasser}, A.~J., {Melis}, C., {et~al.} 2013, \aj, 145, 71

\bibitem[{{Kharchenko} {et~al.}(2007){Kharchenko}, {Scholz}, {Piskunov},
  {R{\"o}ser}, \& {Schilbach}}]{2007AN....328..889K}
{Kharchenko}, N.~V., {Scholz}, R.~D., {Piskunov}, A.~E., {R{\"o}ser}, S., \&
  {Schilbach}, E. 2007, Astronomische Nachrichten, 328, 889

\bibitem[{{King} {et~al.}(2010){King}, {McCaughrean}, {Homeier}, {Allard},
  {Scholz}, \& {Lodieu}}]{2010A&A...510A..99K}
{King}, R.~R., {McCaughrean}, M.~J., {Homeier}, D., {et~al.} 2010, \aap, 510,
  A99

\bibitem[{{Kirkpatrick} {et~al.}(2011){Kirkpatrick}, {Cushing}, {Gelino},
  {Griffith}, {Skrutskie}, {Marsh}, {Wright}, {Mainzer}, {Eisenhardt},
  {McLean}, {Thompson}, {Bauer}, {Benford}, {Bridge}, {Lake}, {Petty},
  {Stanford}, {Tsai}, {Bailey}, {Beichman}, {Bloom}, {Bochanski}, {Burgasser},
  {Capak}, {Cruz}, {Hinz}, {Kartaltepe}, {Knox}, {Manohar}, {Masters},
  {Morales-Calder{\'o}n}, {Prato}, {Rodigas}, {Salvato}, {Schurr}, {Scoville},
  {Simcoe}, {Stapelfeldt}, {Stern}, {Stock}, \& {Vacca}}]{2011ApJS..197...19K}
{Kirkpatrick}, J.~D., {Cushing}, M.~C., {Gelino}, C.~R., {et~al.} 2011, \apjs,
  197, 19

\bibitem[{{Kirkpatrick} {et~al.}(2021){Kirkpatrick}, {Gelino}, {Faherty},
  {Meisner}, {Caselden}, {Schneider}, {Marocco}, {Cayago}, {Smart},
  {Eisenhardt}, {Kuchner}, {Wright}, {Cushing}, {Allers}, {Bardalez Gagliuffi},
  {Burgasser}, {Gagn{\'e}}, {Logsdon}, {Martin}, {Ingalls}, {Lowrance},
  {Abrahams}, {Aganze}, {Gerasimov}, {Gonzales}, {Hsu}, {Kamraj}, {Kiman},
  {Rees}, {Theissen}, {Ammar}, {Andersen}, {Beaulieu}, {Colin}, {Elachi},
  {Goodman}, {Gramaize}, {Hamlet}, {Hong}, {Jonkeren}, {Khalil}, {Martin},
  {Pendrill}, {Pumphrey}, {Rothermich}, {Sainio}, {Stenner}, {Tanner},
  {Th{\'e}venot}, {Voloshin}, {Walla}, {W{\k{e}}dracki}, \& {Backyard Worlds:
  Planet 9 Collaboration}}]{2021ApJS..253....7K}
{Kirkpatrick}, J.~D., {Gelino}, C.~R., {Faherty}, J.~K., {et~al.} 2021, \apjs,
  253, 7

\bibitem[{{Kirkpatrick} {et~al.}(2019){Kirkpatrick}, {Martin}, {Smart},
  {Cayago}, {Beichman}, {Marocco}, {Gelino}, {Faherty}, {Cushing}, {Schneider},
  {Mace}, {Tinney}, {Wright}, {Lowrance}, {Ingalls}, {Vrba}, {Munn}, {Dahm}, \&
  {McLean}}]{2019ApJS..240...19K}
{Kirkpatrick}, J.~D., {Martin}, E.~C., {Smart}, R.~L., {et~al.} 2019, \apjs,
  240, 19

\bibitem[{{Konopacky} {et~al.}(2010){Konopacky}, {Ghez}, {Barman}, {Rice},
  {Bailey}, {White}, {McLean}, \& {Duch{\^e}ne}}]{2010ApJ...711.1087K}
{Konopacky}, Q.~M., {Ghez}, A.~M., {Barman}, T.~S., {et~al.} 2010, \apj, 711,
  1087

\bibitem[{{Kordopatis} {et~al.}(2013){Kordopatis}, {Gilmore}, {Steinmetz},
  {Boeche}, {Seabroke}, {Siebert}, {Zwitter}, {Binney}, {de Laverny},
  {Recio-Blanco}, {Williams}, {Piffl}, {Enke}, {Roeser}, {Bijaoui}, {Wyse},
  {Freeman}, {Munari}, {Carrillo}, {Anguiano}, {Burton}, {Campbell}, {Cass},
  {Fiegert}, {Hartley}, {Parker}, {Reid}, {Ritter}, {Russell}, {Stupar},
  {Watson}, {Bienaym{\'e}}, {Bland-Hawthorn}, {Gerhard}, {Gibson}, {Grebel},
  {Helmi}, {Navarro}, {Conrad}, {Famaey}, {Faure}, {Just}, {Kos},
  {Matijevi{\v{c}}}, {McMillan}, {Minchev}, {Scholz}, {Sharma}, {Siviero}, {de
  Boer}, \& {{\v{Z}}erjal}}]{2013AJ....146..134K}
{Kordopatis}, G., {Gilmore}, G., {Steinmetz}, M., {et~al.} 2013, \aj, 146, 134

\bibitem[{{Kunder} {et~al.}(2017){Kunder}, {Kordopatis}, {Steinmetz},
  {Zwitter}, {McMillan}, {Casagrande}, {Enke}, {Wojno}, {Valentini},
  {Chiappini}, {Matijevi{\v{c}}}, {Siviero}, {de Laverny}, {Recio-Blanco},
  {Bijaoui}, {Wyse}, {Binney}, {Grebel}, {Helmi}, {Jofre}, {Antoja}, {Gilmore},
  {Siebert}, {Famaey}, {Bienaym{\'e}}, {Gibson}, {Freeman}, {Navarro},
  {Munari}, {Seabroke}, {Anguiano}, {{\v{Z}}erjal}, {Minchev}, {Reid},
  {Bland-Hawthorn}, {Kos}, {Sharma}, {Watson}, {Parker}, {Scholz}, {Burton},
  {Cass}, {Hartley}, {Fiegert}, {Stupar}, {Ritter}, {Hawkins}, {Gerhard},
  {Chaplin}, {Davies}, {Elsworth}, {Lund}, {Miglio}, \&
  {Mosser}}]{2017AJ....153...75K}
{Kunder}, A., {Kordopatis}, G., {Steinmetz}, M., {et~al.} 2017, \aj, 153, 75

\bibitem[{{Lafarga} {et~al.}(2020){Lafarga}, {Ribas}, {Lovis}, {Perger},
  {Zechmeister}, {Bauer}, {K{\"u}rster}, {Cort{\'e}s-Contreras}, {Morales},
  {Herrero}, {Rosich}, {Baroch}, {Reiners}, {Caballero}, {Quirrenbach},
  {Amado}, {Alacid}, {B{\'e}jar}, {Dreizler}, {Hatzes}, {Henning}, {Jeffers},
  {Kaminski}, {Montes}, {Pedraz}, {Rodr{\'\i}guez-L{\'o}pez}, \&
  {Schmitt}}]{2020A&A...636A..36L}
{Lafarga}, M., {Ribas}, I., {Lovis}, C., {et~al.} 2020, \aap, 636, A36

\bibitem[{{Lalitha} {et~al.}(2019){Lalitha}, {Baroch}, {Morales}, {Passegger},
  {Bauer}, {Cardona Guill{\'e}n}, {Dreizler}, {Oshagh}, {Reiners}, {Ribas},
  {Caballero}, {Quirrenbach}, {Amado}, {B{\'e}jar}, {Colom{\'e}},
  {Cort{\'e}s-Contreras}, {Galad{\'\i}-Enr{\'\i}quez}, {Gonz{\'a}lez-Cuesta},
  {Guenther}, {Hagen}, {Henning}, {Herrero}, {Husser}, {Jeffers}, {Kaminski},
  {K{\\"u}rster}, {Lafarga}, {Lodieu}, {L{\'o}pez-Gonz{\'a}lez}, {Montes},
  {Perger}, {Rosich}, {Rodr{\'\i}guez}, {Rodr{\'\i}guez-L{\'o}pez}, {Schmitt},
  {Tal-Or}, \& {Zechmeister}}]{2019A&A...627A.116L}
{Lalitha}, S., {Baroch}, D., {Morales}, J.~C., {et~al.} 2019, \aap, 627, A116

\bibitem[{{Lamman} {et~al.}(2020){Lamman}, {Baranec}, {Berta-Thompson}, {Law},
  {Schonhut-Stasik}, {Ziegler}, {Salama}, {Jensen-Clem}, {Duev}, {Riddle},
  {Kulkarni}, {Winters}, \& {Irwin}}]{2020AJ....159..139L}
{Lamman}, C., {Baranec}, C., {Berta-Thompson}, Z.~K., {et~al.} 2020, \aj, 159,
  139

\bibitem[{{Lane} {et~al.}(2001){Lane}, {Zapatero Osorio}, {Britton},
  {Mart{\'\i}n}, \& {Kulkarni}}]{2001ApJ...560..390L}
{Lane}, B.~F., {Zapatero Osorio}, M.~R., {Britton}, M.~C., {Mart{\'\i}n},
  E.~L., \& {Kulkarni}, S.~R. 2001, \apj, 560, 390

\bibitem[{{Laureijs} {et~al.}(2011){Laureijs}, {Amiaux}, {Arduini},
  {Augu{\`e}res}, {Brinchmann}, {Cole}, {Cropper}, {Dabin}, {Duvet}, {Ealet},
  \& et~al.}]{2011arXiv1110.3193L}
{Laureijs}, R., {Amiaux}, J., {Arduini}, S., {et~al.} 2011, arXiv e-prints,
  arXiv:1110.3193

\bibitem[{{Law} {et~al.}(2008){Law}, {Hodgkin}, \&
  {Mackay}}]{2008MNRAS.384..150L}
{Law}, N.~M., {Hodgkin}, S.~T., \& {Mackay}, C.~D. 2008, \mnras, 384, 150

\bibitem[{{Lazorenko} \& {Sahlmann}(2018)}]{2018A&A...618A.111L}
{Lazorenko}, P.~F. \& {Sahlmann}, J. 2018, \aap, 618, A111

\bibitem[{{Leggett} {et~al.}(2012){Leggett}, {Saumon}, {Marley}, {Lodders},
  {Canty}, {Lucas}, {Smart}, {Tinney}, {Homeier}, {Allard}, {Burningham},
  {Day-Jones}, {Fegley}, {Ishii}, {Jones}, {Marocco}, {Pinfield}, \&
  {Tamura}}]{2012ApJ...748...74L}
{Leggett}, S.~K., {Saumon}, D., {Marley}, M.~S., {et~al.} 2012, \apj, 748, 74

\bibitem[{{Leinert} {et~al.}(2000){Leinert}, {Allard}, {Richichi}, \&
  {Hauschildt}}]{2000A&A...353..691L}
{Leinert}, C., {Allard}, F., {Richichi}, A., \& {Hauschildt}, P.~H. 2000, \aap,
  353, 691

\bibitem[{{LIFE collaboration} {et~al.}(2021){LIFE collaboration}, {Quanz},
  {Ottiger}, {Fontanet}, {Kammerer}, {Menti}, {Dannert}, {Gheorghe}, {Absil},
  {Airapetian}, {Alei}, {Allart}, {Angerhausen}, {Blumenthal}, {Cabrera},
  {Carri{\'o}n-Gonz{\'a}lez}, {Chauvin}, {Danchi}, {Dandumont}, {Defr{\`e}re},
  {Dorn}, {Ehrenreich}, {Ertel}, {Fridlund}, {Garc{\'\i}a Mu{\~n}oz},
  {Gasc{\'o}n}, {Glauser}, {Grenfell}, {Guidi}, {Hagelberg}, {Helled},
  {Ireland}, {Kopparapu}, {Korth}, {Kraus}, {L{\'e}ger}, {Leedj{\"a}rv},
  {Lichtenberg}, {Lillo-Box}, {Linz}, {Liseau}, {Loicq}, {Mahendra}, {Malbet},
  {Mathew}, {Mennesson}, {Meyer}, {Mishra}, {Molaverdikhani}, {Noack}, {Oza},
  {Pall{\'e}}, {Parviainen}, {Quirrenbach}, {Rauer}, {Ribas}, {Rice},
  {Romagnolo}, {Rugheimer}, {Schwieterman}, {Serabyn}, {Sharma}, {Stassun},
  {Szul{\'a}gyi}, {Wang}, {Wunderlich}, \& {Wyatt}}]{2021arXiv210107500L}
{LIFE collaboration}, {Quanz}, S.~P., {Ottiger}, M., {et~al.} 2021, arXiv
  e-prints, arXiv:2101.07500

\bibitem[{{Lindegren} {et~al.}(2021){Lindegren}, {Klioner}, {Hern{\'a}ndez},
  {Bombrun}, {Ramos-Lerate}, {Steidelm{\"u}ller}, {Bastian}, {Biermann}, {de
  Torres}, {Gerlach}, {Geyer}, {Hilger}, {Hobbs}, {Lammers}, {McMillan},
  {Stephenson}, {Casta{\~n}eda}, {Davidson}, {Fabricius}, {Gracia-Abril},
  {Portell}, {Rowell}, {Teyssier}, {Torra}, {Bartolom{\'e}}, {Clotet},
  {Garralda}, {Gonz{\'a}lez-Vidal}, {Torra}, {Abbas}, {Altmann}, {Anglada
  Varela}, {Balaguer-N{\'u}{\~n}ez}, {Balog}, {Barache}, {Becciani}, {Bernet},
  {Bertone}, {Bianchi}, {Bouquillon}, {Brown}, {Bucciarelli}, {Busonero},
  {Butkevich}, {Buzzi}, {Cancelliere}, {Carlucci}, {Charlot}, {Cioni},
  {Crosta}, {Crowley}, {del Peloso}, {del Pozo}, {Drimmel}, {Esquej}, {Fienga},
  {Fraile}, {Gai}, {Garcia-Reinaldos}, {Guerra}, {Hambly}, {Hauser},
  {Jan{\ss}en}, {Jordan}, {Kostrzewa-Rutkowska}, {Lattanzi}, {Liao}, {Licata},
  {Lister}, {L{\"o}ffler}, {Marchant}, {Masip}, {Mignard}, {Mints}, {Molina},
  {Mora}, {Morbidelli}, {Murphy}, {Pagani}, {Panuzzo}, {Pe{\~n}alosa Esteller},
  {Poggio}, {Re Fiorentin}, {Riva}, {Sagrist{\`a} Sell{\'e}s}, {Sanchez
  Gimenez}, {Sarasso}, {Sciacca}, {Siddiqui}, {Smart}, {Souami}, {Spagna},
  {Steele}, {Taris}, {Utrilla}, {van Reeven}, \&
  {Vecchiato}}]{2021A&A...649A...2L}
{Lindegren}, L., {Klioner}, S.~A., {Hern{\'a}ndez}, J., {et~al.} 2021, \aap,
  649, A2

\bibitem[{{Lippincott}(1972)}]{1972AJ.....77..165L}
{Lippincott}, S.~L. 1972, \aj, 77, 165

\bibitem[{{Liu} {et~al.}(2012){Liu}, {Dupuy}, {Bowler}, {Leggett}, \&
  {Best}}]{2012ApJ...758...57L}
{Liu}, M.~C., {Dupuy}, T.~J., {Bowler}, B.~P., {Leggett}, S.~K., \& {Best}, W.
  M.~J. 2012, \apj, 758, 57

\bibitem[{{Looper} {et~al.}(2007){Looper}, {Kirkpatrick}, \&
  {Burgasser}}]{2007AJ....134.1162L}
{Looper}, D.~L., {Kirkpatrick}, J.~D., \& {Burgasser}, A.~J. 2007, \aj, 134,
  1162

\bibitem[{{Lopez-Santiago} {et~al.}(2020){Lopez-Santiago}, {Martino},
  {M{\'\i}guez}, \& {V{\'a}zquez}}]{2020AJ....160..273L}
{Lopez-Santiago}, J., {Martino}, L., {M{\'\i}guez}, J., \& {V{\'a}zquez}, M.~A.
  2020, \aj, 160, 273

\bibitem[{{LSST Science Collaboration} {et~al.}(2009){LSST Science
  Collaboration}, {Abell}, {Allison}, {Anderson}, {Andrew}, {Angel}, {Armus},
  {Arnett}, {Asztalos}, {Axelrod}, {Bailey}, {Ballantyne}, {Bankert},
  {Barkhouse}, {Barr}, {Barrientos}, {Barth}, {Bartlett}, {Becker}, {Becla},
  {Beers}, {Bernstein}, {Biswas}, {Blanton}, {Bloom}, {Bochanski}, {Boeshaar},
  {Borne}, {Bradac}, {Brandt}, {Bridge}, {Brown}, {Brunner}, {Bullock},
  {Burgasser}, {Burge}, {Burke}, {Cargile}, {Chandrasekharan}, {Chartas},
  {Chesley}, {Chu}, {Cinabro}, {Claire}, {Claver}, {Clowe}, {Connolly}, {Cook},
  {Cooke}, {Cooray}, {Covey}, {Culliton}, {de Jong}, {de Vries}, {Debattista},
  {Delgado}, {Dell'Antonio}, {Dhital}, {Di Stefano}, {Dickinson}, {Dilday},
  {Djorgovski}, {Dobler}, {Donalek}, {Dubois-Felsmann}, {Durech},
  {Eliasdottir}, {Eracleous}, {Eyer}, {Falco}, {Fan}, {Fassnacht}, {Ferguson},
  {Fernandez}, {Fields}, {Finkbeiner}, {Figueroa}, {Fox}, {Francke}, {Frank},
  {Frieman}, {Fromenteau}, {Furqan}, {Galaz}, {Gal-Yam}, {Garnavich},
  {Gawiser}, {Geary}, {Gee}, {Gibson}, {Gilmore}, {Grace}, {Green}, {Gressler},
  {Grillmair}, {Habib}, {Haggerty}, {Hamuy}, {Harris}, {Hawley}, {Heavens},
  {Hebb}, {Henry}, {Hileman}, {Hilton}, {Hoadley}, {Holberg}, {Holman},
  {Howell}, {Infante}, {Ivezic}, {Jacoby}, {Jain}, {R}, {Jedicke}, {Jee},
  {Garrett Jernigan}, {Jha}, {Johnston}, {Jones}, {Juric}, {Kaasalainen},
  {Styliani}, {Kafka}, {Kahn}, {Kaib}, {Kalirai}, {Kantor}, {Kasliwal},
  {Keeton}, {Kessler}, {Knezevic}, {Kowalski}, {Krabbendam}, {Krughoff},
  {Kulkarni}, {Kuhlman}, {Lacy}, {Lepine}, {Liang}, {Lien}, {Lira}, {Long},
  {Lorenz}, {Lotz}, {Lupton}, {Lutz}, {Macri}, {Mahabal}, {Mandelbaum},
  {Marshall}, {May}, {McGehee}, {Meadows}, {Meert}, {Milani}, {Miller},
  {Miller}, {Mills}, {Minniti}, {Monet}, {Mukadam}, {Nakar}, {Neill}, {Newman},
  {Nikolaev}, {Nordby}, {O'Connor}, {Oguri}, {Oliver}, {Olivier}, {Olsen},
  {Olsen}, {Olszewski}, {Oluseyi}, {Padilla}, {Parker}, {Pepper}, {Peterson},
  {Petry}, {Pinto}, {Pizagno}, {Popescu}, {Prsa}, {Radcka}, {Raddick},
  {Rasmussen}, {Rau}, {Rho}, {Rhoads}, {Richards}, {Ridgway}, {Robertson},
  {Roskar}, {Saha}, {Sarajedini}, {Scannapieco}, {Schalk}, {Schindler},
  {Schmidt}, {Schmidt}, {Schneider}, {Schumacher}, {Scranton}, {Sebag},
  {Seppala}, {Shemmer}, {Simon}, {Sivertz}, {Smith}, {Allyn Smith}, {Smith},
  {Spitz}, {Stanford}, {Stassun}, {Strader}, {Strauss}, {Stubbs}, {Sweeney},
  {Szalay}, {Szkody}, {Takada}, {Thorman}, {Trilling}, {Trimble}, {Tyson}, {Van
  Berg}, {Vanden Berk}, {VanderPlas}, {Verde}, {Vrsnak}, {Walkowicz},
  {Wandelt}, {Wang}, {Wang}, {Warner}, {Wechsler}, {West}, {Wiecha},
  {Williams}, {Willman}, {Wittman}, {Wolff}, {Wood-Vasey}, {Wozniak}, {Young},
  {Zentner}, \& {Zhan}}]{2009arXiv0912.0201L}
{LSST Science Collaboration}, {Abell}, P.~A., {Allison}, J., {et~al.} 2009,
  arXiv e-prints, arXiv:0912.0201

\bibitem[{{Luhman}(2013)}]{2013ApJ...767L...1L}
{Luhman}, K.~L. 2013, \apjl, 767, L1

\bibitem[{{Luhman}(2014)}]{2014ApJ...786L..18L}
{Luhman}, K.~L. 2014, \apjl, 786, L18

\bibitem[{{Luque} {et~al.}(2019){Luque}, {Pall{\'e}}, {Kossakowski},
  {Dreizler}, {Kemmer}, {Espinoza}, {Burt}, {Anglada-Escud{\'e}}, {B{\'e}jar},
  {Caballero}, {Collins}, {Collins}, {Cort{\'e}s-Contreras},
  {D{\'\i}ez-Alonso}, {Feng}, {Hatzes}, {Hellier}, {Henning}, {Jeffers},
  {Kaltenegger}, {K{\"u}rster}, {Madden}, {Molaverdikhani}, {Montes}, {Narita},
  {Nowak}, {Ofir}, {Oshagh}, {Parviainen}, {Quirrenbach}, {Reffert}, {Reiners},
  {Rodr{\'\i}guez-L{\'o}pez}, {Schlecker}, {Stock}, {Trifonov}, {Winn},
  {Zapatero Osorio}, {Zechmeister}, {Amado}, {Anderson}, {Batalha}, {Bauer},
  {Bluhm}, {Burke}, {Butler}, {Caldwell}, {Chen}, {Crane}, {Dragomir},
  {Dressing}, {Dynes}, {Jenkins}, {Kaminski}, {Klahr}, {Kotani}, {Lafarga},
  {Latham}, {Lewin}, {McDermott}, {Monta{\~n}{\'e}s-Rodr{\'\i}guez}, {Morales},
  {Murgas}, {Nagel}, {Pedraz}, {Ribas}, {Ricker}, {Rowden}, {Seager},
  {Shectman}, {Tamura}, {Teske}, {Twicken}, {Vanderspeck}, {Wang}, \&
  {Wohler}}]{2019A&A...628A..39L}
{Luque}, R., {Pall{\'e}}, E., {Kossakowski}, D., {et~al.} 2019, \aap, 628, A39

\bibitem[{{Lurie} {et~al.}(2015){Lurie}, {Davenport}, {Hawley}, {Wilkinson},
  {Wisniewski}, {Kowalski}, \& {Hebb}}]{2015ApJ...800...95L}
{Lurie}, J.~C., {Davenport}, J. R.~A., {Hawley}, S.~L., {et~al.} 2015, \apj,
  800, 95

\bibitem[{{Luyten}(1979)}]{1979nlcs.book.....L}
{Luyten}, W.~J. 1979, {New Luyten catalogue of stars with proper motions larger
  than two tenths of an arcsecond; and first supplement; NLTT. (Minneapolis
  (1979)); Label 12 = short description; Label 13 = documentation by Warren;
  Label 14 = catalogue}

\bibitem[{{Ma} {et~al.}(2018){Ma}, {Ge}, {Muterspaugh}, {Singer}, {Henry},
  {Gonz{\'a}lez Hern{\'a}ndez}, {Sithajan}, {Jeram}, {Williamson}, {Stassun},
  {Kimock}, {Varosi}, {Schofield}, {Liu}, {Powell}, {Cassette}, {Jakeman},
  {Avner}, {Grieves}, {Barnes}, {Zhao}, {Gilda}, {Grantham}, {Stafford},
  {Savage}, {Bland}, \& {Ealey}}]{2018MNRAS.480.2411M}
{Ma}, B., {Ge}, J., {Muterspaugh}, M., {et~al.} 2018, \mnras, 480, 2411

\bibitem[{{Mahadevan} {et~al.}(2021){Mahadevan}, {Stef{\'a}nsson}, {Robertson},
  {Terrien}, {Ninan}, {Holcomb}, {Halverson}, {Cochran}, {Kanodia}, {Ramsey},
  {Wolszczan}, {Endl}, {Bender}, {Diddams}, {Fredrick}, {Hearty}, {Monson},
  {Metcalf}, {Roy}, \& {Schwab}}]{2021arXiv210202233M}
{Mahadevan}, S., {Stef{\'a}nsson}, G., {Robertson}, P., {et~al.} 2021, arXiv
  e-prints, arXiv:2102.02233

\bibitem[{{Maldonado} {et~al.}(2010){Maldonado}, {Mart{\'\i}nez-Arn{\'a}iz},
  {Eiroa}, {Montes}, \& {Montesinos}}]{2010A&A...521A..12M}
{Maldonado}, J., {Mart{\'\i}nez-Arn{\'a}iz}, R.~M., {Eiroa}, C., {Montes}, D.,
  \& {Montesinos}, B. 2010, \aap, 521, A12

\bibitem[{{Malmquist}(1925)}]{1925MeLuF.106....1M}
{Malmquist}, K.~G. 1925, Meddelanden fran Lunds Astronomiska Observatorium
  Serie I, 106, 1

\bibitem[{{Mamajek} {et~al.}(2013){Mamajek}, {Bartlett}, {Seifahrt}, {Henry},
  {Dieterich}, {Lurie}, {Kenworthy}, {Jao}, {Riedel}, {Subasavage}, {Winters},
  {Finch}, {Ianna}, \& {Bean}}]{2013AJ....146..154M}
{Mamajek}, E.~E., {Bartlett}, J.~L., {Seifahrt}, A., {et~al.} 2013, \aj, 146,
  154

\bibitem[{{Mamajek} \& {Bell}(2014)}]{2014MNRAS.445.2169M}
{Mamajek}, E.~E. \& {Bell}, C. P.~M. 2014, \mnras, 445, 2169

\bibitem[{{Marcy} {et~al.}(2001){Marcy}, {Butler}, {Fischer}, {Vogt},
  {Lissauer}, \& {Rivera}}]{2001ApJ...556..296M}
{Marcy}, G.~W., {Butler}, R.~P., {Fischer}, D., {et~al.} 2001, \apj, 556, 296

\bibitem[{{Marocco} {et~al.}(2021){Marocco}, {Eisenhardt}, {Fowler},
  {Kirkpatrick}, {Meisner}, {Schlafly}, {Stanford}, {Garcia}, {Caselden},
  {Cushing}, {Cutri}, {Faherty}, {Gelino}, {Gonzalez}, {Jarrett}, {Koontz},
  {Mainzer}, {Marchese}, {Mobasher}, {Schlegel}, {Stern}, {Teplitz}, \&
  {Wright}}]{2021ApJS..253....8M}
{Marocco}, F., {Eisenhardt}, P. R.~M., {Fowler}, J.~W., {et~al.} 2021, \apjs,
  253, 8

\bibitem[{{Marsh} {et~al.}(2013){Marsh}, {Wright}, {Kirkpatrick}, {Gelino},
  {Cushing}, {Griffith}, {Skrutskie}, \& {Eisenhardt}}]{2013ApJ...762..119M}
{Marsh}, K.~A., {Wright}, E.~L., {Kirkpatrick}, J.~D., {et~al.} 2013, \apj,
  762, 119

\bibitem[{{Mart{\'\i}n} {et~al.}(2020){Mart{\'\i}n}, {Solano}, {Burgasser},
  {Lodieu}, {S{\'a}nchez B{\'e}jar}, {Bouy}, {Barrado}, {Hu{\'e}lamo},
  {Morales}, {Shalmann}, \& {del Burgo}}]{2020sea..confE.157M}
{Mart{\'\i}n}, E.~L., {Solano}, E., {Burgasser}, A., {et~al.} 2020, in
  Contributions to the XIV.0 Scientific Meeting (virtual) of the Spanish
  Astronomical Society, 157

\bibitem[{{Martinache} {et~al.}(2007){Martinache}, {Lloyd}, {Ireland},
  {Yamada}, \& {Tuthill}}]{2007ApJ...661..496M}
{Martinache}, F., {Lloyd}, J.~P., {Ireland}, M.~J., {Yamada}, R.~S., \&
  {Tuthill}, P.~G. 2007, \apj, 661, 496

\bibitem[{{Martioli} {et~al.}(2020){Martioli}, {H{\'e}brard}, {Correia},
  {Laskar}, \& {Lecavelier des Etangs}}]{2020arXiv201213238M}
{Martioli}, E., {H{\'e}brard}, G., {Correia}, A.~C.~M., {Laskar}, J., \&
  {Lecavelier des Etangs}, A. 2020, arXiv e-prints, arXiv:2012.13238

\bibitem[{{Mason} {et~al.}(2017){Mason}, {Hartkopf}, \&
  {Miles}}]{2017AJ....154..200M}
{Mason}, B.~D., {Hartkopf}, W.~I., \& {Miles}, K.~N. 2017, \aj, 154, 200

\bibitem[{{Mason} {et~al.}(2018){Mason}, {Hartkopf}, {Miles}, {Subasavage},
  {Raghavan}, \& {Henry}}]{2018AJ....155..215M}
{Mason}, B.~D., {Hartkopf}, W.~I., {Miles}, K.~N., {et~al.} 2018, \aj, 155, 215

\bibitem[{{Mason} {et~al.}(1995){Mason}, {McAlister}, {Hartkopf}, \&
  {Shara}}]{1995AJ....109..332M}
{Mason}, B.~D., {McAlister}, H.~A., {Hartkopf}, W.~I., \& {Shara}, M.~M. 1995,
  \aj, 109, 332

\bibitem[{{Mason} {et~al.}(2001){Mason}, {Wycoff}, {Hartkopf}, {Douglass}, \&
  {Worley}}]{2001AJ....122.3466M}
{Mason}, B.~D., {Wycoff}, G.~L., {Hartkopf}, W.~I., {Douglass}, G.~G., \&
  {Worley}, C.~E. 2001, \aj, 122, 3466

\bibitem[{{Mawet} {et~al.}(2019){Mawet}, {Hirsch}, {Lee}, {Ruffio}, {Bottom},
  {Fulton}, {Absil}, {Beichman}, {Bowler}, {Bryan}, {Choquet}, {Ciardi},
  {Christiaens}, {Defr{\`e}re}, {Gomez Gonzalez}, {Howard}, {Huby}, {Isaacson},
  {Jensen-Clem}, {Kosiarek}, {Marcy}, {Meshkat}, {Petigura}, {Reggiani},
  {Ruane}, {Serabyn}, {Sinukoff}, {Wang}, {Weiss}, \&
  {Ygouf}}]{2019AJ....157...33M}
{Mawet}, D., {Hirsch}, L., {Lee}, E.~J., {et~al.} 2019, \aj, 157, 33

\bibitem[{{Mayor} {et~al.}(2009{\natexlab{a}}){Mayor}, {Bonfils}, {Forveille},
  {Delfosse}, {Udry}, {Bertaux}, {Beust}, {Bouchy}, {Lovis}, {Pepe}, {Perrier},
  {Queloz}, \& {Santos}}]{2009A&A...507..487M}
{Mayor}, M., {Bonfils}, X., {Forveille}, T., {et~al.} 2009{\natexlab{a}}, \aap,
  507, 487

\bibitem[{{Mayor} {et~al.}(2009{\natexlab{b}}){Mayor}, {Udry}, {Lovis}, {Pepe},
  {Queloz}, {Benz}, {Bertaux}, {Bouchy}, {Mordasini}, \&
  {Segransan}}]{2009A&A...493..639M}
{Mayor}, M., {Udry}, S., {Lovis}, C., {et~al.} 2009{\natexlab{b}}, \aap, 493,
  639

\bibitem[{{McNamara} {et~al.}(1987){McNamara}, {Ianna}, \&
  {Fredrick}}]{1987AJ.....93.1245M}
{McNamara}, B.~R., {Ianna}, P.~A., \& {Fredrick}, L.~W. 1987, \aj, 93, 1245

\bibitem[{{Modirrousta-Galian} {et~al.}(2020){Modirrousta-Galian}, {Stelzer},
  {Magaudda}, {Maldonado}, {G{\\"u}del}, {Sanz-Forcada}, {Edwards}, \&
  {Micela}}]{2020A&A...641A.113M}
{Modirrousta-Galian}, D., {Stelzer}, B., {Magaudda}, E., {et~al.} 2020, \aap,
  641, A113

\bibitem[{{Montagnier} {et~al.}(2006){Montagnier}, {S{\'e}gransan}, {Beuzit},
  {Forveille}, {Delorme}, {Delfosse}, {Perrier}, {Udry}, {Mayor}, {Chauvin},
  {Lagrange}, {Mouillet}, {Fusco}, {Gigan}, \& {Stadler}}]{2006A&A...460L..19M}
{Montagnier}, G., {S{\'e}gransan}, D., {Beuzit}, J.~L., {et~al.} 2006, \aap,
  460, L19

\bibitem[{{Montes} {et~al.}(2001){Montes}, {L{\'o}pez-Santiago},
  {Fern{\'a}ndez-Figueroa}, \& {G{\'a}lvez}}]{2001A&A...379..976M}
{Montes}, D., {L{\'o}pez-Santiago}, J., {Fern{\'a}ndez-Figueroa}, M.~J., \&
  {G{\'a}lvez}, M.~C. 2001, \aap, 379, 976

\bibitem[{{Morales} {et~al.}(2019){Morales}, {Mustill}, {Ribas}, {Davies},
  {Reiners}, {Bauer}, {Kossakowski}, {Herrero}, {Rodr{\'\i}guez},
  {L{\'o}pez-Gonz{\'a}lez}, {Rodr{\'\i}guez-L{\'o}pez}, {B{\'e}jar},
  {Gonz{\'a}lez-Cuesta}, {Luque}, {Pall{\'e}}, {Perger}, {Baroch}, {Johansen},
  {Klahr}, {Mordasini}, {Anglada-Escud{\'e}}, {Caballero},
  {Cort{\'e}s-Contreras}, {Dreizler}, {Lafarga}, {Nagel}, {Passegger},
  {Reffert}, {Rosich}, {Schweitzer}, {Tal-Or}, {Trifonov}, {Zechmeister},
  {Quirrenbach}, {Amado}, {Guenther}, {Hagen}, {Henning}, {Jeffers},
  {Kaminski}, {K{\\"u}rster}, {Montes}, {Seifert}, {Abell{\'a}n}, {Abril},
  {Aceituno}, {Aceituno}, {Alonso-Floriano}, {Ammler-von Eiff}, {Antona},
  {Arroyo-Torres}, {Azzaro}, {Barrado}, {Becerril-Jarque}, {Ben{\'\i}tez},
  {Berdi{\~n}as}, {Bergond}, {Brinkm{\\"o}ller}, {del Burgo}, {Burn},
  {Calvo-Ortega}, {Cano}, {C{\'a}rdenas}, {Cardona Guill{\'e}n}, {Carro},
  {Casal}, {Casanova}, {Casasayas-Barris}, {Chaturvedi}, {Cifuentes}, {Claret},
  {Colom{\'e}}, {Czesla}, {D{\'\i}ez-Alonso}, {Dorda}, {Emsenhuber},
  {Fern{\'a}ndez}, {Fern{\'a}ndez-Mart{\'\i}n}, {Ferro}, {Fuhrmeister},
  {Galad{\'\i}-Enr{\'\i}quez}, {Gallardo Cava}, {Garc{\'i}a Vargas},
  {Garcia-Piquer}, {Gesa}, {Gonz{\'a}lez-{\'A}lvarez}, {Gonz{\'a}lez
  Hern{\'a}ndez}, {Gonz{\'a}lez-Peinado}, {Gu{\`a}rdia}, {Guijarro}, {de
  Guindos}, {Hatzes}, {Hauschildt}, {Hedrosa}, {Hermelo}, {Hern{\'a}ndez
  Arabi}, {Hern{\'a}ndez Otero}, {Hintz}, {Holgado}, {Huber}, {Huke},
  {Johnson}, {de Juan}, {Kehr}, {Kemmer}, {Kim}, {Kl{\\"u}ter}, {Klutsch},
  {Labarga}, {Labiche}, {Lalitha}, {Lamp{\'o}n}, {Lara}, {Launhardt},
  {L{\'a}zaro}, {Lizon}, {Llamas}, {Lodieu}, {L{\'o}pez del Fresno}, {L{\'o}pez
  Salas}, {L{\'o}pez-Santiago}, {Mag{\'a}n Madinabeitia}, {Mall}, {Mancini},
  {Mandel}, {Marfil}, {Mar{\'\i}n Molina}, {Mart{\'\i}n},
  {Mart{\'\i}n-Fern{\'a}ndez}, {Mart{\'\i}n-Ruiz},
  {Mart{\'\i}nez-Rodr{\'\i}guez}, {Marvin}, {Mirabet}, {Moya}, {Naranjo},
  {Nelson}, {Nortmann}, {Nowak}, {Ofir}, {Pascual}, {Pavlov}, {Pedraz},
  {P{\'e}rez Medialdea}, {P{\'e}rez-Calpena}, {Perryman}, {Rabaza}, {Ram{\'o}n
  Ballesta}, {Rebolo}, {Redondo}, {Rix}, {Rodler}, {Rodr{\'\i}guez Trinidad},
  {Sabotta}, {Sadegi}, {Salz}, {S{\'a}nchez-Blanco}, {S{\'a}nchez Carrasco},
  {S{\'a}nchez-L{\'o}pez}, {Sanz-Forcada}, {Sarkis}, {Sarmiento},
  {Sch{\\"a}fer}, {Schlecker}, {Schmitt}, {Sch{\\"o}fer}, {Solano}, {Sota},
  {Stahl}, {Stock}, {Stuber}, {St{\\"u}rmer}, {Su{\'a}rez}, {Tabernero},
  {Tulloch}, {Veredas}, {Vico-Linares}, {Vilardell}, {Wagner}, {Winkler},
  {Wolthoff}, {Yan}, \& {Zapatero Osorio}}]{2019Sci...365.1441M}
{Morales}, J.~C., {Mustill}, A.~J., {Ribas}, I., {et~al.} 2019, Science, 365,
  1441

\bibitem[{{Morbey} \& {Griffin}(1987)}]{1987ApJ...317..343M}
{Morbey}, C.~L. \& {Griffin}, R.~F. 1987, \apj, 317, 343

\bibitem[{{Morel}(2018)}]{2018A&A...615A.172M}
{Morel}, T. 2018, \aap, 615, A172

\bibitem[{{Morin} {et~al.}(2008){Morin}, {Donati}, {Petit}, {Delfosse},
  {Forveille}, {Albert}, {Auri{\`e}re}, {Cabanac}, {Dintrans}, {Fares},
  {Gastine}, {Jardine}, {Ligni{\`e}res}, {Paletou}, {Ramirez Velez}, \&
  {Th{\'e}ado}}]{2008MNRAS.390..567M}
{Morin}, J., {Donati}, J.~F., {Petit}, P., {et~al.} 2008, \mnras, 390, 567

\bibitem[{{Morin} {et~al.}(2010){Morin}, {Donati}, {Petit}, {Delfosse},
  {Forveille}, \& {Jardine}}]{2010MNRAS.407.2269M}
{Morin}, J., {Donati}, J.~F., {Petit}, P., {et~al.} 2010, \mnras, 407, 2269

\bibitem[{{Motalebi} {et~al.}(2015){Motalebi}, {Udry}, {Gillon}, {Lovis},
  {S{\'e}gransan}, {Buchhave}, {Demory}, {Malavolta}, {Dressing}, {Sasselov},
  {Rice}, {Charbonneau}, {Collier Cameron}, {Latham}, {Molinari}, {Pepe},
  {Affer}, {Bonomo}, {Cosentino}, {Dumusque}, {Figueira}, {Fiorenzano},
  {Gettel}, {Harutyunyan}, {Haywood}, {Johnson}, {Lopez}, {Lopez-Morales},
  {Mayor}, {Micela}, {Mortier}, {Nascimbeni}, {Philips}, {Piotto}, {Pollacco},
  {Queloz}, {Sozzetti}, {Vanderburg}, \& {Watson}}]{2015A&A...584A..72M}
{Motalebi}, F., {Udry}, S., {Gillon}, M., {et~al.} 2015, \aap, 584, A72

\bibitem[{{Newcomb}(1904)}]{1904MNRAS..64..570N}
{Newcomb}, S. 1904, \mnras, 64, 570

\bibitem[{{Newton} {et~al.}(2014){Newton}, {Charbonneau}, {Irwin},
  {Berta-Thompson}, {Rojas-Ayala}, {Covey}, \& {Lloyd}}]{2014AJ....147...20N}
{Newton}, E.~R., {Charbonneau}, D., {Irwin}, J., {et~al.} 2014, \aj, 147, 20

\bibitem[{{Nidever} {et~al.}(2002){Nidever}, {Marcy}, {Butler}, {Fischer}, \&
  {Vogt}}]{2002ApJS..141..503N}
{Nidever}, D.~L., {Marcy}, G.~W., {Butler}, R.~P., {Fischer}, D.~A., \& {Vogt},
  S.~S. 2002, \apjs, 141, 503

\bibitem[{{Oppenheimer} {et~al.}(2001){Oppenheimer}, {Golimowski}, {Kulkarni},
  {Matthews}, {Nakajima}, {Creech-Eakman}, \& {Durrance}}]{2001AJ....121.2189O}
{Oppenheimer}, B.~R., {Golimowski}, D.~A., {Kulkarni}, S.~R., {et~al.} 2001,
  \aj, 121, 2189

\bibitem[{{Paloque}(1939)}]{1939AnTou..15...87P}
{Paloque}, E. 1939, Annales de l'Observatoire Astron. et Meteo. de Toulouse,
  15, 87

\bibitem[{{Pauli} {et~al.}(2006){Pauli}, {Napiwotzki}, {Heber}, {Altmann}, \&
  {Odenkirchen}}]{2006A&A...447..173P}
{Pauli}, E.~M., {Napiwotzki}, R., {Heber}, U., {Altmann}, M., \& {Odenkirchen},
  M. 2006, \aap, 447, 173

\bibitem[{{Pe{\~n}a Ram{\'\i}rez} {et~al.}(2012){Pe{\~n}a Ram{\'\i}rez},
  {B{\'e}jar}, {Zapatero Osorio}, {Petr-Gotzens}, \&
  {Mart{\'\i}n}}]{2012ApJ...754...30P}
{Pe{\~n}a Ram{\'\i}rez}, K., {B{\'e}jar}, V.~J.~S., {Zapatero Osorio}, M.~R.,
  {Petr-Gotzens}, M.~G., \& {Mart{\'\i}n}, E.~L. 2012, \apj, 754, 30

\bibitem[{{Pearce} {et~al.}(2021){Pearce}, {Beust}, {Faramaz}, {Booth},
  {Krivov}, {L{\\"o}hne}, \& {Poblete}}]{2021MNRAS.tmp..759P}
{Pearce}, T.~D., {Beust}, H., {Faramaz}, V., {et~al.} 2021, \mnras
  [\eprint[arXiv]{2103.04977}]

\bibitem[{{Pepe} {et~al.}(2011){Pepe}, {Lovis}, {S{\'e}gransan}, {Benz},
  {Bouchy}, {Dumusque}, {Mayor}, {Queloz}, {Santos}, \&
  {Udry}}]{2011A&A...534A..58P}
{Pepe}, F., {Lovis}, C., {S{\'e}gransan}, D., {et~al.} 2011, \aap, 534, A58

\bibitem[{{Perger} {et~al.}(2021){Perger}, {Ribas}, {Anglada-Escud{\'e}},
  {Morales}, {Amado}, {Caballero}, {Quirrenbach}, {Reiners}, {B{\'e}jar},
  {Dreizler}, {Galad{\'\i}-Enr{\'\i}quez}, {Hatzes}, {Henning}, {Jeffers},
  {Kaminski}, {K{\\"u}rster}, {Lafarga}, {Montes}, {Pal{\'e}},
  {Rodr{\'\i}guez-L{\'o}pez}, {Schweitzer}, {Zapatero Osorio}, \&
  {Zechmeister}}]{2021arXiv210310216P}
{Perger}, M., {Ribas}, I., {Anglada-Escud{\'e}}, G., {et~al.} 2021, arXiv
  e-prints, arXiv:2103.10216

\bibitem[{{Perger} {et~al.}(2019){Perger}, {Scandariato}, {Ribas}, {Morales},
  {Affer}, {Azzaro}, {Amado}, {Anglada-Escud{\'e}}, {Baroch}, {Barrado},
  {Bauer}, {B{\'e}jar}, {Caballero}, {Cort{\'e}s-Contreras}, {Damasso},
  {Dreizler}, {Gonz{\'a}lez-Cuesta}, {Gonz{\'a}lez Hern{\'a}ndez}, {Guenther},
  {Henning}, {Herrero}, {Jeffers}, {Kaminski}, {K{\\"u}rster}, {Lafarga},
  {Leto}, {L{\'o}pez-Gonz{\'a}lez}, {Maldonado}, {Micela}, {Montes},
  {Pinamonti}, {Quirrenbach}, {Rebolo}, {Reiners}, {Rodr{\'\i}guez},
  {Rodr{\'\i}guez-L{\'o}pez}, {Schmitt}, {Sozzetti}, {Su{\'a}rez
  Mascare{\~n}o}, {Toledo-Padr{\'o}n}, {Zanmar S{\'a}nchez}, {Zapatero Osorio},
  \& {Zechmeister}}]{2019A&A...624A.123P}
{Perger}, M., {Scandariato}, G., {Ribas}, I., {et~al.} 2019, \aap, 624, A123

\bibitem[{{Perryman} {et~al.}(1997){Perryman}, {Lindegren}, {Kovalevsky},
  {Hog}, {Bastian}, {Bernacca}, {Creze}, {Donati}, {Grenon}, {Grewing}, {van
  Leeuwen}, {van der Marel}, {Mignard}, {Murray}, {Le Poole}, {Schrijver},
  {Turon}, {Arenou}, {Froeschle}, \& {Petersen}}]{1997A&A...323L..49P}
{Perryman}, M.~A.~C., {Lindegren}, L., {Kovalevsky}, J., {et~al.} 1997, \aap,
  500, 501

\bibitem[{{Pettersen}(2006)}]{2006MNRAS.368.1392P}
{Pettersen}, B.~R. 2006, \mnras, 368, 1392

\bibitem[{{Phan-Bao} {et~al.}(2006){Phan-Bao}, {Bessell}, {Mart{\'\i}n},
  {Simon}, {Guibert}, {Forveille}, {Delfosse}, {Crifo}, {Epchtein}, {Wood}, \&
  {Tajahmady}}]{2006MNRAS.366L..40P}
{Phan-Bao}, N., {Bessell}, M.~S., {Mart{\'\i}n}, E.~L., {et~al.} 2006, \mnras,
  366, L40

\bibitem[{{Pinamonti} {et~al.}(2018){Pinamonti}, {Damasso}, {Marzari},
  {Sozzetti}, {Desidera}, {Maldonado}, {Scandariato}, {Affer}, {Lanza},
  {Bignamini}, {Bonomo}, {Borsa}, {Claudi}, {Cosentino}, {Giacobbe},
  {Gonz{\'a}lez-{\'A}lvarez}, {Gonz{\'a}lez Hern{\'a}ndez}, {Gratton}, {Leto},
  {Malavolta}, {Martinez Fiorenzano}, {Micela}, {Molinari}, {Pagano}, {Pedani},
  {Perger}, {Piotto}, {Rebolo}, {Ribas}, {Su{\'a}rez Mascare{\~n}o}, \&
  {Toledo-Padr{\'o}n}}]{2018A&A...617A.104P}
{Pinamonti}, M., {Damasso}, M., {Marzari}, F., {et~al.} 2018, \aap, 617, A104

\bibitem[{{Plavchan} {et~al.}(2020){Plavchan}, {Barclay}, {Gagn{\'e}}, {Gao},
  {Cale}, {Matzko}, {Dragomir}, {Quinn}, {Feliz}, {Stassun}, {Crossfield},
  {Berardo}, {Latham}, {Tieu}, {Anglada-Escud{\'e}}, {Ricker}, {Vanderspek},
  {Seager}, {Winn}, {Jenkins}, {Rinehart}, {Krishnamurthy}, {Dynes}, {Doty},
  {Adams}, {Afanasev}, {Beichman}, {Bottom}, {Bowler}, {Brinkworth}, {Brown},
  {Cancino}, {Ciardi}, {Clampin}, {Clark}, {Collins}, {Davison},
  {Foreman-Mackey}, {Furlan}, {Gaidos}, {Geneser}, {Giddens}, {Gilbert},
  {Hall}, {Hellier}, {Henry}, {Horner}, {Howard}, {Huang}, {Huber}, {Kane},
  {Kenworthy}, {Kielkopf}, {Kipping}, {Klenke}, {Kruse}, {Latouf}, {Lowrance},
  {Mennesson}, {Mengel}, {Mills}, {Morton}, {Narita}, {Newton}, {Nishimoto},
  {Okumura}, {Palle}, {Pepper}, {Quintana}, {Roberge}, {Roccatagliata},
  {Schlieder}, {Tanner}, {Teske}, {Tinney}, {Vanderburg}, {von Braun}, {Walp},
  {Wang}, {Wang}, {Weigand}, {White}, {Wittenmyer}, {Wright}, {Youngblood},
  {Zhang}, \& {Zilberman}}]{2020Natur.582..497P}
{Plavchan}, P., {Barclay}, T., {Gagn{\'e}}, J., {et~al.} 2020, \ at, 582, 497

\bibitem[{{Pourbaix} \& {Boffin}(2016)}]{2016A&A...586A..90P}
{Pourbaix}, D. \& {Boffin}, H. M.~J. 2016, \aap, 586, A90

\bibitem[{{Pourbaix} {et~al.}(2004){Pourbaix}, {Tokovinin}, {Batten}, {Fekel},
  {Hartkopf}, {Levato}, {Morrell}, {Torres}, \& {Udry}}]{2004A&A...424..727P}
{Pourbaix}, D., {Tokovinin}, A.~A., {Batten}, A.~H., {et~al.} 2004, \aap, 424,
  727

\bibitem[{{Poveda} {et~al.}(1994){Poveda}, {Herrera}, {Allen}, {Cordero}, \&
  {Lavalley}}]{1994RMxAA..28...43P}
{Poveda}, A., {Herrera}, M.~A., {Allen}, C., {Cordero}, G., \& {Lavalley}, C.
  1994, \rmxaa, 28, 43

\bibitem[{{Pozuelos} {et~al.}(2020){Pozuelos}, {Su{\'a}rez}, {de El{\'\i}a},
  {Berdi{\~n}as}, {Bonfanti}, {Dugaro}, {Gillon}, {Jehin}, {G{\\"u}nther}, {Van
  Grootel}, {Garcia}, {Thuillier}, {Delrez}, \&
  {Rod{\'o}n}}]{2020A&A...641A..23P}
{Pozuelos}, F.~J., {Su{\'a}rez}, J.~C., {de El{\'\i}a}, G.~C., {et~al.} 2020,
  \aap, 641, A23

\bibitem[{{Prieur} {et~al.}(2014){Prieur}, {Scardia}, {Pansecchi}, {Argyle},
  {Zanutta}, \& {Aristidi}}]{2014AN....335..817P}
{Prieur}, J.~L., {Scardia}, M., {Pansecchi}, L., {et~al.} 2014, Astronomische
  Nachrichten, 335, 817

\bibitem[{{Provencal} {et~al.}(2002){Provencal}, {Shipman}, {Koester},
  {Wesemael}, \& {Bergeron}}]{2002ApJ...568..324P}
{Provencal}, J.~L., {Shipman}, H.~L., {Koester}, D., {Wesemael}, F., \&
  {Bergeron}, P. 2002, \apj, 568, 324

\bibitem[{{Raghavan} {et~al.}(2010){Raghavan}, {McAlister}, {Henry}, {Latham},
  {Marcy}, {Mason}, {Gies}, {White}, \& {ten Brummelaar}}]{2010ApJS..190....1R}
{Raghavan}, D., {McAlister}, H.~A., {Henry}, T.~J., {et~al.} 2010, \apjs, 190,
  1

\bibitem[{{Rauer} {et~al.}(2014){Rauer}, {Catala}, {Aerts}, {Appourchaux},
  {Benz}, {Brandeker}, {Christensen-Dalsgaard}, {Deleuil}, {Gizon}, {Goupil},
  {G{\"u}del}, {Janot-Pacheco}, {Mas-Hesse}, {Pagano}, {Piotto}, {Pollacco},
  {Santos}, {Smith}, {Su{\'a}rez}, {Szab{\'o}}, {Udry}, {Adibekyan}, {Alibert},
  {Almenara}, {Amaro-Seoane}, {Eiff}, {Asplund}, {Antonello}, {Barnes},
  {Baudin}, {Belkacem}, {Bergemann}, {Bihain}, {Birch}, {Bonfils}, {Boisse},
  {Bonomo}, {Borsa}, {Brand{\~a}o}, {Brocato}, {Brun}, {Burleigh}, {Burston},
  {Cabrera}, {Cassisi}, {Chaplin}, {Charpinet}, {Chiappini}, {Church},
  {Csizmadia}, {Cunha}, {Damasso}, {Davies}, {Deeg}, {D{\'\i}az}, {Dreizler},
  {Dreyer}, {Eggenberger}, {Ehrenreich}, {Eigm{\"u}ller}, {Erikson}, {Farmer},
  {Feltzing}, {de Oliveira Fialho}, {Figueira}, {Forveille}, {Fridlund},
  {Garc{\'\i}a}, {Giommi}, {Giuffrida}, {Godolt}, {Gomes da Silva}, {Granzer},
  {Grenfell}, {Grotsch-Noels}, {G{\"u}nther}, {Haswell}, {Hatzes},
  {H{\'e}brard}, {Hekker}, {Helled}, {Heng}, {Jenkins}, {Johansen},
  {Khodachenko}, {Kislyakova}, {Kley}, {Kolb}, {Krivova}, {Kupka}, {Lammer},
  {Lanza}, {Lebreton}, {Magrin}, {Marcos-Arenal}, {Marrese}, {Marques},
  {Martins}, {Mathis}, {Mathur}, {Messina}, {Miglio}, {Montalban}, {Montalto},
  {Monteiro}, {Moradi}, {Moravveji}, {Mordasini}, {Morel}, {Mortier},
  {Nascimbeni}, {Nelson}, {Nielsen}, {Noack}, {Norton}, {Ofir}, {Oshagh},
  {Ouazzani}, {P{\'a}pics}, {Parro}, {Petit}, {Plez}, {Poretti}, {Quirrenbach},
  {Ragazzoni}, {Raimondo}, {Rainer}, {Reese}, {Redmer}, {Reffert},
  {Rojas-Ayala}, {Roxburgh}, {Salmon}, {Santerne}, {Schneider}, {Schou},
  {Schuh}, {Schunker}, {Silva-Valio}, {Silvotti}, {Skillen}, {Snellen}, {Sohl},
  {Sousa}, {Sozzetti}, {Stello}, {Strassmeier}, {{\v{S}}vanda}, {Szab{\'o}},
  {Tkachenko}, {Valencia}, {Van Grootel}, {Vauclair}, {Ventura}, {Wagner},
  {Walton}, {Weingrill}, {Werner}, {Wheatley}, \&
  {Zwintz}}]{2014ExA....38..249R}
{Rauer}, H., {Catala}, C., {Aerts}, C., {et~al.} 2014, Experimental Astronomy,
  38, 249

\bibitem[{{Reid} {et~al.}(2004){Reid}, {Cruz}, {Allen}, {Mungall}, {Kilkenny},
  {Liebert}, {Hawley}, {Fraser}, {Covey}, {Lowrance}, {Kirkpatrick}, \&
  {Burgasser}}]{2004AJ....128..463R}
{Reid}, I.~N., {Cruz}, K.~L., {Allen}, P., {et~al.} 2004, \aj, 128, 463

\bibitem[{{Reid} \& {Gizis}(1997)}]{1997AJ....113.2246R}
{Reid}, I.~N. \& {Gizis}, J.~E. 1997, \aj, 113, 2246

\bibitem[{{Reid} {et~al.}(1995){Reid}, {Hawley}, \&
  {Gizis}}]{1995AJ....110.1838R}
{Reid}, I.~N., {Hawley}, S.~L., \& {Gizis}, J.~E. 1995, \aj, 110, 1838

\bibitem[{{Reid} {et~al.}(2000){Reid}, {Kirkpatrick}, {Gizis}, {Dahn}, {Monet},
  {Williams}, {Liebert}, \& {Burgasser}}]{2000AJ....119..369R}
{Reid}, I.~N., {Kirkpatrick}, J.~D., {Gizis}, J.~E., {et~al.} 2000, \aj, 119,
  369

\bibitem[{{Reid} {et~al.}(2002){Reid}, {Kirkpatrick}, {Liebert}, {Gizis},
  {Dahn}, \& {Monet}}]{2002AJ....124..519R}
{Reid}, I.~N., {Kirkpatrick}, J.~D., {Liebert}, J., {et~al.} 2002, \aj, 124,
  519

\bibitem[{{Reid} \& {Menten}(2020)}]{2020AN....341..860R}
{Reid}, M.~J. \& {Menten}, K.~M. 2020, Astronomische Nachrichten, 341, 860

\bibitem[{{Reiners} \& {Basri}(2009)}]{2009ApJ...705.1416R}
{Reiners}, A. \& {Basri}, G. 2009, \apj, 705, 1416

\bibitem[{{Riaz} {et~al.}(2006){Riaz}, {Gizis}, \&
  {Harvin}}]{2006AJ....132..866R}
{Riaz}, B., {Gizis}, J.~E., \& {Harvin}, J. 2006, \aj, 132, 866

\bibitem[{{Ribas} {et~al.}(2018){Ribas}, {Tuomi}, {Reiners}, {Butler},
  {Morales}, {Perger}, {Dreizler}, {Rodr{\'\i}guez-L{\'o}pez}, {Gonz{\'a}lez
  Hern{\'a}ndez}, {Rosich}, {Feng}, {Trifonov}, {Vogt}, {Caballero}, {Hatzes},
  {Herrero}, {Jeffers}, {Lafarga}, {Murgas}, {Nelson}, {Rodr{\'\i}guez},
  {Strachan}, {Tal-Or}, {Teske}, {Toledo-Padr{\'o}n}, {Zechmeister},
  {Quirrenbach}, {Amado}, {Azzaro}, {B{\'e}jar}, {Barnes}, {Berdi{\~n}as},
  {Burt}, {Coleman}, {Cort{\'e}s-Contreras}, {Crane}, {Engle}, {Guinan},
  {Haswell}, {Henning}, {Holden}, {Jenkins}, {Jones}, {Kaminski}, {Kiraga},
  {K{\\"u}rster}, {Lee}, {L{\'o}pez-Gonz{\'a}lez}, {Montes}, {Morin}, {Ofir},
  {Pall{\'e}}, {Rebolo}, {Reffert}, {Schweitzer}, {Seifert}, {Shectman},
  {Staab}, {Street}, {Su{\'a}rez Mascare{\~n}o}, {Tsapras}, {Wang}, \&
  {Anglada-Escud{\'e}}}]{2018Natur.563..365R}
{Ribas}, I., {Tuomi}, M., {Reiners}, A., {et~al.} 2018, \ at, 563, 365

\bibitem[{{Ricker} {et~al.}(2015){Ricker}, {Winn}, {Vanderspek}, {Latham},
  {Bakos}, {Bean}, {Berta-Thompson}, {Brown}, {Buchhave}, {Butler}, {Butler},
  {Chaplin}, {Charbonneau}, {Christensen-Dalsgaard}, {Clampin}, {Deming},
  {Doty}, {De Lee}, {Dressing}, {Dunham}, {Endl}, {Fressin}, {Ge}, {Henning},
  {Holman}, {Howard}, {Ida}, {Jenkins}, {Jernigan}, {Johnson}, {Kaltenegger},
  {Kawai}, {Kjeldsen}, {Laughlin}, {Levine}, {Lin}, {Lissauer}, {MacQueen},
  {Marcy}, {McCullough}, {Morton}, {Narita}, {Paegert}, {Palle}, {Pepe},
  {Pepper}, {Quirrenbach}, {Rinehart}, {Sasselov}, {Sato}, {Seager},
  {Sozzetti}, {Stassun}, {Sullivan}, {Szentgyorgyi}, {Torres}, {Udry}, \&
  {Villasenor}}]{2015JATIS...1a4003R}
{Ricker}, G.~R., {Winn}, J.~N., {Vanderspek}, R., {et~al.} 2015, Journal of
  Astronomical Telescopes, Instruments, and Systems, 1, 014003

\bibitem[{{Riedel} {et~al.}(2011){Riedel}, {Murphy}, {Henry}, {Melis}, {Jao},
  \& {Subasavage}}]{2011AJ....142..104R}
{Riedel}, A.~R., {Murphy}, S.~J., {Henry}, T.~J., {et~al.} 2011, \aj, 142, 104

\bibitem[{{Rivera} {et~al.}(2010){Rivera}, {Laughlin}, {Butler}, {Vogt},
  {Haghighipour}, \& {Meschiari}}]{2010ApJ...719..890R}
{Rivera}, E.~J., {Laughlin}, G., {Butler}, R.~P., {et~al.} 2010, \apj, 719, 890

\bibitem[{{Rivera} {et~al.}(2005){Rivera}, {Lissauer}, {Butler}, {Marcy},
  {Vogt}, {Fischer}, {Brown}, {Laughlin}, \& {Henry}}]{2005ApJ...634..625R}
{Rivera}, E.~J., {Lissauer}, J.~J., {Butler}, R.~P., {et~al.} 2005, \apj, 634,
  625

\bibitem[{{Roberts} {et~al.}(2016){Roberts}, {Mason}, {Aguilar}, {Carson},
  {Crepp}, {Beichman}, {Brenner}, {Burruss}, {Cady}, {Luszcz-Cook}, {Dekany},
  {Hillenbrand}, {Hinkley}, {King}, {Lockhart}, {Nilsson}, {Oppenheimer},
  {Parry}, {Pueyo}, {Rice}, {Sivaramakrishnan}, {Soummer}, {Vasisht}, {Veicht},
  {Wang}, {Zhai}, \& {Zimmerman}}]{2016AJ....151..169R}
{Roberts}, Lewis~C., J., {Mason}, B.~D., {Aguilar}, J., {et~al.} 2016, \aj,
  151, 169

\bibitem[{{Robertson} {et~al.}(2014){Robertson}, {Mahadevan}, {Endl}, \&
  {Roy}}]{2014Sci...345..440R}
{Robertson}, P., {Mahadevan}, S., {Endl}, M., \& {Roy}, A. 2014, Science, 345,
  440

\bibitem[{{Robin} {et~al.}(2012){Robin}, {Luri}, {Reyl{\'e}}, {Isasi}, {Grux},
  {Blanco-Cuaresma}, {Arenou}, {Babusiaux}, {Belcheva}, {Drimmel}, {Jordi},
  {Krone-Martins}, {Masana}, {Mauduit}, {Mignard}, {Mowlavi},
  {Rocca-Volmerange}, {Sartoretti}, {Slezak}, \&
  {Sozzetti}}]{2012A&A...543A.100R}
{Robin}, A.~C., {Luri}, X., {Reyl{\'e}}, C., {et~al.} 2012, \aap, 543, A100

\bibitem[{{Rodriguez} {et~al.}(2015){Rodriguez}, {Duch{\^e}ne}, {Tom},
  {Kennedy}, {Matthews}, {Greaves}, \& {Butner}}]{2015MNRAS.449.3160R}
{Rodriguez}, D.~R., {Duch{\^e}ne}, G., {Tom}, H., {et~al.} 2015, \mnras, 449,
  3160

\bibitem[{{Ross}(1926)}]{1926AJ.....36..124R}
{Ross}, F.~E. 1926, \aj, 36, 124

\bibitem[{{Sahu} {et~al.}(2017){Sahu}, {Anderson}, {Casertano}, {Bond},
  {Bergeron}, {Nelan}, {Pueyo}, {Brown}, {Bellini}, {Levay}, {Sokol}, {aff1},
  {Dominik}, {Calamida}, {Kains}, \& {Livio}}]{2017Sci...356.1046S}
{Sahu}, K.~C., {Anderson}, J., {Casertano}, S., {et~al.} 2017, Science, 356,
  1046

\bibitem[{{Schneider} {et~al.}(2019){Schneider}, {Shkolnik}, {Allers}, {Kraus},
  {Liu}, {Weinberger}, \& {Flagg}}]{2019AJ....157..234S}
{Schneider}, A.~C., {Shkolnik}, E.~L., {Allers}, K.~N., {et~al.} 2019, \aj,
  157, 234

\bibitem[{{Scholz}(2014)}]{2014A&A...561A.113S}
{Scholz}, R.~D. 2014, \aap, 561, A113

\bibitem[{{Scholz} \& {Bell}(2018)}]{2018RNAAS...2...33S}
{Scholz}, R.-D. \& {Bell}, C. P.~M. 2018, Research Notes of the American
  Astronomical Society, 2, 33

\bibitem[{{Scholz} {et~al.}(2018){Scholz}, {Meusinger}, \&
  {Jahrei{\ss}}}]{2018A&A...613A..26S}
{Scholz}, R.~D., {Meusinger}, H., \& {Jahrei{\ss}}, H. 2018, \aap, 613, A26

\bibitem[{{Sebastian} {et~al.}(2021){Sebastian}, {Gillon}, {Ducrot},
  {Pozuelos}, {Garcia}, {G{\"u}nther}, {Delrez}, {Queloz}, {Demory}, {Triaud},
  {Burgasser}, {de Wit}, {Burdanov}, {Dransfield}, {Jehin}, {McCormac},
  {Murray}, {Niraula}, {Pedersen}, {Rackham}, {Sohy}, {Thompson}, \& {Van
  Grootel}}]{2021A&A...645A.100S}
{Sebastian}, D., {Gillon}, M., {Ducrot}, E., {et~al.} 2021, \aap, 645, A100

\bibitem[{{Shkolnik} {et~al.}(2012){Shkolnik}, {Anglada-Escud{\'e}}, {Liu},
  {Bowler}, {Weinberger}, {Boss}, {Reid}, \& {Tamura}}]{2012ApJ...758...56S}
{Shkolnik}, E.~L., {Anglada-Escud{\'e}}, G., {Liu}, M.~C., {et~al.} 2012, \apj,
  758, 56

\bibitem[{{Silvestri} {et~al.}(2002){Silvestri}, {Oswalt}, \&
  {Hawley}}]{2002AJ....124.1118S}
{Silvestri}, N.~M., {Oswalt}, T.~D., \& {Hawley}, S.~L. 2002, \aj, 124, 1118

\bibitem[{{Sion} {et~al.}(2009){Sion}, {Holberg}, {Oswalt}, {McCook}, \&
  {Wasatonic}}]{2009AJ....138.1681S}
{Sion}, E.~M., {Holberg}, J.~B., {Oswalt}, T.~D., {McCook}, G.~P., \&
  {Wasatonic}, R. 2009, \aj, 138, 1681

\bibitem[{{Skrutskie} {et~al.}(2006){Skrutskie}, {Cutri}, {Stiening},
  {Weinberg}, {Schneider}, {Carpenter}, {Beichman}, {Capps}, {Chester},
  {Elias}, {Huchra}, {Liebert}, {Lonsdale}, {Monet}, {Price}, {Seitzer},
  {Jarrett}, {Kirkpatrick}, {Gizis}, {Howard}, {Evans}, {Fowler}, {Fullmer},
  {Hurt}, {Light}, {Kopan}, {Marsh}, {McCallon}, {Tam}, {Van Dyk}, \&
  {Wheelock}}]{2006AJ....131.1163S}
{Skrutskie}, M.~F., {Cutri}, R.~M., {Stiening}, R., {et~al.} 2006, \aj, 131,
  1163

\bibitem[{{Skrutskie} {et~al.}(1987){Skrutskie}, {Forrest}, \&
  {Shure}}]{1987BAAS...19.1128S}
{Skrutskie}, M.~F., {Forrest}, W.~J., \& {Shure}, M.~A. 1987, in Bulletin of
  the American Astronomical Society, Vol.~19, 1128

\bibitem[{{Smart} {et~al.}(2017){Smart}, {Marocco}, {Caballero}, {Jones},
  {Barrado}, {Beam{\'\i}n}, {Pinfield}, \& {Sarro}}]{2017MNRAS.469..401S}
{Smart}, R.~L., {Marocco}, F., {Caballero}, J.~A., {et~al.} 2017, \mnras, 469,
  401

\bibitem[{{Smart} {et~al.}(2013){Smart}, {Tinney}, {Bucciarelli}, {Marocco},
  {Abbas}, {Andrei}, {Bernardi}, {Burningham}, {Cardoso}, {Costa}, {Crosta},
  {Dapr{\'a}}, {Day-Jones}, {Goldman}, {Jones}, {Lattanzi}, {Leggett}, {Lucas},
  {Mendez}, {Penna}, {Pinfield}, {Smith}, {Sozzetti}, \&
  {Vecchiato}}]{2013MNRAS.433.2054S}
{Smart}, R.~L., {Tinney}, C.~G., {Bucciarelli}, B., {et~al.} 2013, \mnras, 433,
  2054

\bibitem[{{S{\"o}derhjelm}(1999)}]{1999A&A...341..121S}
{S{\"o}derhjelm}, S. 1999, \aap, 341, 121

\bibitem[{{Solano} {et~al.}(2019){Solano}, {Mart{\'\i}n}, {Caballero},
  {Rodrigo}, {Angulo}, {Alcaniz}, {Borges Fernandes}, {Cenarro},
  {Crist{\'o}bal-Hornillos}, {Dupke}, {Alfaro}, {Ederoclite},
  {Jim{\'e}nez-Esteban}, {Hernandez-Jimenez}, {Hern{\'a}ndez-Monteagudo},
  {Lopes de Oliveira}, {L{\'o}pez-Sanjuan}, {Mar{\'\i}n-Franch}, {Mendes de
  Oliveira}, {Moles}, {Orsi}, {Schmidtobreick}, {Sobral}, {Sodr{\'e}},
  {Varela}, \& {V{\'a}zquez Rami{\'o}}}]{2019A&A...627A..29S}
{Solano}, E., {Mart{\'\i}n}, E.~L., {Caballero}, J.~A., {et~al.} 2019, \aap,
  627, A29

\bibitem[{{Soubiran} {et~al.}(2008){Soubiran}, {Bienaym{\'e}}, {Mishenina}, \&
  {Kovtyukh}}]{2008A&A...480...91S}
{Soubiran}, C., {Bienaym{\'e}}, O., {Mishenina}, T.~V., \& {Kovtyukh}, V.~V.
  2008, \aap, 480, 91

\bibitem[{{Soubiran} {et~al.}(2018){Soubiran}, {Jasniewicz}, {Chemin},
  {Zurbach}, {Brouillet}, {Panuzzo}, {Sartoretti}, {Katz}, {Le Campion},
  {Marchal}, {Hestroffer}, {Th{\'e}venin}, {Crifo}, {Udry}, {Cropper},
  {Seabroke}, {Viala}, {Benson}, {Blomme}, {Jean-Antoine}, {Huckle}, {Smith},
  {Baker}, {Damerdji}, {Dolding}, {Fr{\'e}mat}, {Gosset}, {Guerrier}, {Guy},
  {Haigron}, {Jan{\ss}en}, {Plum}, {Fabre}, {Lasne}, {Pailler}, {Panem},
  {Riclet}, {Royer}, {Tauran}, {Zwitter}, {Gueguen}, \&
  {Turon}}]{2018A&A...616A...7S}
{Soubiran}, C., {Jasniewicz}, G., {Chemin}, L., {et~al.} 2018, \aap, 616, A7

\bibitem[{{Sozzetti} \& {de Bruijne}(2018)}]{2018haex.bookE..81S}
{Sozzetti}, A. \& {de Bruijne}, J. 2018, {Space Astrometry Missions for
  Exoplanet Science: Gaia and the Legacy of Hipparcos}, ed. H.~J. {Deeg} \&
  J.~A. {Belmonte}, 81

\bibitem[{{Sperauskas} {et~al.}(2016){Sperauskas},
  {Barta{\v{s}}i{\={u}}t{\.{e}}}, {Boyle}, {Deveikis}, {Raudeli{\={u}}nas}, \&
  {Upgren}}]{2016A&A...596A.116S}
{Sperauskas}, J., {Barta{\v{s}}i{\={u}}t{\.{e}}}, S., {Boyle}, R.~P., {et~al.}
  2016, \aap, 596, A116

\bibitem[{{Stock} {et~al.}(2020){Stock}, {Nagel}, {Kemmer}, {Passegger},
  {Reffert}, {Quirrenbach}, {Caballero}, {Czesla}, {B{\'e}jar}, {Cardona},
  {D{\'\i}ez-Alonso}, {Herrero}, {Lalitha}, {Schlecker}, {Tal-Or},
  {Rodr{\'\i}guez}, {Rodr{\'\i}guez-L{\'o}pez}, {Ribas}, {Reiners}, {Amado},
  {Bauer}, {Bluhm}, {Cort{\'e}s-Contreras}, {Gonz{\'a}lez-Cuesta}, {Dreizler},
  {Hatzes}, {Henning}, {Jeffers}, {Kaminski}, {K{\\"u}rster}, {Lafarga},
  {L{\'o}pez-Gonz{\'a}lez}, {Montes}, {Morales}, {Pedraz}, {Sch{\\"o}fer},
  {Schweitzer}, {Trifonov}, {Zapatero Osorio}, \&
  {Zechmeister}}]{2020A&A...643A.112S}
{Stock}, S., {Nagel}, E., {Kemmer}, J., {et~al.} 2020, \aap, 643, A112

\bibitem[{{Su{\'a}rez Mascare{\~n}o} {et~al.}(2017){Su{\'a}rez Mascare{\~n}o},
  {Gonz{\'a}lez Hern{\'a}ndez}, {Rebolo}, {Velasco}, {Toledo-Padr{\'o}n},
  {Affer}, {Perger}, {Micela}, {Ribas}, {Maldonado}, {Leto}, {Zanmar Sanchez},
  {Scandariato}, {Damasso}, {Sozzetti}, {Esposito}, {Covino}, {Maggio},
  {Lanza}, {Desidera}, {Rosich}, {Bignamini}, {Claudi}, {Benatti}, {Borsa},
  {Pedani}, {Molinari}, {Morales}, {Herrero}, \&
  {Lafarga}}]{2017A&A...605A..92S}
{Su{\'a}rez Mascare{\~n}o}, A., {Gonz{\'a}lez Hern{\'a}ndez}, J.~I., {Rebolo},
  R., {et~al.} 2017, \aap, 605, A92

\bibitem[{{Tal-Or} {et~al.}(2018){Tal-Or}, {Zechmeister}, {Reiners}, {Jeffers},
  {Sch{\"o}fer}, {Quirrenbach}, {Amado}, {Ribas}, {Caballero}, {Aceituno},
  {Bauer}, {B{\'e}jar}, {Czesla}, {Dreizler}, {Fuhrmeister}, {Hatzes},
  {Johnson}, {K{\"u}rster}, {Lafarga}, {Montes}, {Morales}, {Reffert},
  {Sadegi}, {Seifert}, \& {Shulyak}}]{2018A&A...614A.122T}
{Tal-Or}, L., {Zechmeister}, M., {Reiners}, A., {et~al.} 2018, \aap, 614, A122

\bibitem[{{Tamazian} {et~al.}(2008){Tamazian}, {Docobo}, \&
  {Balega}}]{2008msah.conf...71T}
{Tamazian}, V.~S., {Docobo}, J.~A., \& {Balega}, Y.~Y. 2008, in Multiple Stars
  Across the H-R Diagram, ed. S.~{Hubrig}, M.~{Petr-Gotzens}, \&
  A.~{Tokovinin}, 71

\bibitem[{{Tamazian} \& {Malkov}(2014)}]{2014AcA....64..359T}
{Tamazian}, V.~S. \& {Malkov}, O.~Y. 2014, \actaa, 64, 359

\bibitem[{{Tanner} {et~al.}(2010){Tanner}, {Gelino}, \&
  {Law}}]{2010PASP..122.1195T}
{Tanner}, A.~M., {Gelino}, C.~R., \& {Law}, N.~M. 2010, \pasp, 122, 1195

\bibitem[{{Taylor}(2005)}]{Taylor2005}
{Taylor}, M.~B. 2005, in Astronomical Society of the Pacific Conference Series,
  Vol. 347, Astronomical Data Analysis Software and Systems XIV, ed.
  P.~{Shopbell}, M.~{Britton}, \& R.~{Ebert}, 29

\bibitem[{{Terrien} {et~al.}(2015{\natexlab{a}}){Terrien}, {Mahadevan},
  {Bender}, {Deshpande}, \& {Robertson}}]{2015ApJ...802L..10T}
{Terrien}, R.~C., {Mahadevan}, S., {Bender}, C.~F., {Deshpande}, R., \&
  {Robertson}, P. 2015{\natexlab{a}}, \apjl, 802, L10

\bibitem[{{Terrien} {et~al.}(2015{\natexlab{b}}){Terrien}, {Mahadevan},
  {Deshpande}, \& {Bender}}]{2015ApJS..220...16T}
{Terrien}, R.~C., {Mahadevan}, S., {Deshpande}, R., \& {Bender}, C.~F.
  2015{\natexlab{b}}, \apjs, 220, 16

\bibitem[{{Tifft}(1955)}]{1955AJ.....60..144T}
{Tifft}, W.~G. 1955, \aj, 60, 144

\bibitem[{{Tinney} {et~al.}(2011){Tinney}, {Butler}, {Jones}, {Wittenmyer},
  {O'Toole}, {Bailey}, \& {Carter}}]{2011ApJ...727..103T}
{Tinney}, C.~G., {Butler}, R.~P., {Jones}, H. R.~A., {et~al.} 2011, \apj, 727,
  103

\bibitem[{{Tinney} {et~al.}(2014){Tinney}, {Faherty}, {Kirkpatrick}, {Cushing},
  {Morley}, \& {Wright}}]{2014ApJ...796...39T}
{Tinney}, C.~G., {Faherty}, J.~K., {Kirkpatrick}, J.~D., {et~al.} 2014, \apj,
  796, 39

\bibitem[{{Tokovinin} {et~al.}(2019){Tokovinin}, {Everett}, {Horch}, {Torres},
  \& {Latham}}]{2019AJ....158..167T}
{Tokovinin}, A., {Everett}, M.~E., {Horch}, E.~P., {Torres}, G., \& {Latham},
  D.~W. 2019, \aj, 158, 167

\bibitem[{{Tokovinin} {et~al.}(2015){Tokovinin}, {Mason}, {Hartkopf}, {Mendez},
  \& {Horch}}]{2015AJ....150...50T}
{Tokovinin}, A., {Mason}, B.~D., {Hartkopf}, W.~I., {Mendez}, R.~A., \&
  {Horch}, E.~P. 2015, \aj, 150, 50

\bibitem[{{Tomkin} \& {Pettersen}(1986)}]{1986AJ.....92.1424T}
{Tomkin}, J. \& {Pettersen}, B.~R. 1986, \aj, 92, 1424

\bibitem[{{Toonen} {et~al.}(2017){Toonen}, {Hollands}, {G{\"a}nsicke}, \&
  {Boekholt}}]{2017A&A...602A..16T}
{Toonen}, S., {Hollands}, M., {G{\"a}nsicke}, B.~T., \& {Boekholt}, T. 2017,
  \aap, 602, A16

\bibitem[{{Torres} {et~al.}(2006){Torres}, {Quast}, {da Silva}, {de La Reza},
  {Melo}, \& {Sterzik}}]{2006A&A...460..695T}
{Torres}, C.~A.~O., {Quast}, G.~R., {da Silva}, L., {et~al.} 2006, \aap, 460,
  695

\bibitem[{{Torres} {et~al.}(2010){Torres}, {Andersen}, \&
  {Gim{\'e}nez}}]{2010A&ARv..18...67T}
{Torres}, G., {Andersen}, J., \& {Gim{\'e}nez}, A. 2010, \aapr, 18, 67

\bibitem[{{Trifonov} {et~al.}(2021){Trifonov}, {Caballero}, {Morales},
  {Seifahrt}, {Ribas}, {Reiners}, {Bean}, {Luque}, {Parviainen}, {Pall{\'e}},
  {Stock}, {Zechmeister}, {Amado}, {Anglada-Escud{\'e}}, {Azzaro}, {Barclay},
  {B{\'e}jar}, {Bluhm}, {Casasayas-Barris}, {Cifuentes}, {Collins}, {Collins},
  {Cort{\'e}s-Contreras}, {de Leon}, {Dreizler}, {Dressing}, {Esparza-Borges},
  {Espinoza}, {Fausnaugh}, {Fukui}, {Hatzes}, {Hellier}, {Henning}, {Henze},
  {Herrero}, {Jeffers}, {Jenkins}, {Jensen}, {Kaminski}, {Kasper},
  {Kossakowski}, {K{\\"u}rster}, {Lafarga}, {Latham}, {Mann}, {Molaverdikhani},
  {Montes}, {Montet}, {Murgas}, {Narita}, {Oshagh}, {Passegger}, {Pollacco},
  {Quinn}, {Quirrenbach}, {Ricker}, {Rodr{\'\i}guez L{\'o}pez}, {Sanz-Forcada},
  {Schwarz}, {Schweitzer}, {Seager}, {Shporer}, {Stangret}, {St{\\"u}rmer},
  {Tan}, {Tenenbaum}, {Twicken}, {Vanderspek}, \& {Winn}}]{2021Sci...371.1038T}
{Trifonov}, T., {Caballero}, J.~A., {Morales}, J.~C., {et~al.} 2021, Science,
  371, 1038

\bibitem[{{Trifonov} {et~al.}(2018){Trifonov}, {K{\"u}rster}, {Zechmeister},
  {Tal-Or}, {Caballero}, {Quirrenbach}, {Amado}, {Ribas}, {Reiners}, {Reffert},
  {Dreizler}, {Hatzes}, {Kaminski}, {Launhardt}, {Henning}, {Montes},
  {B{\'e}jar}, {Mundt}, {Pavlov}, {Schmitt}, {Seifert}, {Morales}, {Nowak},
  {Jeffers}, {Rodr{\'\i}guez-L{\'o}pez}, {del Burgo}, {Anglada-Escud{\'e}},
  {L{\'o}pez-Santiago}, {Mathar}, {Ammler-von Eiff}, {Guenther}, {Barrado},
  {Gonz{\'a}lez Hern{\'a}ndez}, {Mancini}, {St{\\"u}rmer}, {Abril}, {Aceituno},
  {Alonso-Floriano}, {Antona}, {Anwand-Heerwart}, {Arroyo-Torres}, {Azzaro},
  {Baroch}, {Bauer}, {Becerril}, {Ben{\'\i}tez}, {Berdi{\~n}as}, {Bergond},
  {Bl{\\"u}mcke}, {Brinkm{\\"o}ller}, {Cano}, {C{\'a}rdenas V{\'a}zquez},
  {Casal}, {Cifuentes}, {Claret}, {Colom{\'e}}, {Cort{\'e}s-Contreras},
  {Czesla}, {D{\'\i}ez-Alonso}, {Feiz}, {Fern{\'a}ndez}, {Ferro},
  {Fuhrmeister}, {Galad{\'\i}-Enr{\'\i}quez}, {Garcia-Piquer}, {Garc{\'i}a
  Vargas}, {Gesa}, {G{\'o}mez Galera}, {Gonz{\'a}lez-Peinado},
  {Gr{\\"o}zinger}, {Grohnert}, {Gu{\`a}rdia}, {Guijarro}, {de Guindos},
  {Guti{\'e}rrez-Soto}, {Hagen}, {Hauschildt}, {Hedrosa}, {Helmling},
  {Hermelo}, {Hern{\'a}ndez Arab{\'\i}}, {Hern{\'a}ndez Casta{\~n}o},
  {Hern{\'a}ndez Hernando}, {Herrero}, {Huber}, {Huke}, {Johnson}, {de Juan},
  {Kim}, {Klein}, {Kl{\\"u}ter}, {Klutsch}, {Lafarga}, {Lamp{\'o}n}, {Lara},
  {Laun}, {Lemke}, {Lenzen}, {L{\'o}pez del Fresno}, {L{\'o}pez-Gonz{\'a}lez},
  {L{\'o}pez-Puertas}, {L{\'o}pez Salas}, {Luque}, {Mag{\'a}n Madinabeitia},
  {Mall}, {Mandel}, {Marfil}, {Mar{\'\i}n Molina}, {Maroto Fern{\'a}ndez},
  {Mart{\'\i}n}, {Mart{\'\i}n-Ruiz}, {Marvin}, {Mirabet}, {Moya},
  {Moreno-Raya}, {Nagel}, {Naranjo}, {Nortmann}, {Ofir}, {Oreiro}, {Pall{\'e}},
  {Panduro}, {Pascual}, {Passegger}, {Pedraz}, {P{\'e}rez-Calpena}, {P{\'e}rez
  Medialdea}, {Perger}, {Perryman}, {Pluto}, {Rabaza}, {Ram{\'o}n}, {Rebolo},
  {Redondo}, {Reinhardt}, {Rhode}, {Rix}, {Rodler}, {Rodr{\'\i}guez},
  {Rodr{\'\i}guez Trinidad}, {Rohloff}, {Rosich}, {Sadegi},
  {S{\'a}nchez-Blanco}, {S{\'a}nchez Carrasco}, {S{\'a}nchez-L{\'o}pez},
  {Sanz-Forcada}, {Sarkis}, {Sarmiento}, {Sch{\\"a}fer}, {Schiller},
  {Sch{\\"o}fer}, {Schweitzer}, {Solano}, {Stahl}, {Strachan}, {Su{\'a}rez},
  {Tabernero}, {Tala}, {Tulloch}, {Veredas}, {Vico Linares}, {Vilardell},
  {Wagner}, {Winkler}, {Wolthoff}, {Xu}, {Yan}, \& {Zapatero
  Osorio}}]{2018A&A...609A.117T}
{Trifonov}, T., {K{\"u}rster}, M., {Zechmeister}, M., {et~al.} 2018, \aap, 609,
  A117

\bibitem[{{Tuomi} {et~al.}(2013{\natexlab{a}}){Tuomi}, {Anglada-Escud{\'e}},
  {Gerlach}, {Jones}, {Reiners}, {Rivera}, {Vogt}, \&
  {Butler}}]{2013A&A...549A..48T}
{Tuomi}, M., {Anglada-Escud{\'e}}, G., {Gerlach}, E., {et~al.}
  2013{\natexlab{a}}, \aap, 549, A48

\bibitem[{{Tuomi} {et~al.}(2014){Tuomi}, {Jones}, {Barnes},
  {Anglada-Escud{\'e}}, \& {Jenkins}}]{2014MNRAS.441.1545T}
{Tuomi}, M., {Jones}, H. R.~A., {Barnes}, J.~R., {Anglada-Escud{\'e}}, G., \&
  {Jenkins}, J.~S. 2014, \mnras, 441, 1545

\bibitem[{{Tuomi} {et~al.}(2019){Tuomi}, {Jones}, {Butler}, {Arriagada},
  {Vogt}, {Burt}, {Laughlin}, {Holden}, {Shectman}, {Crane}, {Thompson},
  {Keiser}, {Jenkins}, {Berdi{\~n}as}, {Diaz}, {Kiraga}, \&
  {Barnes}}]{2019arXiv190604644T}
{Tuomi}, M., {Jones}, H.~R.~A., {Butler}, R.~P., {et~al.} 2019, arXiv e-prints,
  arXiv:1906.04644

\bibitem[{{Tuomi} {et~al.}(2013{\natexlab{b}}){Tuomi}, {Jones}, {Jenkins},
  {Tinney}, {Butler}, {Vogt}, {Barnes}, {Wittenmyer}, {O'Toole}, {Horner},
  {Bailey}, {Carter}, {Wright}, {Salter}, \& {Pinfield}}]{2013A&A...551A..79T}
{Tuomi}, M., {Jones}, H.~R.~A., {Jenkins}, J.~S., {et~al.} 2013{\natexlab{b}},
  \aap, 551, A79

\bibitem[{{Valenti} \& {Fischer}(2005)}]{2005ApJS..159..141V}
{Valenti}, J.~A. \& {Fischer}, D.~A. 2005, \apjs, 159, 141

\bibitem[{{van Altena} {et~al.}(1995){van Altena}, {Lee}, \&
  {Hoffleit}}]{1995yCat.1174....0V}
{van Altena}, W.~F., {Lee}, J.~T., \& {Hoffleit}, D. 1995, VizieR Online Data
  Catalog, I/174

\bibitem[{{van Biesbroeck}(1961)}]{1961AJ.....66..528V}
{van Biesbroeck}, G. 1961, \aj, 66, 528

\bibitem[{{van de Kamp} \& {Lippincott}(1951)}]{1951PASP...63..141V}
{van de Kamp}, P. \& {Lippincott}, S.~L. 1951, \pasp, 63, 141

\bibitem[{{van Leeuwen}(2007)}]{2007A&A...474..653V}
{van Leeuwen}, F. 2007, \aap, 474, 653

\bibitem[{{Vigan} {et~al.}(2012){Vigan}, {Bonnefoy}, {Chauvin}, {Moutou}, \&
  {Montagnier}}]{2012A&A...540A.131V}
{Vigan}, A., {Bonnefoy}, M., {Chauvin}, G., {Moutou}, C., \& {Montagnier}, G.
  2012, \aap, 540, A131

\bibitem[{{Vogt} {et~al.}(2015){Vogt}, {Burt}, {Meschiari}, {Butler}, {Henry},
  {Wang}, {Holden}, {Gapp}, {Hanson}, {Arriagada}, {Keiser}, {Teske}, \&
  {Laughlin}}]{2015ApJ...814...12V}
{Vogt}, S.~S., {Burt}, J., {Meschiari}, S., {et~al.} 2015, \apj, 814, 12

\bibitem[{{Vogt} {et~al.}(2010){Vogt}, {Wittenmyer}, {Butler}, {O'Toole},
  {Henry}, {Rivera}, {Meschiari}, {Laughlin}, {Tinney}, {Jones}, {Bailey},
  {Carter}, \& {Batygin}}]{2010ApJ...708.1366V}
{Vogt}, S.~S., {Wittenmyer}, R.~A., {Butler}, R.~P., {et~al.} 2010, \apj, 708,
  1366

\bibitem[{{von Struve}(1840)}]{1840AN.....17..177V}
{von Struve}, O.~W. 1840, Astronomische Nachrichten, 17, 177

\bibitem[{{Ward-Duong} {et~al.}(2015){Ward-Duong}, {Patience}, {De Rosa},
  {Bulger}, {Rajan}, {Goodwin}, {Parker}, {McCarthy}, \&
  {Kulesa}}]{2015MNRAS.449.2618W}
{Ward-Duong}, K., {Patience}, J., {De Rosa}, R.~J., {et~al.} 2015, \mnras, 449,
  2618

\bibitem[{{Wehinger} \& {Wyckoff}(1966)}]{1966AJ.....71Q.185W}
{Wehinger}, P.~A. \& {Wyckoff}, S. 1966, \aj, 71, 185

\bibitem[{{Wenger} {et~al.}(2000){Wenger}, {Ochsenbein}, {Egret}, {Dubois},
  {Bonnarel}, {Borde}, {Genova}, {Jasniewicz}, {Lalo{\"e}}, {Lesteven}, \&
  {Monier}}]{2000A&AS..143....9W}
{Wenger}, M., {Ochsenbein}, F., {Egret}, D., {et~al.} 2000, \aaps, 143, 9

\bibitem[{{Wilson}(1953)}]{1953GCRV..C......0W}
{Wilson}, R.~E. 1953, Carnegie Institute Washington D.C. Publication, 0

\bibitem[{{Winters} {et~al.}(2019{\natexlab{a}}){Winters}, {Henry}, {Jao},
  {Subasavage}, {Chatelain}, {Slatten}, {Riedel}, {Silverstein}, \&
  {Payne}}]{2019AJ....157..216W}
{Winters}, J.~G., {Henry}, T.~J., {Jao}, W.-C., {et~al.} 2019{\natexlab{a}},
  \aj, 157, 216

\bibitem[{{Winters} {et~al.}(2011){Winters}, {Henry}, {Jao}, {Subasavage},
  {Finch}, \& {Hambly}}]{2011AJ....141...21W}
{Winters}, J.~G., {Henry}, T.~J., {Jao}, W.-C., {et~al.} 2011, \aj, 141, 21

\bibitem[{{Winters} {et~al.}(2018){Winters}, {Irwin}, {Newton}, {Charbonneau},
  {Latham}, {Han}, {Muirhead}, {Berlind}, {Calkins}, \&
  {Esquerdo}}]{2018AJ....155..125W}
{Winters}, J.~G., {Irwin}, J., {Newton}, E.~R., {et~al.} 2018, \aj, 155, 125

\bibitem[{{Winters} {et~al.}(2019{\natexlab{b}}){Winters}, {Medina}, {Irwin},
  {Charbonneau}, {Astudillo-Defru}, {Horch}, {Eastman}, {Vrijmoet}, {Henry},
  {Diamond-Lowe}, {Winston}, {Barclay}, {Bonfils}, {Ricker}, {Vanderspek},
  {Latham}, {Seager}, {Winn}, {Jenkins}, {Udry}, {Twicken}, {Teske},
  {Tenenbaum}, {Pepe}, {Murgas}, {Muirhead}, {Mink}, {Lovis}, {Levine},
  {L{\'e}pine}, {Jao}, {Henze}, {Fur{\'e}sz}, {Forveille}, {Figueira},
  {Esquerdo}, {Dressing}, {D{\'\i}az}, {Delfosse}, {Burke}, {Bouchy},
  {Berlind}, \& {Almenara}}]{2019AJ....158..152W}
{Winters}, J.~G., {Medina}, A.~A., {Irwin}, J.~M., {et~al.} 2019{\natexlab{b}},
  \aj, 158, 152

\bibitem[{{Wittenmyer} {et~al.}(2014){Wittenmyer}, {Tuomi}, {Butler}, {Jones},
  {Anglada-Escud{\'e}}, {Horner}, {Tinney}, {Marshall}, {Carter}, {Bailey},
  {Salter}, {O'Toole}, {Wright}, {Crane}, {Schectman}, {Arriagada}, {Thompson},
  {Minniti}, {Jenkins}, \& {Diaz}}]{2014ApJ...791..114W}
{Wittenmyer}, R.~A., {Tuomi}, M., {Butler}, R.~P., {et~al.} 2014, \apj, 791,
  114

\bibitem[{{Woitas} {et~al.}(2000){Woitas}, {Leinert}, {Jahrei{\ss}}, {Henry},
  {Franz}, \& {Wasserman}}]{2000A&A...353..253W}
{Woitas}, J., {Leinert}, C., {Jahrei{\ss}}, H., {et~al.} 2000, \aap, 353, 253

\bibitem[{{Woitas} {et~al.}(2003){Woitas}, {Tamazian}, {Docobo}, \&
  {Leinert}}]{2003A&A...406..293W}
{Woitas}, J., {Tamazian}, V.~S., {Docobo}, J.~A., \& {Leinert}, C. 2003, \aap,
  406, 293

\bibitem[{{Wolf}(1917)}]{1917AN....204..345W}
{Wolf}, M. 1917, Astronomische Nachrichten, 204, 345

\bibitem[{{Wood} \& {Linsky}(2010)}]{2010ApJ...717.1279W}
{Wood}, B.~E. \& {Linsky}, J.~L. 2010, \apj, 717, 1279

\bibitem[{{Wright} {et~al.}(2010){Wright}, {Eisenhardt}, {Mainzer}, {Ressler},
  {Cutri}, {Jarrett}, {Kirkpatrick}, {Padgett}, {McMillan}, {Skrutskie},
  {Stanford}, {Cohen}, {Walker}, {Mather}, {Leisawitz}, {Gautier}, {McLean},
  {Benford}, {Lonsdale}, {Blain}, {Mendez}, {Irace}, {Duval}, {Liu}, {Royer},
  {Heinrichsen}, {Howard}, {Shannon}, {Kendall}, {Walsh}, {Larsen}, {Cardon},
  {Schick}, {Schwalm}, {Abid}, {Fabinsky}, {Naes}, \&
  {Tsai}}]{2010AJ....140.1868W}
{Wright}, E.~L., {Eisenhardt}, P. R.~M., {Mainzer}, A.~K., {et~al.} 2010, \aj,
  140, 1868

\bibitem[{{Wright} {et~al.}(2013){Wright}, {Skrutskie}, {Kirkpatrick},
  {Gelino}, {Griffith}, {Marsh}, {Jarrett}, {Nelson}, {Borish}, {Mace},
  {Mainzer}, {Eisenhardt}, {McLean}, {Tobin}, \&
  {Cushing}}]{2013AJ....145...84W}
{Wright}, E.~L., {Skrutskie}, M.~F., {Kirkpatrick}, J.~D., {et~al.} 2013, \aj,
  145, 84

\bibitem[{{Zechmeister} {et~al.}(2019){Zechmeister}, {Dreizler}, {Ribas},
  {Reiners}, {Caballero}, {Bauer}, {B{\'e}jar}, {Gonz{\'a}lez-Cuesta},
  {Herrero}, {Lalitha}, {L{\'o}pez-Gonz{\'a}lez}, {Luque}, {Morales},
  {Pall{\'e}}, {Rodr{\'\i}guez}, {Rodr{\'\i}guez L{\'o}pez}, {Tal-Or},
  {Anglada-Escud{\'e}}, {Quirrenbach}, {Amado}, {Abril}, {Aceituno},
  {Aceituno}, {Alonso-Floriano}, {Ammler-von Eiff}, {Antona Jim{\'e}nez},
  {Anwand-Heerwart}, {Arroyo-Torres}, {Azzaro}, {Baroch}, {Barrado},
  {Becerril}, {Ben{\'\i}tez}, {Berdi{\~n}as}, {Bergond}, {Bluhm},
  {Brinkm{\\"o}ller}, {del Burgo}, {Calvo Ortega}, {Cano}, {Cardona
  Guill{\'e}n}, {Carro}, {C{\'a}rdenas V{\'a}zquez}, {Casal},
  {Casasayas-Barris}, {Casanova}, {Chaturvedi}, {Cifuentes}, {Claret},
  {Colom{\'e}}, {Cort{\'e}s-Contreras}, {Czesla}, {D{\'\i}ez-Alonso}, {Dorda},
  {Fern{\'a}ndez}, {Fern{\'a}ndez-Mart{\'\i}n}, {Fuhrmeister}, {Fukui},
  {Galad{\'\i}-Enr{\'\i}quez}, {Gallardo Cava}, {Garcia de la Fuente},
  {Garcia-Piquer}, {Garc{\'i}a Vargas}, {Gesa}, {G{\'o}ngora Rueda},
  {Gonz{\'a}lez-{\'A}lvarez}, {Gonz{\'a}lez Hern{\'a}ndez},
  {Gonz{\'a}lez-Peinado}, {Gr{\\"o}zinger}, {Gu{\`a}rdia}, {Guijarro}, {de
  Guindos}, {Hatzes}, {Hauschildt}, {Hedrosa}, {Helmling}, {Henning},
  {Hermelo}, {Hern{\'a}ndez Arabi}, {Hern{\'a}ndez Casta{\~n}o}, {Hern{\'a}ndez
  Otero}, {Hintz}, {Huke}, {Huber}, {Jeffers}, {Johnson}, {de Juan},
  {Kaminski}, {Kemmer}, {Kim}, {Klahr}, {Klein}, {Kl{\\"u}ter}, {Klutsch},
  {Kossakowski}, {K{\\"u}rster}, {Labarga}, {Lafarga}, {Llamas}, {Lamp{\'o}n},
  {Lara}, {Launhardt}, {L{\'a}zaro}, {Lodieu}, {L{\'o}pez del Fresno},
  {L{\'o}pez-Puertas}, {L{\'o}pez Salas}, {L{\'o}pez-Santiago}, {Mag{\'a}n
  Madinabeitia}, {Mall}, {Mancini}, {Mandel}, {Marfil}, {Mar{\'\i}n Molina},
  {Maroto Fern{\'a}ndez}, {Mart{\'\i}n}, {Mart{\'\i}n-Fern{\'a}ndez},
  {Mart{\'\i}n-Ruiz}, {Marvin}, {Mirabet}, {Monta{\~n}{\'e}s-Rodr{\'\i}guez},
  {Montes}, {Moreno-Raya}, {Nagel}, {Naranjo}, {Narita}, {Nortmann}, {Nowak},
  {Ofir}, {Oshagh}, {Panduro}, {Parviainen}, {Pascual}, {Passegger}, {Pavlov},
  {Pedraz}, {P{\'e}rez-Calpena}, {P{\'e}rez Medialdea}, {Perger}, {Perryman},
  {Rabaza}, {Ram{\'o}n Ballesta}, {Rebolo}, {Redondo}, {Reffert}, {Reinhardt},
  {Rhode}, {Rix}, {Rodler}, {Rodr{\'\i}guez Trinidad}, {Rosich}, {Sadegi},
  {S{\'a}nchez-Blanco}, {S{\'a}nchez Carrasco}, {S{\'a}nchez-L{\'o}pez},
  {Sanz-Forcada}, {Sarkis}, {Sarmiento}, {Sch{\\"a}fer}, {Schmitt},
  {Sch{\\"o}fer}, {Schweitzer}, {Seifert}, {Shulyak}, {Solano}, {Sota},
  {Stahl}, {Stock}, {Strachan}, {Stuber}, {St{\\"u}rmer}, {Su{\'a}rez},
  {Tabernero}, {Tala Pinto}, {Trifonov}, {Veredas}, {Vico Linares},
  {Vilardell}, {Wagner}, {Wolthoff}, {Xu}, {Yan}, \& {Zapatero
  Osorio}}]{2019A&A...627A..49Z}
{Zechmeister}, M., {Dreizler}, S., {Ribas}, I., {et~al.} 2019, \aap, 627, A49

\bibitem[{{Zuckerman} {et~al.}(2001){Zuckerman}, {Song}, {Bessell}, \&
  {Webb}}]{2001ApJ...562L..87Z}
{Zuckerman}, B., {Song}, I., {Bessell}, M.~S., \& {Webb}, R.~A. 2001, \apjl,
  562, L87

\end{thebibliography}

\begin{appendix} 
\label{app}
\section{References in the 10\,pc catalogue}
\paragraph{Astrometry.}
\cite{
2006AJ....132.2360H,
2007A&A...474..653V,
2008ApJ...689L..53B,
2010A&ARv..18...67T,
2011ApJS..197...19K,
2012ApJ...748...74L,
2012ApJ...752...56F,
2012ApJS..201...19D,
2013ApJ...762..119M,
2013MNRAS.433.2054S,
2014ApJ...784..156D,
2014ApJ...796...39T,
2016A&A...586A..90P,
2016AJ....152..141B,
2018A&A...618A.111L,
2018yCat.1345....0G,
2019ApJS..240...19K,
2020AJ....159..257B,
2020yCat.1350....0G,
2021ApJS..253....8M,
2021ApJS..253....7K}.

\paragraph{Line-of-sight velocities.}
\cite{
1953GCRV..C......0W,
1993AJ....105.1033A,
1996AJ....112.2799H,
2001A&A...379..976M,
2002AJ....123.3356G,
2002AJ....124.1118S,
2002ApJS..141..503N,
2004MNRAS.349.1069K,
2004A&A...424..727P,
2005ApJS..159..141V,
2006A&A...447..173P,
2006A&A...460..695T,
2006AstL...32..759G,
2007AN....328..889K,
2008A&A...480...91S,
2008MNRAS.390..567M,
2009ApJ...705.1416R,
2009yCat.2294....0A,
2010A&A...521A..12M,
2010ApJ...723..684B,
2010MNRAS.407.2269M,
2011AJ....142..104R,
2012ApJ...758...56S,
2013AJ....145...71K,
2013AJ....146..134K,
2013AJ....146..154M,
2014AJ....147...20N,
2014MNRAS.439.3094B,
2015AJ....149..104B,
2015AJ....150..179G,
2015ApJ...802L..10T,
2015ApJS..220...16T,
2015ApJS..220...18B,
2016A&A...586A..90P,
2016A&A...593A.127K,
2016A&A...596A.116S,
2016AJ....152..123G,
2016ApJS..225...10F,
2017AJ....153...75K,
2018A&A...614A..76J,
2018A&A...615A..31D,
2018yCat.1345....0G,
2018A&A...616A...7S,
2018A&A...619A..81H,
2019AJ....157..234S,
2019AJ....158..152W,
2020A&A...636A..36L}.

\paragraph{Spectral types.}
\cite{
1955AJ.....60..144T,
1957MNRAS.117..534E,
1967AJ.....72.1334C,
1984ApJS...55..657C,
1985ApJS...59..197B,
1986AJ.....92.1424T,
1989ApJS...71..245K,
1995AJ....109..332M,
1995AJ....110.1838R,
2000AJ....119..369R,
2000AJ....120.2082G,
2001AJ....121.2148G,
2001ApJ...560..390L,
2002AJ....123.2002H,
2002AJ....124..519R,
2002ApJ...564..466G,
2002ApJ...568..324P,
2003AJ....125..850B,
2003AJ....126.2048G,
2004A&A...415..265H,
2004AJ....128..463R,
2005ApJ...634.1336B,
2006A&A...460..695T,
2006A&A...460L..19M,
2006AJ....131.2722C,
2006AJ....132..161G,
2006AJ....132..866R,
2006AJ....132.2360H,
2006ApJ...637.1067B,
2006MNRAS.366L..40P,
2007AJ....133..439C,
2007AJ....134.1162L,
2007PhDT.........2B,
2008MNRAS.384..150L,
2008msah.conf...71T,
2009AJ....138.1681S,
2009ApJ...704..975J,
2010A&A...518A..39D,
2010AJ....139.2448B,
2010APJ...710...45B,
2010ApJ...710.1142B
2010MNRAS.408L..56L,
2011AJ....141..203A,
2011AJ....142..171G,
2011APJ...743...50C,
2011ApJ...743..138G,
2011ApJS..197...19K,
2011MNRAS.414L..90B,
2012A&A...540A.131V,
2012A&A...541A.163S,
2012A&A...545A..85R,
2013A&A...556A..15R,
2012AJ....144...99D,
2012ApJ...745...26B,
2012APJ...753..156K,
2013ApJ...776..128K,
2012ApJ...758...57L,
2013A&A...557A..43B,
2013AJ....145...84W,
2013AJ....145..102L,
2013ApJ...772..129B,
2013ApJ...776..126C,
2013ApJ...777...84B,
2013APJS..205....6M,
2013yCat....1.2023S,
2014A&A...567A..43S,
2014AJ....147...20N,
2014AJ....147...21J,
2014AJ....147...26D,
2014AJ....147...34S,
2014AJ....147..113C,
2014AJ....148...91L,
2014MNRAS.443.2561G,
2014yCat.5144....0M,
2015A&A...577A.128A,
2015AJ....149..104B,
2015AJ....149..106D,
2015ApJ...800...95L,
2015ApJ...804...92S,
2015ApJS..219...19L,
2016AJ....151..169R,
2016ApJS..224...36K,
2017A&A...600A..19P,
2017AJ....154...32S,
2017ApJ...842..118L,
2018A&A...613A..26S,
2018AJ....155..265H,
2018ApJ...867..109M,
2018ApJS..236...28T,
2018ApJS..238...29P,
2019A&A...625A..68S,
2019AJ....157...63K,
2019AJ....157..101M,
2020ApJ...899..123M,
2020ApJ...904..112B,
2020MNRAS.497..130T,
2021ApJS..253....7K}.

\paragraph{System information.}
\cite{
1803RSPT...93..339H,
1939AnTou..15...87P,
1951PASP...63..141V,
1956AJ.....61..405E,
1957AJ.....62..379F,
1966AJ.....71Q.185W,
1972AJ.....77..165L,
1987A&AS...71...57B,
1987AJ.....93.1245M,
1987BAAS...19.1128S,
1988AJ.....95.1841G,
1989A&A...225..369C,
1994RMxAA..28...43P,
1995AJ....109..332M,
1995ApJ...444L.101G,
1996AJ....111..365G,
1997AJ....113.2246R,
1998A&A...338..455F,
1998AJ....115.2579G,
1998AJ....116.1440H,
1998ApJ...497..935H,
1999A&A...341..121S,
1999MNRAS.308..111K,
2000A&A...353..253W,
2000A&A...353..691L,
2000ApJ...531L..57B,
2001AJ....121.2189O,
2002A&A...383..548G,
2002ApJS..141..503N,
2002PASP..114..224H,
2003A&A...406..293W,
2003AJ....125.1530G,
2004A&A...424..727P,
2004A&A...425..997B,
2004AJ....128.1733G,
2005ApJ...634..625R,
2006A&A...460L..19M,
2006AJ....132..994D,
2006ApJ...646..505B,
2006MNRAS.368.1392P,
2007A&A...468..721B,
2007A&A...474..293B,
2007ApJ...661..496M,
2008A&A...482..631E,
2008A&A...488..667K,
2008MNRAS.389..869E,
2009A&A...507..251C,
2009A&A...507..487M,
2010A&A...510A..99K,
2010AJ....139.2308F,
2010ApJ...708.1366V,
2010ApJ...710.1142B,
2010ApJ...711.1087K,
2010ApJ...717.1279W,
2010ApJ...719..890R,
2010ApJS..190....1R,
2010MNRAS.404.1952B,
2011A&A...534A..58P,
2011AJ....141...21W,
2011AJ....142...57G,
2011ApJ...727..103T,
2012A&A...540A.131V,
2012ApJ...751L..16A,
2012ApJ...758...57L,
2012ApJ...760...55B,
2013A&A...549A.109B,
2013A&A...553A...8D,
2013A&A...556A.126A,
2013AJ....145...84W,
2013AJ....146..154M,
2013ApJ...767L...1L,
2013MNRAS.429..859J,
2014AcA....64..359T,
2014AJ....147...26D,
2014AN....335..817P,
2014ApJ...781..103K,
2014ApJ...791..114W,
2014ApJ...794...51H,
2014MNRAS.441.1545T,
2014Sci...345..440R,
2015AJ....149..104B,
2015AJ....149..106D,
2015AJ....150...50T,
2015ApJ...800...22F,
2019MNRAS.490.5002F,
2020ApJS..246...11F,
2015ApJ...800...95L,
2015ApJ...813..106B,
2015MNRAS.449.2618W,
2015MNRAS.449.3160R,
2015MNRAS.453.1439J,
2016AJ....151..169R,
2016AJ....152..141B,
2016MNRAS.459.3551B,
2017A&A...602A..16T,
2017A&A...602A..88A,
2017A&A...605A..92S,
2017A&A...605L..11A,
2017AJ....154..200M,
2017NatAs...1E..56G,
2017Sci...356.1046S,
2018A&A...609A.117T,
2018A&A...613A..25B,
2018A&A...613A..26S,
2018A&A...614A..76J,
2018A&A...615A.172M,
2018A&A...617A.104P,
2018A&A...618A.115K,
2018AJ....155..265H,
2018ApJ...856...39C,
2019A&A...624A.123P,
2019A&A...625A..17D,
2019A&A...627A..49Z,
2019A&A...627A.116L,
2019A&A...628A..39L,
2019AJ....157...33M,
2019AJ....157..216W,
2019AJ....158..152W,
2019AJ....158..167T,
2019MNRAS.482.4096D,
2019MNRAS.487..268J,
2019Sci...365.1441M,
2020A&A...637A..93G,
2020A&A...640A..50B,
2020A&A...641A.113M,
2020A&A...643A.112S,
2020AJ....160..273L,
2020Sci...368.1477J,
2021AJ....161...10B,
2020MNRAS.493..536D,
2020Natur.582..497P,
2021arXiv210202233M,
2021Sci...371.1038T,
2021arXiv210310216P}.

\paragraph{Other in {\tt COMMENT}.}
\cite{
1976ApJS...30..273A,
1987ApJ...317..343M,
1999A&A...344..897D,
2010PASP..122.1195T,
2012ApJ...754...44J,
2012AJ....143...42H,
2012AJ....144...64D,
2014MNRAS.443L..89A,
2015ApJS..216....7B,
2015ApJ...814...12V,
2016ApJ...821...74J, 
2016MNRAS.463.1592C,
2017ApJS..231...15D,
2017A&A...597A..47C,
2017A&A...605A.103F,
2017AJ....154..135F,
2020ApJS..250...29F,
2018AJ....155..215M,
2018Natur.563..365R,
2018AJ....155..125W,
2018MNRAS.480.2411M,
2019arXiv190604644T,
2019ESS.....410203D,
2020SciA....6.7467D,
2020A&A...641A..23P,
2020AJ....159..139L,
2020arXiv201213238M}.

\section{Illustration}


   \begin{figure*}[h]
   \label{fig:map}
   \centering
   \includegraphics[width=\textwidth,clip=]{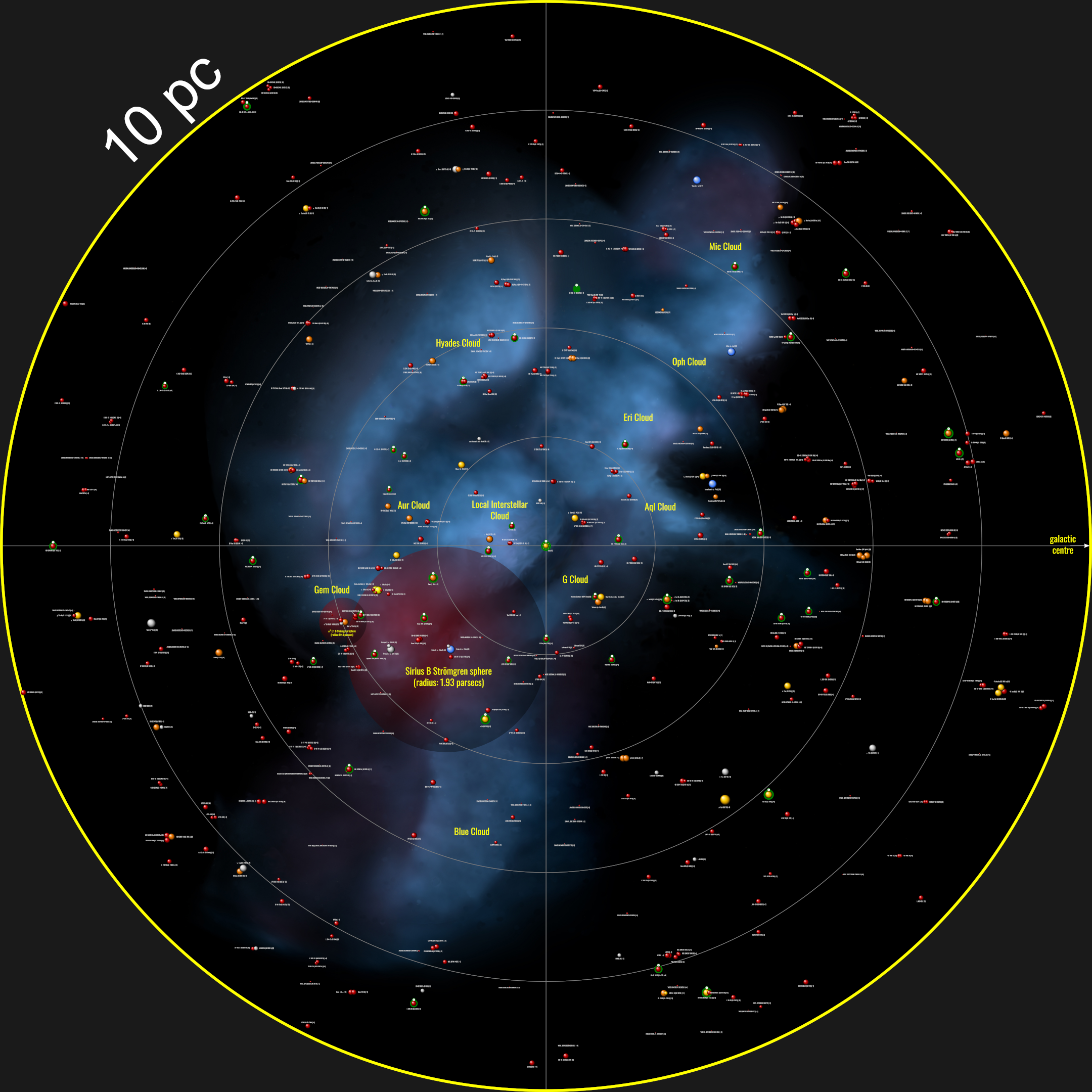}
   \caption{Orthographic projection from above the galactic plane. Guide circles are shown every two parsecs. Distance above or below the galactic plane, in pc, are given in square brackets after the star label. Green circles show the number of confirmed planets. A higher resolution, zoomable map is available at \url{https://gruze.org/galaxymap/10pc/}.}
              \label{fig:map}%
    \end{figure*}

%
\end{appendix}

\end{document}